\documentclass[11pt]{article} 
\usepackage{hyperref} 
\usepackage{graphicx} 
\usepackage{graphics} 
\usepackage{dsfont} 
\usepackage{epsfig} 
\usepackage{amsmath,amssymb,amsthm,amscd} 
\usepackage[parfill]{parskip}

\newcommand{\nn}{{\nonumber}} 
\newcommand{\Tr}{\textrm{Tr}}

\def\Mgn[#1]#2{{\overline{\cal M}_{#1,#2}}} 
\newcommand{\ba}{\begin{eqnarray*}} 
\newcommand{\ea}{\end{eqnarray*}} 
\newcommand{\ban}{\begin{eqnarray}} 
\newcommand{\ean}{\end{eqnarray}} 
\newcommand{\be}{\begin{equation}} 
\newcommand{\ee}{\end{equation}} 
\newcommand{\ben}{\begin{equation}} 
\newcommand{\een}{\end{equation}} 
\numberwithin{equation}{section}

\newcommand{\IZ}{\mathbb{Z}} 
\newcommand{\IC}{\mathbb{C}} 
\newcommand{\IP}{\mathbb{P}}

\newcommand{\del}{\partial}

\newcommand{\Aut}{{\rm Aut}}

\newcommand{\re}{{\rm Re \,}}

\newcommand{\Spin}{{\rm Spin}} 
\newcommand{\Res}{{\rm Res}\,}

\newcommand{\cO}{{\cal O}} 
\newcommand{\K}{{\cal K}} 
\newcommand{\M}{{\cal M}} 
 
\newcommand{\tm}{\tilde{m}} 
\newcommand{\tu}{\tilde{u}} 
 
\newcommand{\PN}{P_N} 
\newcommand{\q}{\tilde{q}}

\newcommand{\aaa}{e_1(\tau_{uv})} 
\newcommand{\bbb}{e_2(\tau_{uv})} 
\newcommand{\ccc}{e_3(\tau_{uv})} 
\newcommand{\AAA}{e_1(\tau)} 
\newcommand{\BBB}{e_2(\tau)} 
\newcommand{\CCC}{e_3(\tau)}

\begin{document} 
\begin{titlepage} 
 
\begin{flushright} 
\parbox[t]{1.8in}{IPMU10-0162\\  
BONN-TH-2011-13\\ 
LPTENS 11/38} 
\end{flushright}

\vskip 2.5 cm 
\centerline{\Large \bf The $\Omega$ deformed B-model for rigid $N=2$ theories}     
\vskip 0.5 cm 
\renewcommand{\thefootnote}{\fnsymbol{footnote}} 
\vskip 20pt \centerline{  
{\large \rm Min-xin Huang\footnote{minxin.huang@ipmu.jp}, Amir-Kian Kashani-Poor\footnote{kashani@lpt.ens.fr} 
  and Albrecht Klemm\footnote{aklemm@th.physik.uni-bonn.de}  } } \vskip .5cm \vskip 20pt 
 
\begin{center} 
{\em  $^*$ Institute for the Physics and Mathematics of the Universe, 
  \\[.1 mm]  University of Tokyo, Kashiwa, Chiba 277-8582, Japan}\\ [3mm] 
{\em  $^\dagger$ Laboratoire de Physique Th\'eorique\footnote{Unit\'e Mixte du CNRS et de l'\'Ecole Normale Sup\'erieure associ\'ee \`a l'Universit\'e Pierre et Marie Curie 6, UMR 8549}, \'Ecole Normale Sup\'erieure, 
  \\[.1 mm] 24 rue Lhomond, 75231 Paris Cedex 05, France    }\\ [3mm] 
{\em  $^\ddagger$ Bethe Center for Theoretical Physics, Physikalisches 
  Institut \\ [.1 cm] 
Universit\"at Bonn, Nussallee 12, 53115 Bonn, Germany} 
\end{center}

\setcounter{footnote}{0} 
\renewcommand{\thefootnote}{\arabic{footnote}} 
\vskip 60pt 
\begin{abstract} 
We give an interpretation of the $\Omega$ deformed $B$-model  
that leads naturally to the generalized holomorphic anomaly  
equations. Direct integration of the latter calculates  
topological amplitudes of four dimensional rigid $N=2$  
theories explicitly in general $\Omega$-backgrounds in  
terms of modular forms. These amplitudes encode the refined BPS spectrum  
as well as new gravitational couplings in the effective action  
of $N=2$ supersymmetric theories. The rigid $N=2$ field theories we focus on are the conformal rank  
one $N=2$ Seiberg-Witten theories.  The failure of holomorphicity  
is milder in the conformal cases, but fixing the holomorphic ambiguity is only possible upon mass deformation. Our formalism  
applies irrespectively of whether a Lagrangian  
formulation exists. In the class of rigid $N=2$ theories arising from compactifications on local Calabi-Yau manifolds, we consider the theory of local $\mathbb{P}^2$.  
We calculate motivic Donaldson-Thomas invariants for this geometry and make  
predictions for generalized Gromov-Witten invariants at  
the orbifold point.  
\end{abstract} 
 
\end{titlepage} 
\vfill \eject

 
\newpage

\baselineskip=16pt 
 
\tableofcontents

\section{Introduction} 
\label{introduction}  
Theories with $N=2$ rigid supersymmetry provide examples in  
which non-perturbative properties of 4d quantum theories 
can be studied exactly. Their topological sector  
describes the exact low energy gauge coupling, the masses and the  
stability properties of BPS states. This data is encoded  
geometrically, typically by a Seiberg-Witten curve  
${\cal C}_g$ and the Seiberg-Witten meromorphic  
differential $\lambda$~\cite{SW1,SW2}, which makes  
it easily extractable and has led to the discovery of  
many new phenomena, e.g. 4d QFTs with no Lagrangian description.  
Irrespective of the existence of such a description, the  
geometrical data of $N=2$ rigid supersymmetric QFTs can be  
constructed from non-compact Calabi-Yau manifolds $M$,  
where the role of $\lambda$ is played by the Calabi-Yau  
$(n,0)$-form $\Omega$. If ${\cal C}_g$ exists, it can be  
derived from $M$, and $\lambda$ obtained from $\Omega$. A non-compact CY manifold is a non-compact K\"ahler manifold, with a non-vanishing holomorphic $(n,0)$-form  
which is sufficiently regular at infinity.     
For $n=3$, this geometric engineering approach relates the  
topological sector of the Type II string on $M$ with the one  
of a 4d gauge theory. The genus one sector determines the gauge  
coupling and the properties of the BPS spectrum, while  
the moduli dependent coefficients of the higher genus  
expansion of the topological string partition function  
compute exact gravitational couplings of the 4d field theory.  
 
Recently, much attention has been devoted to a refinement of the genus expansion which takes the form  
\begin{equation} 
\log Z(t,m,\epsilon_1,\epsilon_2)=  \sum_{n,g=0}^\infty (\epsilon_1+\epsilon_2)^{n} 
(\epsilon_1\epsilon_2)^{g-1}F^{(\frac{n}{2},g)}(t,m) \,. 
\label{expansion1} 
\end{equation} 
This expression appeared first as an equivariant instanton partition function of $N=2$ $SU(N)$ gauge  
theories in~\cite{Nekrasov,NO} and was generalized to arbitrary gauge groups in~\cite{MR2199008,MR2214246}.  
The parameter $t$ collectively denotes flat coordinates  on the vector multiplet moduli 
space, $m$ the bare hypermultiplet masses, and $\epsilon_1, \epsilon_2$ are the equivariant  
rotation parameters acting on the four dimensional spacetime in the so-called 
$\Omega$-background~\cite{Nekrasov,Nekrasov:2011bc}. Note that $\epsilon_1, \epsilon_2$  
have mass dimension~1. Denoting  
$\epsilon_1 \epsilon_2=g_s^2$ and $s=(\epsilon_1+\epsilon_2)^2$,  
one can think of the $\Omega$ deformation as opening up a new direction in the parameter space of the theory, parametrized by $s$. We argue  
in this paper that this direction corresponds  
to insertion of a field $\phi$ in the topological $B$-model. The generalization of the  
holomorphic anomaly equations of~\cite{BCOV} proposed in~\cite{HK3} follows from this interpretation. We develop the direct integration 
approach~\cite{Yamaguchi:2004bt,Huang:2006hq,GKMW,AL,HK2,HK3} to the B-model further to explicitly calculate 
(\ref{expansion1}) in terms of  modular functions. The formalism applies to all $N=2$ 
rigid theories. We solve  a selection of such theories, comprising both 
non-conformal and conformal QFTs, including an example without an effective action  
description, as well as a topological string theory on a non-compact Calabi-Yau manifold. 
 
The spacetime interpretation of the amplitudes $F^{(n,g)}$ occurring in  
(\ref{expansion1}) is that they compute gravitational couplings beyond  
the graviphoton-curvature coupling captured by the conventional  
$F^{(0,g)}$~\cite{Antoniadis:2010iq,Nakayama:2011be,Antoniadis:2011hq}.  
A fluxbrane realization of the $\Omega$-deformation has been given  
in~\cite{Hellerman:2011mv}.   
 
In the A-model, (\ref{expansion1}) can be interpreted as a  
supersymmetric index which counts refined BPS numbers for $D2/D0$ bound  
states on $M$ associated to $D2$ branes with charge $\beta\in H^{comp}_2(M,\mathbb{Z})$  
labelled by the powers of $e^t$~\cite{IKV,Awata}.  
The spacetime spin quantum numbers $Spin(4)=SU(2)_+\times SU(2)_-$ are encoded in the powers of  
$\epsilon_\pm=\frac{1}{2}(\epsilon_1 \pm \epsilon_2)$. Wall crossing properties  
of these invariants, which are related to motivic Donaldson-Thomas invariants,  
have been studied e.g. in~\cite{MR2403807,Dimofte:2009bv}. Supersymmetry is  
not compatible with the $\epsilon_+$ rotation symmetry unless it is  
twisted with an extra $U(1)_R$-symmetry~\cite{Nekrasov,Nekrasov:2011bc},  
which should be realized as an isometry of $M$~\cite{ACDKV}. Since non-compact  
Calabi-Yau manifolds, as opposed to compact ones, can have such an isometry, it  
is on these spaces that the index can be defined. By relating the refined BPS  
numbers $N^\beta_{j_-j_+}$ to modular functions, our formalism  
allows us to calculate them efficiently, see section~\ref{DT}. 
 
(\ref{expansion1}) has interesting limits. In the $\epsilon_1=-\epsilon_2=i g_s$ limit, only the  
$F^{(0,g)}$'s contribute to (\ref{expansion1}), reproducing  
the ordinary genus expansion of the topological string. The $\epsilon_1=0$, $\epsilon_2=\hbar\neq 0$ limit collapses  
the genus expansion to the $g=0$ sector with, in the interpretation developed in this paper, $n$ insertions of a field $\phi$.  
This is called the Nekrasov-Shatashvili limit. At $\epsilon_1=0$,  
the 4d super-Poincar\'e algebra is only partially broken and  
dimensional reduction leads to a two dimensional theory 
with a 2d super-Poincar\'e algebra. Living at genus zero, $W(t,m,\hbar)=\lim_{\epsilon_1\rightarrow 0}  
\epsilon_1 \log(Z(t,m,\epsilon_1,\epsilon_2))$ is readily computed in the present formalism and calculates the twisted  
superpotential of the 2d theory. It satisfies  
$\frac{1}{2 \pi i}\partial_{t_i} W(t,m,\hbar)= n_i$ with $n_i\in \mathbb{Z}$  
and is identified with the Yang-Yang function of a quantum  
integrable system~\cite{Nekrasov:2011bc,NS,Chen:2011sj}. Relations to  
non-commutative Riemann surfaces have been studied in~\cite{Fucito:2011pn,ACDKV}. 
Properties of the NS limit will be investigated in this paper  
using relations beween modular functions to all orders  in $t$ 
in section \ref{NSLimit}.      
 
An explicit link between topological strings and matrix models was  
established in~\cite{remodelling}. Such matrix models possess deformations  
closely related to the $\epsilon$ deformations studied here, taking the  
form $Z=\frac{1}{N! (2\pi)^N}\int d^N\lambda [{\cal D}(\lambda)]^{- \frac{\epsilon_1}{\epsilon_2}} e^{-{1\over \hbar} \sum_{i} V(\lambda_i)}$,  
e.g  with measure ${\cal D}(\lambda)=\prod_{i<j}(\lambda_i-\lambda_j)^2$ for  
the hermitian case~\cite{Dijkgraaf:2009pc}. These refined $N$-matrix  
ensembles reduce to familiar ensembles in certain limits:  
$\epsilon_1=-\frac{1}{2} \epsilon_2$ corresponds to the standard orthogonal  
ensemble, and $\epsilon_1=-2 \epsilon_2$  to the symplectic  
ensemble~\cite{Chekhov:2006rq}. While the remodelling description  
has not yet been extended to a refined version of the B-model (see  
\cite{Brini:2010fc} for a discussion), the compatibility of  
the refined recursion and the generalized holomorphic anomaly  
equations we will present and study in this paper has already  
been established in the case of hyperelliptic curves $y^2 = p(x)$ when 
the Seiberg-Witten differential is $ydx$~\cite{workinprogress2}. Given  
matrix expressions for sums over partitions, it is possible to write Nekrasov's partition  
function as a matrix integral~\cite{Sulkowski:2009ne}. Relations  
to integrable systems have been found~\cite{Bonelli:2011na}.

As a final application of (\ref{expansion1}), we cite the Alday, Gaiotto,  
Tachikawa correspondence~\cite{AGT} for $SU(2)$ quivers, which relates gauge theory at $(\epsilon_1,\epsilon_2)$ to Liouville conformal field theory at central charge $c= 1+ 6 Q^2$, $Q= b+1/b$,  with parameter $b = \sqrt{\epsilon_1/\epsilon_2}$. Four dimensional $SU(N)$ quiver theories are expected to arise as the worldvolume theory of $N$  
M5 branes on a Riemann surface $\Sigma_g$. It has been argued that for such theories, a correspondence exists to 2d $A_{N-1}$ Toda  
theory~\cite{Wyllard:2009hg}. However, in this 4d/2d  
correspondence, both sides are hard to compute. Already in  
$SU(3)$ quiver theories, the conformal theory with $E_6$  flavor  
symmetry~\cite{Minahan:1996cj} occurs as a building block in  
the pant decomposition of $\Sigma_g$, just as the $SU(2)$ $N_f=4$  
theories served as building blocks for $SU(2)$ quivers~\cite{gaiotto}. 
The $E_n$ conformal theories are associated by geometric  
engineering to local del Pezzo surfaces. Since they do not admit a  
Lagrangian description, they are currently inaccessible by any other means.  
We solve a two parameter deformation of the $E_6$-theory in  
section~\ref{E_n}.   
 
Regarding 4d gauge theories, the mass-deformed  
conformal theories are the most interesting and challenging  
cases, and will therefore be the main focus of the gauge theory portion of this paper.  
Gauge theories with fundamental matter become conformal when the number of flavors $N_f$ equals  
twice the rank of the gauge group. The asymptotically free theories with  
smaller $N_f$ can be reproduced from the conformal theory by taking multi-scaling limits in which the masses  
and the inverse coupling are sent to infinity,  
holding combinations of these quantities fixed;  e.g. the $\tau\rightarrow i\infty, m_4\rightarrow  
\infty$ limit with $\lim_{\tau\rightarrow i\infty,m_4\rightarrow  
\infty} e^{2 \pi i\tau} m_4=\Lambda_3$ 
describes the flow from the conformal $N_f=4$ $SU(2)$  
Seiberg-Witten theory to the asymptotically free $N_f=3$ theory,  
with $\Lambda_3$ the QCD scale of the latter. Deriving the Seiberg-Witten curves for the mass-deformed conformal  
theories is more challenging than for the asymptotically free  
theories, as the curve depends on an additional dimensionless  
parameter, the UV coupling. The definition of this coupling is  
ambiguous, and we re-discuss the curves proposed in~\cite{SW2}  
with regard to this ambiguity. The conformal limits of these theories at vanishing mass deformation prove to have several peculiarities.   
Technically most far reaching is the fact that the gap conditions, which are necessary boundary  
conditions to solve the holomorphic anomaly equations by fixing the holomorphic ambiguity,  
rely on the non-conformal light spectrum at nodal singularities of the Seiberg-Witten curve, and are trivial  
in the case of conformal theories. We are hence required to mass-deform the conformal  
theories to apply our techniques of direct integration, and then recover the conformal amplitudes in the massless limit. Interestingly, this leads to different results for  
the $N=4$ and $N_f=4$ theory, even though they have the same Seiberg-Witten curve and  
(up to a normalization) differential. In the Nekrasov-Shatashvili limit, however, i.e. at genus 0, the amplitudes of the two theories are related by a simple rescaling of the $a$ parameter.

While the UV parameters for the gauge coupling as well as the  
mass parameters depend on various choices, some of them have  
been unfortunate in the literature as they unnecessarily break the  
underlying symmetries. 
 
We find a systematics in the  
breaking of holomorphicity in the infrared, summarized  
in Table \ref{TabX} below.    
 
\begin{table}[h!] 
\centering{{ 
\begin{tabular}[h]{|c|c|l|c|l|} 
\hline  
Ampl.&n& Theory&max. power anhol. Gen.& Recursion\\   
$F^{(n,g)}$&$\phi$&$\Omega$-background: generic& $X^{3g+2n-3}$& (\ref{gen_hol_ano})  \\  
$F^{(n,g)}$&$\phi$&$\Omega$-background: conformal&$X^{g+n-1}$& (\ref{gen_hol_ano})$\rightarrow$(\ref{gen_hol_ano-cf2})  \\ 
$F^{(g)}$&$-$&Top. String: generic \cite{Huang:2006hq,GKMW}&$X^{3g-3}$& BCOV  \\ 
$F^{(g)}$&$-$&Top. String: Enriques-CY \cite{GKMW} &$X_{\rm basis}^{g-1},X_{\rm fibre}^{2g-1}$& BCOV  \\ 
$F^{(n,g)}$&$\phi_b$&Top. String: ell. del Pezzo \cite{Hosono:2002xj} &$X^{n+g-1}$& (1.4) in \cite{Hosono:2002xj}  \\ 
$F^{(n,0)}$&$\phi_m$&$m$ deform. $N=2^*$ \cite{MMW} &$X^{n-1}$& (2.27) in \cite{MMW}\\ 
$F^{(g)}$&$-$& Hurwitz \# on $T^2$, 2d QCD \cite{Dijkgraaf} &$X^{3g-3}$& unknown\\ 
\hline 
\end{tabular}}} 
\caption{Breaking of holomorphicity (modularity) by the almost holomorphic  
(quasimodular) generator $X$, for the amplitudes $F^{(n,g)}$ of various physical  
theories. The insertions of the operator $\phi_*$ are counted by $n$. In the case  
relevant for the $\Omega$-deformation, $\phi_*$ is the operator $\phi$ discussed in  
sect~\ref{interpretationofphi}. For the theory discussed in~\cite{Hosono:2002xj},  
$\phi_b$ corresponds to the modulus of the base of the elliptic fibered del Pezzo.  
This is closely related to the insertions of the mass operator $\phi_m$  treated in \cite{MMW}. } 
\label{TabX} 
\end{table} 
 
Beside $SU(2)$ $N=2$ gauge theory with $N_f=4$ massive flavors and $D_4$ flavor symmetry, the more exotic  
theory with $E_6$ flavor symmetry and the  $N=2^*$ theory, we also study conformal limits of  
the massive $N_f=1,2,3$ theories with $A_0,A_1$ and $A_2$ flavor symmetry.    
We explain in general why the leading power in the anholomorphic generator grows  
more slowly in the conformal as compared to the non-conformal theories. In this sense,  
the breaking of holomorphicity is weaker in conformal theories.

\section{The holomorphic anomaly equations}  
\label{holomorphicanomaly} 
In this section, we discuss and generalize the holomorphic anomaly  
equations, which where first derived from the worldsheet point  
of view in~\cite{BCOV}. They were  
interpreted as describing the transformation of a wave function under symplectic  
basis changes in $H_3(M,\mathbb{R})$ in~\cite{Witten:1993ed}, with the partition function $Z=e^F$ playing the role of the wave function. The relation between the holomorphic  
anomaly and target space modularity was developed in~\cite{Yamaguchi:2004bt,ABK}  
and led, upon imposing suitable boundary conditions, to the direct  
integration method~\cite{Huang:2006hq,GKMW,HK2,HK3}. 
 
\subsection{The BCOV holomorphic anomaly equations}   
 
Naively, BRST invariance guarantees that the topological string partition function is a holomorphic  
function on the moduli space of the theory. The argument for the B-model  
is the following: the partition function is an integral over the compactified  
moduli space $\overline{\cal M}_g$ of genus $g$ Riemann surfaces $\Sigma_g$, 
\begin{equation} 
F^{g}(t)= \int_{{\overline{{\cal M}}_g}} {\left\langle \prod_{k=1}^{3 g-3} \beta^k {\bar \beta}^{k}\right\rangle}_g\cdot 
[d m\wedge d \bar m] \,, 
\end{equation} 
where $\beta^k=\int_{\Sigma_g} G^{-} \mu^k$, $\bar \beta^k=\int_{\Sigma_g} \bar G^{-} \bar \mu^k $  
contain the worldsheet Beltrami differentials $\mu^k\in H^1(T\Sigma_g)$ with corresponding  
deformation coordinate $m_k$ and the worldsheet supersymmetry generators $G^-, {\bar G}^-$. The  
contraction of the $m_k,\bar m_k$ with the genus $g$ worldsheet correlator gives a  
real $6g -6$ form on $\overline{{\cal M}}_g$.  
Derivatives with regard to anti-holomorphic variables (on which the action underlying the  
expectation value $\langle \cdot \rangle_g$ depends) lead to the insertion of BRST trivial  
operators, thus suggesting the vanishing of the expectation value. As noted in~\cite{BCOV1, BCOV},  
this argument fails due to contributions from the boundary of moduli space: BRST trivial  
operators correspond to exact forms, and the integral over these receives contributions  
from the boundary of the integration domain.  
 
Rather than being a nuisance, the {\it holomorphic anomaly}, as the anti-holomorphic  
dependence of the partition function was christened in~\cite{BCOV1}, gives rise to  
a recursion relation between the topological string amplitudes at different genera, 
\ban   \label{hae} 
\bar{\partial}_{\bar{\imath i}} F^{g}= \frac{1}{2}\bar{C}_{\bar{\imath }}^{jk}\big{(}D_jD_kF^{g-1} 
+\sum_{h=1}^{g-1}  D_jF^{h}D_kF^{g-h}\big{)} \,,  \quad g>1 \,, 
\ean 
where $\bar C^{ij}_{{\bar \imath}}=e^{2{\cal K}}  G^{j\bar \jmath} G^{k\bar k} C_{\bar \imath\bar \jmath \bar k}$  contains the K\"ahler potential  
${\cal K}$, the metric $G^{j\bar \jmath}$, and the complex conjugate of the-three point function  
$C_{ijk}$. The latter data is determined at genus zero and related by special  
geometry. The recursion  begins with $F^{1}$, which satisfies its own holomorphic  
anomaly equation in terms of special geometry data (see next section), 
\be 
\label{f1anomaly} 
\del_i \bar{\del}_{\bar{j}} F^1 = \frac{1}{2} C_{ijk} C^{jk}_{\bar{j}} - \frac{\chi-1}{24} G_{i \bar{j}}\,. 
\ee 
These equations, fortified with modularity and appropriate boundary conditions, can be used to integrate the  
topological string partition function \cite{Huang:2006hq,GKMW,HK2,Haghighat:2008gw}. The origin of the two terms on the right  
hand side of (\ref{hae}) is easily recognizable: the worldsheet degenerates at the boundary of moduli space.  
Pictorially, this corresponds to cycles of the Riemann surface pinching. If the pinched cycle does not sever  
the Riemann surface in two, we are left with a Riemann surface with genus reduced by 1. This is the  
origin of the first term in (\ref{hae}). If, by contrast, the pinched surface becomes disconnected,  
leaving two surface components of genus $h$ and $g-h$ respectively, this gives rise to the second set of  
terms in the above equation. 
 
\subsection{The generalized holomorphic anomaly equations}  
 
To obtain a similar sets of equations governing the behavior  
of the partition function in an $\Omega$-background, we would  
like to argue that in the expansion of the free energy given in~(\ref{expansion1}), upon setting $s=(\epsilon_1+\epsilon_2)^2$ and $g_s^2=\epsilon_1\epsilon_2$, 
\begin{equation} 
F(s,t,g_s)= \log Z=\sum_{n,g=0}^\infty s^{\frac{n}{2}} g_s^{2g -2} F^{(\frac{n}{2},g)}(t)\ , 
\label{expansion2} 
\end{equation} 
the power of $s$ counts the number of insertions of an operator $\cO$ in the genus $g$ amplitude $F^{(\frac{n}{2},g)}(t)$. For this interpretation to be possible, $n\in 2 \mathbb{Z}$ must hold. This condition also follows from the interpretation of the amplitudes as generating functions for the BPS degeneracies $N^\beta_{j_-j_+}$, as a Schwinger loop calculation (see (\ref{schwingerloope1e2}) in section \ref{fixingtheambiguity}) implies that $F$ is even under simultaneous sign flip of $\epsilon_1$ and $\epsilon_2$~\cite{HK3}. Naively,  $n\in 2 \mathbb{Z}$ is incompatible with the Nekrasov expansion for  
certain gauge theories with flavor. However, it was pointed out  
in~\cite{Shadchin, GNY, OP,IKV,KW2} that the masses $\hat m_i$ in the  
Nekrasov expansion should be redefined in terms of physical masses   
\begin{equation}  
\label{shift} 
\hat m_i = m_i + \frac{\epsilon_1+\epsilon_2}{2} \,. 
\end{equation} 
This redefinition eliminates the odd terms in the  
expansion\footnote{More precisely, this is true for all  
$F^{(\frac{i}{2},j)}$ with odd $i$ save $F^{(\frac{1}{2},0)}$.  
For this latter case, the instanton part vanishes for  $N_f=1,2$, is  
a constant independent of the flat coordinate $a$ for $N_f=3$,  
and for the $N_f=4$ case we observe  
\begin{eqnarray} 
F^{(\frac{1}{2},0)}_{\textrm{instanton}} = \frac{1}{2}\log(1-q) \sum_{i=1}^4m_i \,. 
\end{eqnarray} 
These expressions however do not enter the integration of the holomorphic  
anomaly equations.}. Moreover, it is the amplitudes $F^{(n,g)}$ defined upon shifting the masses that have natural modular properties (natural in the sense that 
assigning complicated transformations to the masses is not necessary). The physical  
masses $m_i$ are also the ones featuring in the AGT correspondence.    
 
In~\cite{KW1} a different generalization of the holomorphic  
anomaly equations referred to as {\it extended} holomorphic  
anomaly equations was proposed, which does generate odd terms and  
reproduces the Nekrasov partition function in terms of the  
$\hat m_i$. It involves the so-called Griffiths infinitesimal  
invariant which appears in open topological string theory.  
However, since the latter has no easy modular interpretation, it is not naturally incorporated in the direct integration formalism,  
which relies on almost holomorphic generators in the real polarisation  
with a non-holomorphic modular completion in the holomorphic  
polarisation~\cite{ABK}. 
 
Without a string theory definition of the partition function (\ref{expansion1}), one cannot follow the same route as~\cite{BCOV}  
to derive holomorphic anomaly equations for the amplitudes $F^{(n,g)}$. Instead, in \cite{HK3}, a very simple generalization of the holomorphic anomaly  
equations was conjectured, then checked by computing the partition function of the topological string on local  
Calabi-Yau threefolds as well of  
asymptotically free massless Seiberg-Witten gauge theories, which arise in a field theory limit of compactifications  
on certain local Calabi-Yau geometries \cite{KKV}. These generalized holomorphic anomaly equations take the form 
\begin{eqnarray} \label{gen_hol_ano} 
\bar{\partial}_{\bar{i}} F^{(n,g)}= \frac{1}{2}\bar{C}_{\bar{i}}^{jk}\big{(}D_jD_kF^{(n,g-1)} 
+{\sum_{m,h} }^{\prime}  D_jF^{(m,h)}D_kF^{(n-m,g-h)}\big{)} \,, \quad n+g>1\,, 
\end{eqnarray} 
where the prime denotes omission of $(m,h)=(0,0)$ and $(m,h)=(n,g)$ in the sum. The covariant derivatives will be explained in the next section. The first  
term on the right hand side is set to zero if $g=0$.  
 
The equations (\ref{gen_hol_ano}) have passed checks in a variety of physical systems in the existing literature:  
topological string theory on the non-compact Calabi-Yau spaces ${\cal O}(-3)\rightarrow  
\mathbb{P}^2$ and  ${\cal O}(-2,-2)\rightarrow \mathbb{P}^1\times \mathbb{P}^1$~\cite{HK3},  
matrix models with the Eynard-Orantin recursion~\cite{workinprogress2},  
and $SU(2)$ Seiberg-Witten theory with $N_f=0,2$~\cite{KW1,HK3}. In the upcoming sections, we will further extend this list, and provide various checks of our results. Here,  
assuming that the conjecture is correct, we would like to extract lessons regarding  
the form an underlying microscopic worldsheet description of the $\Omega$-background  
must take. 
 
The generalized equations are reminiscent of the holomorphic anomaly equations derived in \cite{BCOV} for correlators. Indeed,  consider a genus $g$ amplitude with $n$ field insertions  
\begin{equation} 
F^{(n,g)}(t)= \int_{{\overline{{\cal M}_g}}} {\left\langle {\cal O}^n \prod_{k=1}^{3 g-3} \beta^k {\bar \beta}^{k}\right\rangle}_g\cdot 
[d m\wedge d \bar m] \,. 
\end{equation} 
To preserve conformal invariance, the operator ${\cal O}$ should take the form of a  $2$-form field integrated over the Riemann surface  
${\cal O}=\int_{\Sigma_g} \phi^{(2)}$, where $\phi^{(2)}$ emerges as usual  
by the descent equations  from a $0$-form field $\phi^{(0)}$. These insertions must  
correspond to the appropriate vertex operators inducing the $\Omega$-background deformation  
from the worldsheet point of view. The arguments  
underlying the holomorphic anomaly equations (\ref{hae}) can now be repeated with  
$\overline{\cal{M}}_g$ replaced by the moduli space of punctured Riemann  
surfaces, $\overline{\cal{M}}_{g,n}$. The pinchings that disconnect the  
worldsheet must here be distinguished by how the punctures are  
divided among the two resulting surface components. This gives  
rise to the second sum on the right hand side of (\ref{gen_hol_ano}).  
Note that the covariant derivatives are not modified, indicating that contact terms between the  
operator $\phi^{(0)}$ and the marginal moduli field operators $\phi^{(0)}_i$  
should not exist. Other boundary components of $\overline{\cal{M}}_{g,n}$  
can contribute due to short distant singularities  
$\phi^{(0)}(z){\bar \phi}^{(0)}_{\bar \imath}(w) \sim   
\frac{G_{0\bar \imath}}{|z-w|^2}$ as $z\rightarrow w$, where the index $0$ in the two-point correlator $G$ labels the operator $\phi^{(0)}$. As no such contributions arise in (\ref{gen_hol_ano}), we must also require $G_{0\bar \imath}=0$ as a condition on $\phi^{(0)}$. 
 
For the case $g=0$, we conjecture 
\begin{equation} \label{haeg0} 
F^{(n+1,0)}=\langle \phi^{(0)}(0)\phi^{(0)}(1)\phi^{(0)}(\infty)  
{\cal O}^{n}\rangle_{g=0}\  . 
\end{equation}  
Note in particular that this identification implies that $F^{(1,0)}$ is holomorphic, as the boundary  
of the moduli space $\Mgn[0]{n}$ is due solely to coincident punctures, and 3 points  
can be fixed by $SL(2,\mathbb{C})$ transformations to arbitrary values, as indicated in (\ref{haeg0}).

\subsection{The local limit and a proposal for the insertion $\phi$}  
\label{interpretationofphi} 
A hint towards the nature of the worldsheet insertion ${\phi}$ which induces the $\Omega$-background comes from the observation that this deformation requires taking the local limit of the target space Calabi-Yau manifold. 
 
The theory underlying the topological string is twisted $N=2$ superconformal field theory coupled to topological gravity. Of all the operators in this theory, $U(1)$ charge conservation only permits non-vanishing correlation functions involving the marginal fields of the matter sector and the first gravitational descendant of the puncture operator, the dilaton, in the gravity sector. The coordinate directions on moduli space, denoted by $t_i$ above, correspond to the marginal fields. The dilaton plays an important structural role in the topological string, via its contact terms with the marginal operators: The holomorphic (3,0)-form $\Omega$ on a compact Calabi-Yau manifold $M$ is a section of a line bundle ${\cal L}$ (called the vacuum line bundle in \cite{BCOV}) over the moduli space $\cal M$ of complex structures on $M$. The cohomology class of the K\"ahler form $K$ on this moduli space is the Chern class of the line bundle, hence $K = \frac{i}{2\pi} \partial \bar{\partial} \K$ in terms of the K\"ahler potential ${\K(t,\bar t)}= \log i\int_{M} \Omega \wedge \bar \Omega$. The choice of section induces a metric on the line bundle, with connection $\partial \K$. The physical manifestation of this geometric setup is that holomorphic coordinate transformations on the line bundle ${\cal L}$,  
\begin{equation}  
\Omega \rightarrow e^{-f(t)}\Omega   \,, 
\label{linebundletransformation}   
\end{equation} 
induce K\"ahler gauge transformations on the K\"ahler potential ${\K}$, 
\begin{equation}   
\K(t,t)\rightarrow \K(t,\bar t)-f(t)-\bar f(\bar t)\ . 
\label{kaehlertransformation}  
\end{equation}  
Insertions of operators inducing marginal deformations naively correspond to taking derivatives with regard to the appropriate coordinates. Contact terms of the operators amongst themselves however covariantize this derivative; acting on a correlation function of $k$ operators, one obtains the covariant derivative on the bundle ${\rm Sym}^k T^{(1,0)} {\cal M}$. Contact terms of the insertion with the dilaton operator covariantize the derivative further, with regard to the connection $\partial \K$ on ${\cal L}$ \cite{BCOV}. In total, one obtains 
\begin{equation}  
D_t=\partial_t-(\Gamma_t)^k-(2-2g)\partial_t \K \ . 
\label{fullconnection}  
\end{equation}  
In particular, this reasoning allows one to deduce that the topological string amplitudes $F^g$ on compact Calabi-Yau manifolds are sections of the line bundle ${\cal L}^{2-2g}$. 
 
In the local limit of Calabi-Yau manifolds, the line bundle ${\cal L}$ becomes flat. In this limit, the contact terms between the dilaton and the marginal fields must hence vanish. The dilaton thus becomes a candidate for the field $\phi$. A question we have not addressed in this section is the implication of such insertions for the physical string. The study of this question is under way \cite{work_in_progress3}. 
 
The local Calabi-Yau setting may also offer some insight into the mass shift (\ref{shift}). A general feature of the  
non-compact limit is that the holomorphic (3,0)-form $\Omega$ reduces to a meromorphic form $\lambda$ with a  
distinguished non-vanishing residue. This corresponds to the constant solution of the  
Picard-Fuchs operator ${\cal D}$ (an example of such an operator can be found in section \ref{DT} in equation (\ref{pfp^2})), which we may normalize to  
\begin{equation}  
\int_{\gamma_p} \lambda=\sqrt{s} \,. 
\label{res-s} 
\end{equation} 
The 3-cycle of $M$ which degenerates to $\gamma_p\in H_1({\cal C}_1\setminus \{ p\},\mathbb{Z})$ is the $T^3$ of the Strominger-Yau Zaslow construction, whose period corresponds to $D0$ brane charge at the  large volume point in which the non-compact limit is taken. In the  
geometric engineering limit with bare hypermultiplet masses, other  
residua $\int_{\gamma_i} \lambda=m_i$ appear as limits of $A$-cycles  
in a symplectic basis of $H_3(M,\mathbb{Z})$ whose associated K\"ahler moduli  
$t_i$ become non-dynamical in the field theory limit~\cite{KKV}. From this point of view,  
(\ref{shift}) is just a linear transformation on $H_1({\cal C}_1 
\setminus \{ p,p_i\},\mathbb{Z})$. Since $\gamma_p$ in (\ref{res-s})  
is singled out in the construction, one has a canonical choice of residue to identify with $\sqrt{s}$.  
In order for the action of the symplectic monodromy  
group on $H_3(M,\mathbb{Z})$ to descend naturally to the  
non-compact limit, the classes of $A$-cycles must descend to  
non-intersecting $\gamma_p$ and $\gamma_i$ respectively.  
Up to flavor symmetries, this defines the choice of physical  
masses in the B-model. Starting with a random choice from  
the geometric point of view, redefinitions like (\ref{shift})  
might be necessary in order to make the  
underlying symplectic symmetries of the theory manifest.   
 
Finally, as we will elucidate in section \ref{conf_limit},  
the prepotential of the $\Omega$-undeformed $N=2^*$ theory  
in the treatment of \cite{MNW} exhibits close parallels to  
the $F^{(n,0)}$ amplitudes studied in this paper. The  
deformation in \cite{MNW} is that of $N=4$ theory by  
insertion of a mass operator $\phi_m$, and the  
prepotential in an expansion in the mass can be  
shown to satisfy the same holomorphic anomaly  
equations as $F^{(n,0)}$, see equation (\ref{gen_hol_ano-cf1}).  
The similarity between the two theories is what  
motivated in part the conjecture (\ref{res-s}),  
in particular the square root which mimics the  
relation between the residue $\sim m$ in the $N=2^*$ theory, and the expansion parameter $m^2$ of \cite{MNW}.

\subsection{The wave function transformation of Z}  
\label{wavefunction}  
Following the logic presented above, we can define higher point correlation functions as 
\begin{equation}  
F^{(n,g)}_{i_1,\ldots, i_m} = \int_{{\overline{{\cal M}_g}}}  
{\left\langle {\cal O}^n \prod_{l=1}^m {\cal O}_{i_l} \prod_{k=1}^{3 g-3}  
\beta^k {\bar \beta}^{k}\right\rangle}_g\cdot [d m\wedge d \bar m] \,. 
\end{equation} 
In the following equations, we will assume that the non-compact limit has been taken, such that the operator $\phi$ does not have contact terms with the chiral operators $\phi_{i_k}$. Insertions of $\phi_{i_k}$ hence still correspond to covariant differentiation, also in the presence of $\phi$ insertions, 
\begin{equation}  
\label{correlators} 
F^{(n,g)}_{i_1,\ldots, i_m} =D_{i_1}\ldots  D_{i_m} F^{(n,g)} \,.   
\end{equation} 
As the K\"ahler line bundle becomes trivial  
in the non-compact limit, the covariant derivative is with regard to the bundle $T^{(1,0)}\M$ and its symmetric powers. The correlation functions satisfy a simple generalization of the holomorphic  
anomaly equations of~\cite{BCOV}, 
\begin{eqnarray} 
\frac{\partial}{\partial {{\bar t}_{\bar \imath}}}F^{(n,g)}_{i_1,\ldots, i_m}&=&\frac{1}{2} C^{jk}_{\bar \imath}\left(F^{(n,g-1)}_{i,j,i_1,\ldots, i_m}+\!\!\!\!\!\! 
\sum_{g'+g''=g\atop n'+n''=n,\ m'+m''=m} \!\!\! \!\!\!\frac{1}{m'! m''!} F^{(n',g')}_{i,i_{\sigma(1)},\ldots, i_{\sigma(m')}}F^{(n'',g'')}_{j,i_{\sigma(1)},\ldots, i_{\sigma(m'')}}\right)\nonumber \\ 
&&-(2g-2+m-1) \sum_{r=1}^m G_{\bar \imath i_r}  F^{(n,g)}_{i_1,\ldots,i_{r-1} i_{r+1}\ldots i_m} \ . 
\end{eqnarray} 
Note that in this sum $(m',g')$ run from $(0,0)$ to $(m,g)$ and for $g'=0$ or $g'=g$  either $n>0$ or $m>3$. Defining 
\begin{equation}  
W(g_s,s,x,t,\bar t)= \sum_{n,g=0}^\infty \sum_{m=0}^\infty g_s^{2g-2}  
\frac{1}{m!} F^{(n,g)}_{i_1,\ldots, i_m}  x^{i_1}\dots x^{i_m} s^n \ ,     
\end{equation} 
the holomorphic anomaly equations for the $F^{(n,g)}_{i_1,\ldots, i_n}$  
can be summarized by a heat kernel like equation for $\exp(W)$, 
\begin{equation}  
{\cal D}_{\rm heat}\exp(W)=\left[\frac{\partial}{\partial {{\bar t}_{\bar \imath}}}-\frac{g_s^2}{2} 
\bar C^{jk}_{\bar \imath} \frac{\partial^2}{\partial x^j \partial x^k} - 
G_{\bar \imath j} x^j\left(g_s\frac{\partial }{\partial g_s}+x^k \frac{\partial }{\partial x^k}\right)\right]  
\exp(W)=0 \ .    
\end{equation} 
Applying the heat kernel operator to $\exp(W)$, evaluating at ${\bf x}=0$, and invoking (\ref{correlators}) yields  
\begin{equation}  
\left. { {\cal D}_{\rm heat}\exp(W)}\right|_{{\bf x}=0} 
=\left[\frac{\partial}{\partial {{\bar t}_{\bar \imath}}}-\frac{g_s^2}{2} \bar C^{jk}_{\bar \imath} D_i D_j\right] \Psi=0\ . 
\end{equation} 
As observed in~\cite{Witten:1993ed}, this equation supplemented by the fact  
that $\Psi=\exp(F)$ is a holomorphic function in $\bar t_{\bar \imath}$, considered as a new variable, independent from the $t_{i}$, is equivalent to the infinitesimal wave  
function transformation property of $\Psi=\exp(F)$. In fact, as explained  
in~\cite{Witten:1993ed}, the fact that $ {\cal D}_{\rm heat}$ defines a projectively flat connection on the simply connected space of base points can be used to (projectively) identify all wave functions defined upon such a choice of base point, thereby restoring background independence of the topological string. 
 
We will need the wave function transformation property to extract the generalized orbifold Gromov-Witten invariants of local $\IP^2$ from our modular expressions in section (\ref{orbifoldinvariantsp2}). In particular, the arguments of~\cite{ABK}  that the change from real to holomorphic  
polarisation of the wave function corresponds to the change from  
quasimodular to almost holomorphic functions extend to the refined case, consistent with the modular invariant form of the $F^{(n,g)}$ in terms of  
$\hat E_2$ in holomorphic polarisation.\footnote{This is not true for the results obtained with the holomorphic  
anomaly of~\cite{KW1} as it generally breaks modular invariance.} The holomorphic limit of the counting function ${\cal F}$ in various regions of moduli space is obtained as follows. One uses the wave function transformation to change to the  
symplectic basis appropriate for the definition of local flat coordinates in terms of  
the global symplectic basis of $H_1({\cal C}_g,\mathbb{Z})$. In the new local flat  
coordinates, one uses the wavefunction transformation of $Z$ again to change to the real  
polarisation in which ${\cal F}$ is a holomorphic counting function. 
 
\section{Integrating the holomorphic anomaly equations} 
\label{directintegration} 
In this section, we will discuss the integration of the holomorphic anomaly equations (\ref{gen_hol_ano})  
for rigid $N=2$ theories. The most familiar member of this class are $N=2$ Seiberg-Witten  
gauge theories. The data defining such a theory with gauge group $SU(r+1)$ is a family of Riemann surfaces ${\cal C}_r(u)$  of genus $r$  
parametrized by $u_1,\ldots,u_r$ moduli and a meromorphic (1,0)-differential $\lambda$ of the third kind with the property $\frac{d \lambda}{du_i}=\omega_i$, $i=1,\ldots, r$, with $\{\omega_i\}$ furnishing a basis of the  
holomorphic 1-forms spanning $H^1({\cal C}_r)$.  
 
For ease of exposition, we will use the gauge theory language in this section. The formalism discussed applies however without modification to the  
B-model description of the topological string on local Calabi-Yau geometries, irrespectively of whether the geometry can be reduced to a Riemann surface and the holomorphic (3,0)-form to a meromorphic differential. 
 
We will mainly focus on the gauge group $SU(2)$, and correspondingly on local Calabi-Yau threefolds  
with genus~1 mirror curves, such as for example the total spaces ${\cal O}(-K_S)\rightarrow S$ of the  
anticanonical line bundles over del Pezzo surfaces $S$. The direct integration formalism extends  
also to the higher rank or higher genus case, but concrete computations then require deriving and solving Picard-Fuchs equations, a complication we circumvent in the rank~one~/~genus one case by using well-known  
general formulae for periods of elliptic curves. Aside from $N=2$ gauge theories~\cite{HK1,HK2} and  
topological string theory on non-compact Calabi-Yau geometries \cite{Haghighat:2008gw},  
the original formalism has been applied to matrix models with more than two cuts and polynomial  
potentials~\cite{Klemm:2010tm}. The $F^{(n,g)}$ for the latter theories should also be  
covered by the formalism described below.

\subsection{Rigid special K\"ahler geometry}  
The Coulomb branch or vector multiplet moduli space of $N=2$ supersymmetric Yang-Mills theory is a K\"ahler  
manifold governed by rigid special geometry. In particular, this implies that the K\"ahler metric on this space 
is derived from a holomorphic function $F^{(0,0)}$ called the prepotential. This is the leading 
quantity appearing in (\ref{expansion1}). From the topological string perspective,  
it is determined from the genus zero sector of the theory. The  vector multiplet moduli space is coordinatized by  
the expectation value of the adjoint scalar field  $u=\frac{1}{2}\Tr (\phi^2)$ sitting in the  
$N=2$ vector multiplet. Locally, a flat coordinate $t$ can be introduced on this moduli  
space which plays a crucial role in describing the IR physics.\footnote{The connection vanishes in this coordinate in the holomorphic limit; more on this limit below.} Geometrically, the flat $t$ coordinate is determined in the neighborhood of a singularity by the period of ${\cal C}_1(u)$ with the most  
regular behavior at the singularity. The dual coordinate $t_D$ is geometrically the symplectically dual period. In terms of the prepotential, it is given by\footnote{The different normalizations for theories with and without fundamental matter come about due to the rescaling $n_e \rightarrow 2 n_e$ in the presence of fundamentals, to avoid half integral charges, see \cite{SW1}. While the physics is of course invariant under the choice of normalization of $n_e$, the Seiberg-Witten curve depends sensitively on it, as the monodromy group does. The overall normalization of the prepotential trickles down into the holomorphic anomaly equations, it is coupled to the normalization of the central charge which enters the gap condition. In particular, changing the normalization of the prepotential by rescaling $c_0$, $c_0 \rightarrow k \,c_0$, requires rescaling $t$ in equation (\ref{gap}) by $t \rightarrow \frac{t}{\sqrt{k}}$.}
\be \label{rescaling}
t_D = -\frac{c_0}{2 \pi i}  \frac{\partial F^{(0,0)}}{\partial t}  \quad \mbox{with} \quad
\begin{cases} 
c_0=1 \quad \mbox{theory with fundamental matter} \,,\\ 
c_0=2 \quad \mbox{theory without fundamental matter}  \,.
\end{cases} 
\ee
The flat coordinate defined in the neighborhood of the singular point corresponding to weak coupling of the non-abelian gauge theory is called $a$ by  
Seiberg and Witten, and its dual $a_D$.\footnote{We will distinguish strictly between the weak coupling periods $(a_D,a)$ and general periods $(t_D,t)$ in this subsection. In the following subsections, to conform to existing literature, we will generically use the letter $a$ in our discussions of field theory and the letter $t$ in the context of string theory.} The gauge coupling and theta angle of the gauge theory are conveniently combined into a complex gauge coupling $\tau$, 
\be \label{tau} 
\tau =  \frac{1}{c_0} \left( \frac{\theta}{\pi} + \frac{8 \pi i}{g^2} \right) \,.
\ee 
In terms of the local flat coordinate,  
the exact IR complex gauge coupling of the theory is given by  
\be 
\tau = - \frac{c_0}{2\pi i} \frac{\del^2 F^{(0,0)}}{\del t^2} = \frac{\partial t_D}{\partial t} \,. 
\label{generalgaugecoupling} 
\ee 
The K\"ahler metric on the moduli space follows from the K\"ahler potential $\re (\bar{t} \del_t F^{(0,0)})$ via 
\be \label{PW_metric} 
G_{t \bar{t}} = 2 \del_t \del_{\bar{t}} \re (\bar{t} \del_t F^{(0,0)}) = \frac{4 \pi}{c_0} \tau_2 \,. 
\ee 
The three point function, which already made an appearance in the holomorphic anomaly equations in the  
previous section, is the third derivative of the prepotential, 
\be 
C_{ttt} = \frac{\del^3 F^{(0,0)}}{\del t^3} = - \frac{2 \pi i}{c_0} \frac{\partial \tau}{\partial t} \,. 
\label{general3pointcoupling} 
\ee 
Note that it is this quantity, rather than the prepotential, which most readily is computed from  
the topological string~\cite{BCOV} as a three point correlator on a sphere. As the moduli space  
of the three punctured sphere is a point (hence has no boundary), the three point function and  
hence $F^{(0,0)}$ is a purely holomorphic section over the moduli space.  
 
We take the  existence of $F^{(0,0)}$ with the local equations (\ref{generalgaugecoupling}), (\ref{PW_metric}),  
and (\ref{general3pointcoupling}) as the definition of rigid special geometry. In addition, we have a symplectic electric/magnetic charge lattice spanned locally by $t$ and $t_D$.

\subsection{The family ${\cal C}_1(u)$, its periods, and its degenerations}   \label{family} 
In $N=2$ theories with fundamental matter of bare masses $m_i$, the masses of BPS states of electric, magnetic, and $U(1)$ flavor charge $(n_e,n_m,S_i)$  
follow from the central charge formula\footnote{Note that this formula is written in terms of $(a_D,a)$ rather than $(t_D,t)$, as the charges are defined with regard to the weak coupling point. Due to monodromy, they are not uniquely defined  in the interior of moduli space.}
\be 
\label{centralcharge} 
Z= n_e a + n_m a_D + \sum_i S_i \frac{m_i}{\sqrt{2}} \,, 
\ee 
by $m=|Z|$. The flat parameter $t$ and its dual $t_D$ in the limit of vanishing bare masses $m_i$ can be identified with the periods of the meromorphic 1-form $\lambda$ along an appropriately chosen\footnote{A different choice of basis will generically be required in the neighborhood of each singular point in moduli space in order to guarantee that the regular period $t$ is the period along the $A$ cycle.} symplectic basis $(\Sigma_A,\Sigma_B)$ of $H_1({\cal C}_1,\mathbb{Z})$, 
\begin{equation}  
t=\int_{\Sigma_A} \lambda, \qquad t_D=\int_{\Sigma_B} \lambda \,. 
\end{equation} 
Upon considering $m_i\neq0$, the bare masses appear linearly and with an integer structure in   
the residua of the meromorphic form $\lambda$ as  
${\rm Res}\,\lambda=\frac{n^i m_i} {2 \pi i \sqrt{2}}$  
with  $n^i\in \mathbb{Z}$. The integrals (\ref{centralcharge}) then no longer merely depend on the homology class of the integration path. 
 
The fundamental relation 
\be 
\frac{d\lambda}{du} = \omega \,, 
\ee 
with $\omega$ denoting the holomorphic 1-form of the curve, together with the special geometry relation (\ref{generalgaugecoupling}) determining the IR gauge coupling of the theory yields 
\be 
\tau = \frac{d t_D}{dt} = \frac{dt_D}{du} /\frac{dt}{du} = \dfrac{\int_{\Sigma_B} \omega}{\int_{\Sigma_A} \omega} \,. 
\ee 
The ratio of two symplectically dual periods of the holomorphic 1-form $\omega$ of the curve takes values in the upper half-plane, ensuring the positivity of the gauge coupling by the above identification. This ratio determines the complex structure of the curve up to $SL(2,\IZ)$ transformations. Note that unlike $(t_D, t)$, the derivatives $(\frac{dt_D}{du}, \frac{dt}{du})$ do not pick up path dependence upon considering $m_i \neq 0$. 
 
The periods of the holomorphic 1-form on elliptic curves are readily calculable starting from the Weierstrass form of the  
curve  
\begin{eqnarray}  \label{curve_equation} 
y^2 = 4x^3 -g_2(u) x- g_3(u) \, . 
\end{eqnarray} 
We have indicated only the dependence on the global coordinate  
$u$ of the moduli space of the theory explicitly. In Seiberg-Witten theory, the curve depends in addition on either the UV parameter $q_{uv}=\exp(2 \pi i \tau_{uv})$ in conformal cases  
or the renormalization scale $\Lambda$ in asymptotically free cases, and potentially the bare masses $m_i$ of the matter hypermultiplets in the theory. A quartic curve 
\be 
\label{quarticweierstrass1}  
y^2 = a \,x^4 + 4 b\, x^3+6 c \,x^2 + 4 d\, x +e   
\ee 
can be brought to Weierstrass form via a variable redefinition \cite{FrickeKlein}, such that  
\ban 
g_2 &=& ae - 4b d +3c^2 \,, \nn\\ 
g_3 &=& ace + 2bcd-ad^2 -b^2 e - c^3 \,.  \label{quarticweierstrass2}  
\ean 
Note that $g_2$ and $g_3$ are not absolute invariants of a curve; under a rescaling of $x$ by $r$, they scale as 
\be 
g_2 \rightarrow r^2 g_2 \,, \quad g_3 \rightarrow r^3 g_3  \,. 
\ee 
For all ensuing computations involving $g_2$ and $g_3$, care must be taken to work with a consistent normalization throughout. 
 
As introduced above, a parameter $\tau$ equal to the ratio of two symplectically dual periods of the holomorphic 1-form $\omega$ completely specifies the complex structure of an elliptic curve. Two such tau parameters related by an  
$SL(2,\IZ)$ transformation describe the same complex structure. A more easily  
accessible quantity encoding the complex structure is the $J$-invariant of the  
curve. When the latter is given in Weierstrass form (\ref{curve_equation}),  
the $J$-invariant is computed via 
\begin{eqnarray}  
\label{jdefu}  
J= \frac{g_2(u)^3}{\Delta(u)}\,, 
\end{eqnarray} 
where $\Delta$ is the discriminant of the curve, 
\begin{eqnarray} 
\Delta(u) = g_2(u)^3-27g_3(u)^2 \,. 
\end{eqnarray} 
Unlike $g_2$ and $g_3$, $J$ is an absolute invariant of a curve. It is related to the tau parameter via a quotient of Eisenstein series of weight 4 and 6, 
\begin{eqnarray}  
\label{jdeftau} \label{relation1} 
J(\tau)= \frac{E_4(\tau)^3}{E_4(\tau)^3-E_6(\tau)^2} \,. 
\end{eqnarray} 
As the RHS is a quotient of modular forms of weight 12, $J$ in fact descends to a function on the $SL(2,\IZ)$ orbits of $\tau$.  
As such, it contains less information than the gauge coupling of the theory. 
 
At weak coupling, corresponding to $\tau \rightarrow i \infty$, the $J$-invariant has the expansion 
\begin{equation} 
\label{jexp} 
j(\tau)=1728 J(\tau)=\frac{1}{q}+744+196884 q+21493760 q^2+864299970 q^3+{\cal O}(q^4) \,, 
\end{equation} 
where we have introduced the quantity $q = \exp(2 \pi i\,\tau)$. 
 
The moduli space of the theory is parametrized globally by the parameter $u$. Locally, $\tau$ or $t$ can serve as coordinates, but both undergo monodromy upon circling points in moduli space at which the curve degenerates. As $t$ as well as its dual coordinate $t_D$ are periods of the Seiberg-Witten differential, they satisfy the Picard-Fuchs differential equations.  
For Seiberg-Witten theory with $N_f=1,2,3$ massive flavors, these degree $3$ differential equations can be found in \cite{Ohta1, Ohta2}. However it is more convenient to use the well-known formulae for the periods of the holomorphic differential $\omega$ involving modular forms and compute the  
relevant periods of $\lambda$ by integration. At degeneration points $u_0$ in the moduli space at which $J(u_0)=\infty$, i.e. $\Delta(u_0)=0$, $\omega$ develops a period with a logarithmic singularity, while the dual period is finite \cite{FrickeKlein}. Such points hence exhibit monodromy of infinite order. Physically, this is an indication that a particle in the spectrum of the theory is becoming massless. The finite period is uniquely determined, and obeys the equation \cite{BS}  
\begin{eqnarray} \label{nonlogperiod} \label{relation2} 
\frac{dt}{du} =c_1 \sqrt{\frac{g_2(u)}{g_3(u)} \frac{E_6(\tau)}{E_4(\tau)}}=3^\frac{1}{4} c_1 \sqrt[4]{\frac{E_4(\tau)}{g_2(u)}}  \,. 
\end{eqnarray}  
The form of this equation reflects a general fact about the periods of elliptic curves: the  
normalized periods, which are the appropriate integrals over $\sqrt{\frac{g_3}{g_2}} \,\omega$,  
are invariants of the curve, in that they only depend on its complex structure, not its embedding in $\IC \IP^2$. One can derive a second order differential equation in $J$ for these normalized periods \cite{FrickeKlein} and obtain (\ref{relation2}) as one solution. To identify this solution as the constant period, it is enough to note that it cannot develop a logarithmic singularity. We note furthermore that this period is non-vanishing at the singular points $u_0$ if we rule out singularities for which not only $\Delta=0$, but also $g_2=g_3=0$, as the zeros of the Eisenstein series $E_4$ and $E_6$ in the fundamental domain are of unit norm (in particular, for $E_4$ and $E_6$, they lie at $\tau = \exp(2 \pi i/6)$ and $\tau=i$ respectively), hence do not lie in the $SL(2,\IZ)$ orbit of $\tau = i \infty$, the values the effective coupling can take at $J = \infty$. In integrating (\ref{nonlogperiod}), we can hence arrange the integration constant such that $t \sim (u-u_0)$, consistent with the physical requirement that a particle becomes massless at this point. At the weak coupling point $u \rightarrow \infty$, the physical boundary condition is\footnote{The factor of 2 between the two cases is the same as the one in equation (\ref{rescaling}).}
\be 
a =  
\begin{cases} 
\sqrt{\frac{u}{2}}  \quad \mbox{theory with fundamental matter} \,,\\ 
\sqrt{2u}  \quad \mbox{theory without fundamental matter}  \,. 
\end{cases} 
\ee 
The two expressions for $\frac{dt}{du}$ in (\ref{nonlogperiod}) hence imply $g_2 \sim u^2$, $g_3 \sim u^3$ at $u \rightarrow \infty$ as a condition on the Seiberg-Witten curve for any $SU(2)$ theory. The constant $c_1$, reflecting the normalization of $\lambda$, can be fixed by requiring the correct proportionality constant between $a$ and $u$ at $u \rightarrow \infty$. As pointed out above, care must be taken to fix the normalization of $g_2$ and $g_3$, as these are not absolute invariants of the curve.

An example for the period with logarithmic behavior for $|J|>1$ and $\arg(1-J)< \pi$ is \cite{BS}  
\begin{equation}  
\frac{dt_D}{du} = -\frac{1}{2 \pi i} \frac{dt}{du}(\log(J)+3 \log(12))+[(1-J)J^{-1}]^\frac{1}{4} w_1(J^{-1}) + n \frac{dt}{du}\,, 
\end{equation} 
where  $w_1(x)$ is a holomorphic power series.\footnote{The above expression at $n=0$ is a period of $\omega$ linearly independent from $\frac{dt}{du}$. We have added the $n$-dependent term (with $n$ to be determined) to justify calling this period the dual period to $\frac{dt}{du}$.}

\subsection{The amplitudes $F^{(n,g)}$ for $g+n=0$} 
\label{g=0} 
Relations (\ref{jdefu}), (\ref{jdeftau}), and (\ref{relation2}) are sufficient to determine the prepotential of the theory at weak coupling as a function of $a$, up to two unphysical integration constants 
as follows: Equating $J$ in equations (\ref{jdefu}) and (\ref{jdeftau}), one obtains $\tau$ as a function of the UV parameters of the theory.  
Substituting into (\ref{relation2}) and integrating yields the period $t$ as a function of these parameters. Solving this  
relation to obtain $u(t)$ and plugging into $\tau$, we can obtain the prepotential by integrating twice with regard to $t$, 
\begin{equation}  
F^{(0,0)} \sim \int^t dt \int^t dt \ \tau   \,. 
\end{equation} 
Note that generically, all of these steps will only be possible computationally to a given order in $q_{uv}$ or $\Lambda$. 
 
\subsection{The amplitudes $F^{(n,g)}$ for $g+n=1$} 
\label{genus1} 
For the genus one case, $F^{(0,1)}$ follows from the genus one holomorphic anomaly  
equation~(\ref{f1anomaly}), while $F^{(1,0)}$, as we have argued above, should have no holomorphic anomaly, 
\begin{eqnarray}  
F^{(0,1)} &=& -\frac{1}{2} \log (G_{u\bar u} |\Delta|^\frac{1}{3})   \label{genus1a}\ , \\ 
F^{(1,0)} &=& \frac{1}{24}\log (\Delta) \label{genus1b}\ . 
\end{eqnarray} 
Here, $\Delta$ is the discriminant of the curve ${\cal C}_1$. 
The first expression comes from integrating (\ref{f1anomaly}) using the simplification  
of rigid special K\"ahler geometry. Such expressions first appeared in the context  
of local Calabi-Yau manifolds in~\cite{Klemm:1999gm}.  A derivation from the point  
of view of gauge theory instantons can be found in~\cite{MR2199008}. The exponent of $\Delta$ is fixed  
by the boundary condition (second line of (\ref{Schwinger})) at single zeros of the discriminant $\Delta=(u-u_0)\hat\Delta$, using the fact that the local coordinate   
goes like $t\sim (u-u_0)+O((u-u_0)^2)$ at such points.  In the holomorphic limit, we get 
\be  \label{genus1a_hol}
F^{(0,1)}_{hol}= -\frac{1}{2} \log\left({d a \over du} 
\right)-\frac{1}{12} \log(\Delta) \,. 
\ee 
The exponent of $\Delta$ in the second equation follows similarly from the boundary condition (\ref{Schwinger}).

\subsection{The amplitudes $F^{(n,g)}$ for $g+n>1$ and modularity}  \label{gnb1} 
As $u$ is a global coordinate on the moduli space of our theory, by definition, physical quantities should be invariant under the monodromies that arise by circling singularities in this moduli space (otherwise, the correct moduli space would be a cover of the $u$-plane). The partition functions $F^{(n,g)}$ for $n+g>1$ ($n+g=0$ and $n+g=1$ are special cases, as it is their derivatives that are physical) should hence be modular forms of weight 0 with regard to the monodromy group $\Gamma \subset SL(2,\IZ)$. $u$ by definition must be invariant under $\Gamma$ as well.\footnote{There is a sense in which $u$ is a weight 2 modular form in the conformal theories and their mass deformations. We discuss this at length in section \ref{conformal_cases}.}The naive assumption of holomorphicity is known to be invalid (either by recourse to string theory or to path integral regularization arguments \cite{Jan1,Jan2}). We hence make the assumption that the $F^{(n,g)}$ are almost holomorphic forms (see e.g. \cite{Zagier} for this concept) of the group $\Gamma$. These are functions on the upper half plane that transform like modular forms, but are polynomials in $\frac{1}{\tau_2}$ with holomorphic coefficients. A prominent example is the modular completion $\hat{E}_2$ of the quasi modular form $E_2$ of weight 2, 
\be 
\hat{E}_2(\tau, \bar{\tau}) = E_2(\tau) - \frac{3}{\pi \tau_2} \,. 
\ee 
In fact, it is not hard to show that the polynomial dependence on $\frac{1}{\tau_2}$ can be replaced by a polynomial dependence on $\hat{E}_2$, with coefficients that are holomorphic modular forms (see Prop. 20 in \cite{Zagier}). 
 
The assumption that the $F^{(n,g)}$ are almost holomorphic forms is consistent with the holomorphic anomaly equations. This can be seen as follows: 
 
Note first that upon retaining only the constant piece with regard to the variable $\frac{1}{\tau_2}$ of an almost holomorphic form, all other terms can be reproduced by modularity. Hence, the ring thus obtained, called the ring of quasimodular holomorphic forms, is canonically isomorphic to the ring of almost holomorphic modular forms \cite{Zagier}. In the presentation in which almost holomorphic forms are written as polynomials in $\hat{E}_2$, this isomorphism is induced by mapping $\hat{E_2}$ to $E_2$. This map from almost holomorphic to quasimodular forms is referred to in the physics literature as taking the holomorphic limit $\bar \tau \rightarrow \infty$ (the mnemonic rooted in the occurrence of $\bar{\tau}$ in the denominator of $\frac{1}{\tau_2} = \frac{2i}{\tau - \bar{\tau}}$). In this limit, the coordinate $a$ is flat (the connection in this variable vanishes) and therefore a convenient choice to express the holomorphic anomaly equations in. They become \cite{HK2} 
\begin{eqnarray}  \label{hae_hol} 
24 \frac{\del F^{(n,g)} }{\del {E_2}} =   c_0 \big{(} \frac{\del^2 F^{(n,g-1)}}{\del a^2} 
+{\sum_{m,h} }^{\prime} \frac{\del F^{(m,h)}}{\del a} \frac{\del F^{(n-m,g-h)}}{\del a}    \big{)}. 
\end{eqnarray} 
 
To study the general structure of the $F^{(n,g)}$, it proves convenient to introduce the variable 
\begin{eqnarray} 
X=\frac{g_3(u)}{g_2(u)} \frac{{E}_2(\tau)E_4(\tau)}{E_6(\tau)} \,. 
\label{Xdef} 
\end{eqnarray} 
Rewriting (\ref{hae_hol}) in terms of $X$ and the global variable $u$, we obtain 
\ban 
24 \frac{\partial F^{(n,g)} }{\partial X} &=& c_0 \frac{g_2(u)}{g_3(u)} \frac{E_6}{E_4} \Big[ \left(\frac{d u}{d a}\right)^2 \frac{\del^2 F^{(n,g-1)}}{\del u^2} +\frac{d^2 u}{d a^2} \frac{\del F^{(n,g-1)}}{\del u} \nn \\ 
&& +\left(\frac{d u}{d a}\right)^2 {\sum_{m,h} }^{\prime} \frac{\del F^{(m,h)}}{\del u} \frac{\del F^{(n-m,g-h)}}{\del u} \Big] \,.   \label{hol_an_X}  
\ean 
A glance at (\ref{relation2}) reveals the utility of introducing the variable $X$, particularly since using the Ramanujan relations 
\begin{equation}  
q \frac{dE_2}{dq} = \frac{E_2^2 - E_4}{12} \,, \qquad  
q \frac{dE_4}{dq} = \frac{E_2 E_4 - E_6}{3} \,, \qquad 
q \frac{dE_6}{dq} = \frac{E_2 E_6 - E_4^2}{2} \,,  
\end{equation}  
and relations (\ref{relation1}) and (\ref{relation2}), one can show that  
\be  \label{d2uda2} 
\frac{d^2 u}{d a^2}  = \frac{1}{\Delta} \frac{g_3(u)}{g_2(u)} \frac{E_4}{E_6} \,p_1(X)  \,, 
\ee 
where $p_n(X)$ is an $n^{th}$ degree polynomial in $X$ with coefficients that are polynomials of derivatives of $g_2(u)$ and $g_3(u)$. 
 
We are now in a position to discuss the modular properties of the equation (\ref{hol_an_X}). By (\ref{relation2}), $\left(\frac{du}{da}\right)^2$ is modular of weight -2 under the full modular group $SL(2,\IZ)$ (note that $\frac{du}{da}$ is modular only with regard to an index 2 subgroup), as is $\frac{d^2 u}{d a^2}$ by (\ref{d2uda2}). The weights in the equation (\ref{hol_an_X}) thus add up to 0 correctly on both sides, demonstrating the consistency of our identification of the amplitudes $F^{(n,g)}$ as almost holomorphic forms with this equation. From this discussion, one might be tempted to conclude that the symmetry group of the theory is the full modular group $SL(2,\IZ)$. This is incorrect. The holomorphic anomaly equations as written in (\ref{hol_an_X}) depend on both UV parameters ($u$, potentially bare masses, and the UV coupling $\tau_{uv}$ or the dimensional transmutation scale $\Lambda$) and IR parameters (the effective coupling $\tau$, the argument of the Eisenstein series contained in $X$). To draw conclusions regarding the symmetry group, one should re-express it fully in terms of IR parameters. That this reduces the symmetry group to a subgroup of $SL(2,\IZ)$ can be seen explicitly e.g. in the massless asymptotically free cases of $SU(2)$ gauge theory with $N_f<4$ flavors, in which $u$ can be obtained explicitly as a function of $\tau$ (see e.g. \cite{HK2}) and proves to be modular only under the monodromy group $\Gamma \subset SL(2,\IZ)$. 
 
Analogous calculations that led to (\ref{d2uda2}) show that 
\begin{equation} 
\frac{d X}{d u} = \frac{1}{\Delta}\, p_2(X) \,, \qquad \ \  
\frac{d^2 X}{du^2} = \frac{1}{\Delta^2}\, p_3(X)  \,. 
\end{equation}  
Starting from the form of the genus 1 partition functions given above, an easy induction argument shows (for $k\!>\!0$, $k\!=\!0$ will be taken up in section (\ref{fixingtheambiguity})) that the partition functions take the form 
\begin{equation} \label{generalformfng} 
F^{(n,g)}=\frac{1}{\Delta^{2(g+n)-2}(u)} \sum_{k=0}^{3g+2n-3} X^k p^{(n,g)}_k(u), 
\end{equation} 
where $p^{(n,g)}_k(u)$ are polynomials in derivatives of $g_2(u)$ and $g_3(u)$. As presented in (\ref{generalformfng}), $F^{(n,g)}$ is a quasimodular holomorphic form. The map to the ring of almost holomorphic modular forms is induced by
\be
X \mapsto \hat{X} = \frac{g_3(u)}{g_2(u)} \frac{\hat{E}_2(\tau)E_4(\tau)}{E_6(\tau)} \,. 
\ee
Note that the leading power of $E_2$ (and hence $X$) in $F^{(n,g)}$ can be obtained from (\ref{hae_hol}) by inspection: derivatives with regard to $a$ acting on the Eisenstein series act via $\frac{d\tau}{da} \frac{d}{d\tau}$. By the Ramanujan relations, and the fact that  $\frac{d\tau}{da}$ is holomorphic, each such derivative hence increases the power in $E_2$ by 1. The first term in the holomorphic anomaly equations then implies a contribution $3g$ to the leading power, and the second term a contribution $-3$. Taking into account that $F^{(1,0)}$ has no holomorphic anomaly to fix the power in $F^{(2,0)}$ then yields the final result $3g +2n-3$.  
 
The leading power of $X$ can be lower in theories where the leading $p^{(n,g)}_k$ vanish identically -- this happens in the conformal cases studied in this paper, massless $N_f=4$ and $N=4$. The above derivation fails as $\frac{d\tau}{da}=0$ in the conformal cases. Likewise, the leading negative power of the discriminant can be lower if all $p^{(n,g)}_k$ contain powers of the discriminant, as turns out to be the case for mass deformed $N=4$. 
 
All coefficients $p^{(n,g)}_k(u)$ are fixed by the holomorphic anomaly equations save $p_0^{(n,g)}(u)$. This coefficient gives rise to a meromorphic term in $F^{(n,g)}$ annihilated by the antiholomorphic derivative on the left hand side of the holomorphic anomaly equations, referred to as the holomorphic ambiguity. It can be fixed by imposing appropriate boundary conditions on the partition functions: finiteness of the holomorphic ambiguity at large $u$, and the gap conditions at the singularities at $\Delta(u)=0$ in moduli space. 
 
\newpage 

\subsection{The conformal limit of the holomorphic anomaly equations} \label{conf_limit} 
 
The  defining feature of the conformal limit is that the infrared gauge coupling  
$\tau$ becomes independent of the scale  $a$. Since the derivatives in (\ref{hae_hol})  
with regard to $a$ do not act on $\tau$, the argument outlined in  
section~\ref{gnb1} is modified: differentiating with regard to $a$ no longer  
entails an increase in the power of $E_2$ (by two) on the RHS of (\ref{hae_hol}).  
As a consequence, the highest power of $E_2 \sim X$ in $F^{(n,g)}$ grows as  
$F^{(n,g)}\sim E_2^{g+n-1}$ rather than exhibiting the generic growth  
$F^{(n,g)}\sim E_2^{3g+2 n-3}$, see Table \ref{TabX}. 
More precisely, in order to have (\ref{expansion1}) dimensionless with  
$\epsilon_1,\epsilon_2$ of dimension one, the scale dependence of the $F^{(n,g)}$ in the conformal limit can be extracted into an $a$-dependent prefactor,\footnote{A similar argument leading to the correct prefactor proceeds via imposing the correct modular weight.}  
\begin{equation}  
\label{deffng} 
F^{(n,g)}=\left\{\begin{array}{ll}  
\displaystyle{\frac{1}{a^{2 (n+g)-2}} \frac{f^{(n,g)}(\tau)}{n+g-1}} & {\rm if } \,\, (n+g)>1  \,,\\ [3 mm]  
\displaystyle{-2\log(a) f^{(n,g)}(\tau)} & {\rm if }  \,\,(n+g)=1  \,. 
\end{array}\right. 
\end{equation} 
The functions $f^{(n,g)}(\tau)$ carry neither scale nor $a$-dependence, but they can have an interesting 
dependence on the gauge coupling. With the $a$-dependence thus extracted, the holomorphic anomaly equations  
governing the conformal theory becomes algebraic, taking the form 
\begin{eqnarray} 
\label{gen_hol_ano-cf2} 
\frac{\del f^{(n,g)}}{\del \hat{E_2}} = \frac{c_0(n+g-1)}{6}\biggl((n+g-\frac{3}{2}) f^{(n,g-1)} +  {{\sum}^\prime_{n''+n'=n\atop g'+g''=g }}  f^{(n',g')} f^{(n'',g'')}\biggr) \ . 
\end{eqnarray}  
These equations govern the gauge coupling dependence of the $F^{(n,g)}$ in  the massless  
limit of the $N=2^*$ and the $N_f=4$ theories discussed in sections \ref{N=4massless} and  
\ref{Nf=4massless}. For the conformal theories at the simplest Argyres-Douglas points, 
the equations also hold, but are less interesting as the infrared gauge coupling at these points is  
fixed as well.   
 
Considering only the genus 0 equations (the Nekrasov-Shatashvili limit), we obtain 
\begin{equation}  
\label{gen_hol_ano-cf1} 
\frac{\del f^{(n,0)}}{\del \hat{E_2}}= \frac{c_0(n+1)}{6} \sum_{m=1}^{n-1} f^{(m,0)} f^{(n-m,0)}\ . 
\end{equation} 
This is exactly the form of the holomorphic anomaly equations presented in~\cite{MMW} for the mass deformation of $F^{(0,0)}(a,m)$ of the $SU(2)$ $N=2^*$ theory, with $n$ labeling the number of insertions of the mass operator $\phi_m$. The analogous equation for the $N_f=4$ theory was derived in \cite{billo1}. We will comment further on this connection in section~\ref{N=4amplitudesfromcurve}, focussing on the $N=4$ case. What is more, equations (\ref{gen_hol_ano-cf1}) are, up to a  
shift in the label $k$ in $F^{(k,0)}$ by $1$, the holomorphic anomaly equations for the elliptically  
fibered $T_2\rightarrow S\rightarrow \mathbb{P}^1$ del Pezzo surface $S$  
discussed in~\cite{Hosono:2002xj}. $k$ here labels the deformation $\phi_b$ with respect to the   
$\mathbb{P}^1$ base modulus. These holomorphic anomaly equations are closely related to the  
holomorphic anomaly for $N=4$ rank $n$ gauge theories on del Pezzo surfaces. The anholomorphic  
generator here is the completion of a Mock modular rather than just  
an almost holomorphic form~\cite{Vafa:1994tf,Alim:2010cf}. 
 
\subsection{BPS states and fixing the holomorphic ambiguity} 
\label{fixingtheambiguity}  
As we have reviewed above, the singularities of an elliptic curve lie at the zeros of its discriminant.  
At each such point, a period of the Seiberg-Witten differential vanishes.  
The physical meaning of these singularities is that BPS particles in  
the spectrum of the theory are becoming massless. This determines the leading  
behavior of the $F^{(n,g)}$ at these points. As we shall demonstrate in this subsection, imposing this leading behavior at each singular point as boundary conditions is sufficient to fix the holomorphic ambiguity.

\subsubsection{The BPS amplitudes} 
The BPS interpretation of the amplitude is obtained by computing it via a Schwinger loop integral with BPS states running in the  
loop~\cite{Gopakumar:1998jq}\cite{IKV}. The latter are in a representation  
$\mathcal{R}=[j_-,j_+]$ of the little group of the 5D Lorentz group  
$SO(4)\sim SU(2)_+ \times SU(2)_-$ and have a mass $m$ related to their  
charge by the BPS formula. In the BPS saturated amplitude, the BPS states  
couple to insertions of the self- and anti-self-dual part of a background graviphoton field strength   
${G}=\epsilon_1 dx^1\wedge dx^2+\epsilon_2 dx^3 \wedge dx^4$ and  
two insertions of the background curvature 2-form ${R}$ respectively.  
Passing to spinor notation for the field ${G}$, one gets  
$\epsilon_-^2= -\det { G}_{\alpha,\beta}$ and $\epsilon_+^2 =  
\det { G}_{\dot \alpha,\dot \beta}$, with $\epsilon_\pm=\frac{1}{2} 
(\epsilon_1\pm \epsilon_2)$.  The anti-self-dual  
and self-dual parts of the field strength couple to the left and right spin $j_-$ and $j_+$ of the BPS particle respectively. The  
Schwinger loop calculation for these amplitudes yields, with $q_\pm=e^{-2\epsilon_\pm}$, 
\begin{equation} 
{\cal F}^{hol}(\epsilon_\pm)=-\int_\epsilon^{\infty} \frac{ds}{s}\frac{\textrm{Tr} 
  _\mathcal{R} (-1)^{\sigma_+^3+\sigma_-^3} e 
^{-s m} 
q_-^{s \sigma^3_-} q_+^{s\sigma_+^3}}{4\left(\sinh^2\left(\frac{s\epsilon_-}{2}\right)  
- \sinh^2\left(\frac{s \epsilon_+}{2}\right)\right)}\ . 
\label{gapschwingerloop}  
\end{equation} 
The importance of this formula for our purposes is that if something is known about the BPS spectrum  
associated to a geometrical singularity at a special point in the moduli space,  
one can read off the leading behavior of the $F^{(n,g)}$ at this point.  We discuss this  
first for large radius singularities in non-compact Calabi-Yau spaces. Such points can be relevant also for field theory, if embedded in a  
Type II string non-compact Calabi-Yau geometry. Then we address the  
behavior at the conifold, which universally applies for the geometric  
description of field theories and for  string theory on Calabi-Yau spaces.  

At large radius in the A-model on non-compact Calabi-Yau spaces the relevant BPS states  
are D-brane bound states with charge  
$(Q_6,Q_4,Q_2,Q_0)=(1,0,\beta,m)$ where  $\beta\in H_2(M,\IZ)$. A Poisson resummation over $m$ gives rise to a sum (over $k$ in (\ref{schwingerloope1e2})) of delta functions against which the $s$ integral in (\ref{gapschwingerloop}) can be evaluated \cite{Gopakumar:1998jq}.  One can thus write a generating function for the multiplicity of BPS states $N^\beta_{j_-j_+}\in \mathbb{Z}_+$ of given charge $\beta$ and spin representation  $[j_-,j_+]$ as
\begin{equation} 
\begin{array}{rl} 
{\cal F}^{hol}(\epsilon,t)&= 
\displaystyle{\sum_{{2j_-,2j_+=0}\atop {k=1}}^\infty \sum_{\beta\in H_2(M,\mathbb{Z})} (-1)^{2(j_-+j_+)} \frac{N^\beta_{j_-j_+}}{k}   
\frac{\displaystyle{\sum_{m_-=-j_-}^{j_-}} q_-^{k m_-}}{2\sinh\left( \frac{k \epsilon_1}{2}\right)} \frac{\displaystyle{\sum_{m_+=-j_+}^{j_+}} q_+^{k m_+}} 
{2\sinh\left( \frac{k \epsilon_2}{2}\right)}e^{-k\, \beta \cdot t}}\\[8 mm] 
&=\displaystyle{\frac{1}{\epsilon_1\epsilon_2}\sum_{g=0}^\infty \sum_{m=0}^{2g}\epsilon_1^{2g-m}   
\epsilon_2^{m} F_{m,g}=\sum_{n,g=0}^\infty (\epsilon_1+\epsilon_2)^{2n} (\epsilon_1\epsilon_2)^{g-1}  F^{(n,g)} (t)}\ . 
\end{array} 
\label{schwingerloope1e2} 
\end{equation} 
The sum over $m_{\pm}$ is taken in integral increments both for $j_\pm$ integral and half-integral.

It is sometimes convenient to change from the irreducible highest weight representations  
$\left[\frac{i}{2}\right]$ to a basis given by  
\be 
I^n=\left(2 [0]+\left[\frac{1}{2}\right]\right)^{\otimes n}= 
\sum_{i}\left( \left(2n\atop n-i\right)- \left(2n  \atop n-i-2\right) \right)\left[\frac{i}{2}\right] \,, 
\ee 
because  
\be 
{\rm Tr}_{I^n} (-1)^{2 \sigma_3} e^{-2 \sigma_3 s}= (-1)^n \left(2 \sinh \frac{s}{2}\right)^{2n}\,. 
\ee 
The multiplicity of BPS states in the $I_{n_-}\otimes I_{n_+}$ basis also gives rise to integers which we denote $n^\beta_{n_- n_+}$. Unlike the $N^\beta_{j_-j_+}$, these exhibit alternating signs. The $n^\beta_{n_- n_+}$ specialize to the index $n^\beta_g=n^\beta_{g ,0}$, defined by 
\be 
\sum_{g=0}^\infty n^\beta_g I^g=\sum_{j_+} N^\beta_{j_-j_+} (-1)^{2j_+}(2j_++ 1)\left[j_-\right]  \,, 
\ee 
which is usually calculated by the topological string and is invariant under complex structure deformations. 
Formula (\ref{schwingerloope1e2}) can easily be exponentiated upon expanding the $\sinh(x)$. This yields the following expression for the partition function~\cite{IKV}, 
\begin{equation}  
Z=\prod_\beta \prod_{2j_\pm=0}^\infty \prod_{m_\pm=-j_\pm}^{j_\pm}\prod_{m_1,m_2=1}^\infty \left(1-q_-^{m_-} q_+^{m_+} e^{\epsilon_1(m_1-\frac{1}{2})}  
e^{\epsilon_2(m_2-\frac{1}{2})} e^{-\beta \cdot t}\right)^{ (-1)^{2(j_-+j_+)+1} N^\beta_{j_-j_+}}\ . 
\end{equation} 
 
We note that no information about the large radius expansion of the $F^{(n,g)}(t)$ is needed to fix the  
holomorphic ambiguity up to an additive constant to $F^{(n,g)}(t)$. This constant  
is unphysical in the Seiberg-Witten cases and depends on a regularization of the  
Euler number in the non-compact Calabi-Yau cases. Up to this regularization, it is  
nevertheless possible to work it out from the BPS sum (\ref{schwingerloope1e2}).   
 
Define the Bernoulli numbers by their generating function  
$t/(e^t-1)=\sum_{m=0}^\infty B_m \frac{t^m}{m!}$. Note that the $B_m$ vanish for $m$ odd,   
except  $B_{1}=-\frac{1}{2}$.  
Dividing the generating function by $t$ and taking the derivative establishes 
\be 
\frac{t^2}{\left(2 \sin \left(\frac{t}{2}\right)\right)^2}=\sum_{m=0}^\infty (-1)^{m-1} \frac{B_{2m}}{2 m(2m -2)!} t^{2m}\ . 
\label{bernoulliidentity1} 
\ee    
The constant (i.e. $\beta$ independent) term in the usual topological string amplitudes $F^{(0,g)}$ is calculated from  
(\ref{schwingerloope1e2}) by setting $\epsilon_1=-\epsilon_2=i g_s$. In this limit  
the self-dual insertions decouple from (\ref{schwingerloope1e2}) and the right (+) spins contribute  
with their multiplicity and a sign according to their spin statistic. The evaluation of  
(\ref{schwingerloope1e2}) then reduces to the calculation of~\cite{Gopakumar:1998jq}.  
Note that the constant term comes exclusively from the $D0$ brane contribution which takes the universal form \cite{BCOV} 
$N^0_{0,0}=-\frac{\chi(M)}{2}$. Using  (\ref{bernoulliidentity1}) with the argument  
scaled by $m$, $\zeta(x)=\sum_{m=1}^{\infty}\frac{1}{m^x}$, and the regularized  
values $\zeta(-n)=-\frac{B_{n+1}}{n+1}$, we obtain the constant term of $F^{(0,g)}$ from (\ref{schwingerloope1e2}), 
\be 
\label{TSconstant} 
\langle 1\rangle^M_{(g),0}=(-1)^g \frac{\chi(M)}{2} \int_{{\cal M}_g} c_{g-1}^3=(-1)^g\frac{\chi(M)}{2}\frac{|B_{2g} B_{2g-2}|}{2 g (2g-2) (2g-2)!} \,. 
\ee  
On the LHS, we have included the mathematical expression which yields this contribution \cite{BCOV,FP}. 
 
The constant term in $F^{(n,g)}$ comes about by noting a similar expansion for   
\be  
\frac{\epsilon_1 \epsilon_2}{4 \sinh\left(\frac{\epsilon_1}{2}\right) \sinh\left(\frac{\epsilon_2}{2}\right)}=\sum_{g=0}^\infty \sum_{m=0}^g \hat B_{2g} \hat B_{2g-2m} \epsilon_1^{2g-2m} \epsilon_2^{2m} \ , 
\ee   
with  
\be  
\hat B_{m}=\left(\frac{1}{2^{m-1}}-1\right)\frac{B_m}{m!}\ . 
\ee 
It follows that the constant part of $F_{m,g}$ is  
\be 
\langle 1\rangle^M_{m,g,0}=\frac{\chi(M)}{2} \frac{ \hat B_{2g-m} \hat B_{m} B_{2g-2}}{2 g-2}\ .  
\ee 
The constant part $\langle 1\rangle^M_{(n,g),0}$ in  $F^{(n,g)}$ is obtained by  
changing the basis of the symmetric polynomials in  
$\epsilon_1,\epsilon_2$ in the second line of (\ref{schwingerloope1e2}).   
E.g., one has for the Nekrasov-Shatashvili limit   
\be  
\label{NSconstant} 
\langle 1\rangle^M_{(n,0),0}=\frac{\chi(M)}{2}\frac{\hat B_{2n} B_{2n-2}}{2 n-2} 
\ee    
and  as expected $\langle 1\rangle^M_{(0,g),0}=\langle 1\rangle^M_{(g),0}$.   
 
Using the gap condition at the conifold, which we discuss next, the direct integration method gives a  
very efficient method for calculating the $N^\beta_{[j_-,j_+]}\in \mathbb{Z}$,   
which are related to motivic Donaldson-Thomas invariants. We demonstrate  
this in section \ref{DT} for the simplest non-trivial local Calabi-Yau  
space ${\cal O}(-3)\rightarrow \mathbb{P}^2$.

A nodal singularity of ${\cal C}_r$ occurs at the conifold point in moduli space. It corresponds to a cuspidal point in the fundamental region of the modular group.  
Such a singularity arises due to a single dyon of charge $(n_e,n_m,S_i)$ becoming 
massless at this point in moduli space. Geometrically, a period $t=n_e a+ n_m a_D+S_i \frac{m_i}{\sqrt{2}}$, whose local  
expansion can be obtained from (\ref{nonlogperiod}), is vanishing here.   
Expanding the Schwinger loop formula assuming a single dyon of vanishing mass $|t|$ in $\epsilon_1$,  
$\epsilon_2$, and $\frac{1}{t}$ gives us the leading behavior of each $F^{(n,g)}$ near  
the conifold point from the corresponding coefficients of $g_s^2= (\epsilon_1 \epsilon_2)$  
and $s= (\epsilon_1 + \epsilon_2)^{2}$~\cite{HK3} 
\begin{eqnarray}  \label{Schwinger} 
F(s,\lambda,t)&=&\int_0^{\infty} \frac{ds}{s}\frac{\exp(-s t)}{4\sinh(s\epsilon_1/2)\sinh(s\epsilon_2/2)} +\mathcal{O} (t^0) \\  
&=& \big{[}-\frac{1}{12}+\frac{1}{24} (\epsilon_1+\epsilon_2)^2 (\epsilon_1\epsilon_2)^{-1}\big{]}\log(t) \nonumber \nonumber \\ &&  
+ \frac{1}{\epsilon_1\epsilon_2} \sum_{g=0}^\infty \frac{(2g-3)!}{t^{2g-2}}\sum_{m=0}^g \hat B_{2g} \hat B_{2g-2m} \epsilon_1^{2g-2m} \epsilon_2^{2m} +\ldots \nonumber \\ && 
=\big{[}-\frac{1}{12}+\frac{1}{24} s g_s^{-2}\big{]}\log(t) 
+ \big{[} -\frac{1}{240}g_s^2+\frac{7}{1440} s- 
\frac{7}{5760} s^2g_s^{-2} \big{]} \frac{1}{t^2} \nonumber \\ &&  
+ \big{[} \frac{1}{1008}g_s^4-\frac{41}{20160} s g_s^2 +\frac{31}{26880} s^2 -\frac{31}{161280} s^3 g_s^{-2}\big{]} \frac{1}{t^4}   +\mathcal{O} (t^0)  \nonumber\\[ 7 mm] 
&& +  \,\,\mbox{contributions to $2(g+n)-2>4$}\, .  \nonumber 
\end{eqnarray} 
Hence, e.g., 
\be  
F^{(0,2)} = - \frac{1}{240 } \frac{1}{t^2}+\mathcal{O}(t^0),  \quad  
F^{(1,1)} = \frac{7}{1440} \frac{1}{t^2} +\mathcal{O}(t^0),  \quad  
F^{(2,0)} = -\frac{7}{5760 } \frac{1}{t^2} +\mathcal{O}(t^0)  \,. 
\ee 
The leading behavior of (\ref{Schwinger}) is the same as that of the  
$S^1$ compactification of the $c=1$ string\footnote{As was noted for  
$\epsilon_1=-\epsilon_2$ in~\cite{Ghoshal:1995wm}}, where the same  
integral appears~\cite{Gross:1990ub}. 
 
We note that the coefficients of  
the leading singularities at the conifold, $N^{(n,g)}$, diverge in arbitrary  
directions of large $(n,g)$  exponentially at twice the rate  
of the ones at infinity, cf. e.g. (\ref{TSconstant},\ref{NSconstant})\footnote{In the  
NS and the topological string limit one has $N^{(n,0)}=(-1)^{n - 1} (2 n - 3)! \hat B_{2 n}$ and  $N^{(0,g)}=B_{2g}/(2 g (2 g - 2))$  
respectively.}. This suggests that (\ref{expansion1}) is an asymptotic, presumably Borel summable expansion everywhere in moduli space. 
This is different for the conformal cases, where $\chi(M)=0$ and the  
absence of conifold singularities could allow for much better  
convergence properties.       
 
The fact that near singular points in moduli space, the relation 
\be  \label{gap} 
F^{(n,g)} = \frac{N^{(n,g)}}{t^{2(g+n)-2}} + \mathcal{O} (t^0) 
\ee 
holds, i.e. the absence of subleading poles in the $t$ expansion, is referred to as the gap structure  
of the $F^{(n,g)}$ at these points~\cite{HK1,Huang:2006hq}. We will discuss next that this behavior is sufficient to fix the holomorphic ambiguity also for the  
deformed models. For the undeformed models, this discussion was presented in~\cite{Haghighat:2008gw}. 
 
\newpage

\subsubsection{The completeness of the gap boundary condition} 
\label{completeness} 
From (\ref{gap}) and $t \sim u-u_0$ near a zero $u_0$ of the discriminant $\Delta$, we see that the holomorphic ambiguity as a function of $u$ can at worst have a pole of order $2(g+n)-2$ at $u_0$. When $\Delta \sim u-u_0$, we can hence parameterize the ambiguity as 
\be 
F^{(n,g)}_{hol. amb.} = \frac{1}{\Delta(u)^{2(g+n)-2}} \, p(u)  \,, 
\ee 
with $p(u)$ a holomorphic function of $u$. Demanding that the ambiguity be finite in the limit $u \rightarrow \infty$  implies  
that $p(u)$ is in fact a polynomial in $u$ of degree $\left(2(g+n)-2\right) d_\Delta$, where $d_{\Delta}$ denotes the  
degree of the discriminant. There are hence $\left(2(g+n)-2\right) d_\Delta$ coefficients to be fixed to determine $p(u)$ (recall that the constant term of $F^{(n,g)}$ is unphysical),  
and exactly this many boundary conditions, $\left(2(g+n)-2\right)$ via the gap condition at each of the $d_\Delta$ zeros of the discriminant, at our disposal. 
 
To proceed calculationally, one could determine the $d_\Delta$ zeros of $\Delta$, expand the right hand side of (\ref{gap}) as well as $F^{(n,g)'} + F^{(n,g)}_{hol. amb.}$ around each of these zeros, with $F^{(n,g)'}$ the solution of the holomorphic anomaly equations, set the two equal and compare coefficients. Already the first step, determining the zeros of the discriminant, is difficult in general. 
 
Instead, we can multiply both sides by $\Delta(u)^{2(g+n)-2}$, set $u=u_0 + x$, and expand in $x$ to order $(2(g+n)-2)d_\Delta -1$, arriving at an equation 
\be  \label{ambig} 
\sum_i q_i(u_0) x^i = \sum_i r_i(u_0) x^i \,, 
\ee 
where $q_i(u)$ and $r_i(u)$ are rational functions of $u$. Note that this calculation requires expressing $\tau$ in terms of UV quantities. This can be achieved by inverting (\ref{jdefu}), invoking $1/J$ as an expansion parameter at the singular points. Again, if we explicitly knew the roots of $\Delta$, we would set $u_0$ equal to these in (\ref{ambig}), and obtain a system of equations to determine the coefficients of $p(u)$. Lacking this information, we instead mod out both sides of the equation by $\Delta(u_0)$. E.g. for $q_i(u) = \frac{f_i(u)}{g_i(u)}$, this means that we solve the equation 
\be 
f_i(u) = a_i(u) \Delta(u) + b_i(u) g_i(u) 
\ee 
for $a_i(u)$ and $b_i(u)$. This is always possible when $\mbox{gcd}(\Delta(u), g_i(u)) = 1$. We write 
\be 
b_i(u) \equiv q_i(u)   \mod \Delta(u)  \,. 
\ee 
Likewise, set 
\be 
d_i(u) \equiv r_i(u)   \mod \Delta(u)  \,. 
\ee 
As $\Delta(u_0)=0$ by definition of $u_0$, we finally arrive at the following polynomial equation in both $u_0$ and $x$, 
\be \label{final_gap} 
\sum_i b_i(u_0) x^i = \sum_i d_i(u_0) x^i \,. 
\ee 
Equating coefficients of $u_0$ and $x$ now allows us to solve for the coefficients of $p(u)$. 
 
So far, we have assumed that a single particle becomes massless at each singularity. In fact, as moduli are varied, singularities can merge. These higher degeneracies of the Seiberg-Witten curve correspond to roots of the discriminant with higher multiplicity, the order of the root indicating the number of particles becoming massless (the $N=2^*$ theory requires an exception to this rule, as we discuss below). As long as these massless particles are mutually local, this situation requires only a slight modifications of the above analysis: to begin with, the boundary conditions (\ref{Schwinger}) given by the Schwinger calculation at each multiple root $u_{0,i}$ must be multiplied by the degeneracy of this root. Also, for each $u_{0,i}$, the calculation leading to (\ref{final_gap}) must be performed with a reduced discriminant $\Delta_i$ which satisfies $\Delta_i \sim u-u_{0,i}$.

\section{Seiberg-Witten theory: The asymptotically free cases} \label{computationsec}   
\subsection{Exact results} \label{asy_free} 
The Seiberg-Witten curves for $SU(2)$ Seiberg-Witten theory are families of elliptic curves that can be parametrized as follows,  
\begin{equation} 
y^2 = C(x)^2 -G(x) \,, 
\end{equation} 
where the functions $C(x)$ and $G(x)$ for different numbers of flavors are given by  
\begin{eqnarray} \label{SWcurve} 
N_f=0:~~&& C(x)=x^2-u,~~G(x)=\Lambda^4\,, \nonumber \\ 
N_f=1:~~&& C(x)=x^2-u,~~G(x)=\Lambda^3(x+m_1)\,, \nonumber \\ 
N_f=2:~~&& C(x)=x^2-u+\frac{\Lambda^2}{8},~~ G(x)=\Lambda^2(x+m_1)(x+m_2)\,, \nonumber \\ 
N_f=3:~~&& C(x)=x^2-u+\frac{\Lambda}{4}(x+\frac{m_1+m_2+m_3}{2})\,, \nonumber \\ 
&& G(x)=\Lambda(x+m_1)(x+m_2)(x+m_3)\,. \nonumber 
\end{eqnarray} 
Here, $m_i$ are the bare mass parameters of the fundamental matter, and $\Lambda$ is the low energy strong coupling scale which enters the infrared physics via dimensional transmutation for asymptotically free theories. 
 
The $g_2(u)$ and $g_3(u)$ functions for the Seiberg-Witten curves given above are easily computed using (\ref{quarticweierstrass1},\ref{quarticweierstrass2}), 
\begin{eqnarray}  
N_f=0:~~&&  g_2(u)= \frac{4u^2}{3}- \Lambda^4 \,, \nonumber \\ 
&&   g_3(u) = \frac{1}{27}(- 8 u^3 +9 u \Lambda^4) \,, \nonumber \\ 
N_f=1:~~&&  g_2(u)= \frac{4u^2}{3}-m_1 \Lambda^3\,, \nonumber \\ 
&&   g_3(u) = -\frac{8u^3}{27} +\frac{1}{3}m_1u \Lambda^3- \frac{\Lambda^6}{16} \,, \nonumber \\ 
N_f=2:~~&& g_2(u) = \frac{4u^2}{3}-p_2 \Lambda^2+ \frac{\Lambda^4}{16} \,, \nonumber \\ 
&&   g_3(u) = -\frac{8u^3}{27} +\frac{1}{3} p_2 u\Lambda^2 - \frac{1}{16} p_1^2 \Lambda^4+ \frac{1}{24} u \Lambda^4  \,, \nonumber \\ 
&& (  ~p_1\equiv m_1+m_2,  ~p_2\equiv m_1m_2 ~) \nonumber \\ 
N_f=3:~~&&  g_2(u)= \frac{4u^2}{3} -\frac{4u}{3} \frac{\Lambda^2}{4^2} -4p_3  \frac{\Lambda}{4}+(p_1^2-2p_2)  \frac{\Lambda^2}{4^2} +\frac{1}{12}\frac{\Lambda^4}{4^4} \,, \nonumber \\ 
&&   g_3(u) = -\frac{8u^3}{27} -\frac{5}{9}u^2 \frac{\Lambda^2}{4^2}+\frac{1}{9}u \left(\frac{\Lambda^4}{4^4}+(6p_1^2-12p_2) \frac{\Lambda^2}{4^2}+12p_3  \frac{\Lambda}{4}\right)  
\nonumber \\ && ~~  -\frac{p_1^2}{12}\frac{\Lambda^4}{4^4}+\frac{p_2}{6} \frac{\Lambda^4}{4^4} 
-p_2^2 \frac{\Lambda^2}{4^2}+\frac{p_3}{3} \frac{\Lambda^3}{4^3}+2p_1p_3 \frac{\Lambda^2}{4^2}  -\frac{1}{216} \frac{\Lambda^6}{4^6}    \nonumber \\  
&& (~p_1\equiv m_1+m_2+m_3,  ~p_2 \equiv m_1m_2+m_1m_3+m_2m_3,~p_3\equiv m_1m_2m_3~ ) \,.\nonumber  
\end{eqnarray} 
Notice that $g_2(u)$ and $g_3(u)$ are polynomials of degree $2$ and $3$ respectively for all of these theories, consistent with the requirement derived in section \ref{family}. $\Delta$ is a degree $2,3,4,5$ polynomial  
of $u$ respectively for the cases $N_f=0,1,2,3$ (anticipating the upcoming section, the discriminant for the $N_f=4$ will be of degree 6 in $u$, likewise for $N=2^*$). 
 
By computing the $J$-invariant of the curve via (\ref{jdefu}) and comparing  
it with (\ref{jdeftau}) and the expansion (\ref{jexp}), we can easily extract  
the asymptotic behavior of $q$ near the weak coupling, large modulus point $u\sim \infty$  
and the singular points $u\sim u_0$, where $u_0$ is a root of the discriminant, $\Delta(u_0)=0$. These are the loci  
where monopoles or dyons become massless. The asymptotic behavior near the large  
complex modulus point $u\sim \infty$ is 
\begin{equation}  
q\sim u^{N_f-4} \,,\qquad q\sim u^0 \ \ {\rm for } \ \ {N=2^*}  
\end{equation}  
thus reproducing the correct (weak coupling) behavior of the $SU(2)$ gauge coupling, 
\be 
\tau \sim \frac{N_f-4}{2 \pi i} \log \frac{u}{\Lambda^2} \,. 
\ee 
 
Invoking the algorithm introduced in section \ref{directintegration}, we can now obtain the partition function for $n+g>0$ in closed form, in principle for arbitrary high $n+g$. In practice, computing time poses an upper limit. We have computed the $N_f=0,1,2,3$ case with general mass parameters up to  $g+n=4,4,3,2$, though these by no means represent a fixed upper bound. We note that the case of $N_f=1$ and $\epsilon_1+\epsilon_2=0$ has been considered earlier in \cite{HK2}. As an example, we reproduce here the formulae for $N_f=1$ at $g+n=2$.  For better readability, we have set the dynamical scale to a constant, specifically $\Lambda= 2^\frac{2}{3}$ (this choice is to simplify comparison with Nekrasov's results, see below; the analogous choice for $N_f=2,3$ are $\Lambda=2, \Lambda=4$). The dynamical scale can be recovered by dimensional analysis. 
\begin{eqnarray} 
F^{(2,0)} &=& - \frac{X}{ 6\Delta^2 }(6u^2-4m_1^2u -9m_1)^2  +\frac{1}{135\Delta^2}\{ 36 u^5+ 312 m_1^2 u^4-2 m_1 (16 m_1^3+1593 ) u^3 \nonumber \\  
&& +108 (28 m_1^3+45 ) u^2-6 m_1^2 (400 m_1^3+567 ) 
   u+27 m_1 (184 m_1^3-189 ) \}, \\ 
F^{(1,1)} &=&  \frac{3X^2}{\Delta^2} (3 u-4 m_1^2 ) (6 u^2 -4 m_1^2 u-9 m_1)  
\nonumber \\ 
&& +\frac{2X}{\Delta^2} \{ 24 u^4 -32 m_1^2 u^3+16 m_1^4 u^2-3 (40 m_1^3+27) u+m_1^2 (64 m_1^3+189) \} 
\nonumber \\ 
&& +\frac{2}{135\Delta^2} \{ 108 u^5 -744 m_1^2 u^4+14 m_1 (16 m_1^3+243) u^3-9 (272 m_1^3+675) u^2 \nonumber \\ \label{NF=1:11} && 
+6 m_1^2 (400 m_1^3+729)   u-6696 m_1^4+6561 m_1 \},   
\end{eqnarray} 
\begin{eqnarray} 
\label{NF=1:02}  
F^{(0,2)} &=& - \frac{45 X^3}{2 \Delta^2} (3u-4m_1^2)^2  - \frac{3 X^2}{4 \Delta^2}  \{252 u^3 -648 m_1^2 u^2+ (352 m_1^4+54 m_1) u+27 (8 m_1^3-9 )  \}  \nonumber \\ 
&&  - \frac{X}{6 \Delta^2} \{324 u^4  -528 m_1^2 u^3+4 m_1 (76 m_1^3-27) u^2-36 (26 m_1^3+27 ) u \nonumber \\ && +3 m_1^2 (128 m_1^3+729) \}Ê- \frac{1}{405\Delta^2} \{ 684 u^5  -3192 m_1^2 u^4+2 m_1 (656 m_1^3+4293) u^3 \nonumber \\ &&  
-378 (8 m_1^3+45 ) u^2+54 m_1^2 (80 m_1^3+183) 
   u+27 m_1 (729-664 m_1^3) \}  \,. 
\end{eqnarray} 
The formulae become increasingly lengthy as the number of flavors increases. We provide a sample formula for $F^{(2,0)}$ in the $N_f=2$ case. Denoting symmetric polynomials in the mass parameters as  
\begin{eqnarray} 
p_1=m_1+m_2, ~~~ p_2=m_1m_2 \,, 
\end{eqnarray} 
the formula for $F^{(2,0)}$ in the case of $N_f=2$ is  
\begin{eqnarray} 
F^{(2,0)}_{N_f=2} &=& -\frac{X}{24\Delta^2} \{  16 u^3 -12 (p_1^2-2 p_2) u^2+(8 p_2^2-40 p_2-4) u+9 (p_1^2-2 p_2) (2 p_2+1) \}  
 \nonumber \\ && + \frac{2}{135\Delta^2} \{ 32 u^7 +108 (p_1^2-2 p_2) u^6+2 (9 p_1^4-36 p_2 p_1^2-4 (68 p_2^2+233 p_2+77)) u^5 \nonumber \\  && +3   (p_1^2-2 p_2) (52 p_2^2+676 p_2+1207) u^4+[-27 (59 p_2+187) p_1^4 \nonumber \\ && +108 p_2 
   (59 p_2+187) p_1^2-2 (8 p_2^4+2008 p_2^3+7824 p_2^2+1246 p_2-151)] u^3 \nonumber \\ &&  
   +\frac{27}{4} 
   [360 p_1^6-2160 p_2 p_1^4+(224 p_2^3+4312 p_2^2-8 p_2-1) p_1^2 \nonumber \\ &&  +2 p_2 (-224 p_2^3-1432 p_2^2+8 
   p_2+1)] u^2   
   -\frac{3}{8} (4 p_2-1) [567 (2 p_2-3) p_1^4 \nonumber \\ &&  -2268 p_2 (2 
   p_2-3) p_1^2+4 (200 p_2^4+2584 p_2^3+336 p_2^2-503 p_2-25)] u \nonumber \\ &&  -\frac{27}{16} [378 (4 
   p_2-1) p_1^6-2268 p_2 (4 p_2-1) p_1^4 \nonumber \\ && +(-1472 p_2^4+12032 p_2^3-2244 p_2^2+92 p_2-65) 
   p_1^2 \nonumber \\ &&  +2 p_2 (2 p_2+1)^2 (368 p_2^2-352 p_2+65)]   \}\,. 
\end{eqnarray} 
 
\subsection{Argyres-Douglas points} 
 
One very interesting feature of the massive $SU(2)$ cases is the existence of conformal points where mutually  
non-local dyons become simultaneously massless. As specializations of the  $SU(2)$    
models with $N_f=1,2,3$, these conformal theories exhibit $U(1)$ effective gauge symmetry and $A_0, A_1, A_2$ flavor symmetry. The  
degeneration of the curves has been analysed in~\cite{APSW}. The corresponding points  
in the moduli space are called Argyres-Douglas points.   
The general decoupling of the scale at these points 
should imply the vanishing of the highest powers of $X$, cutting down the generic leading $X^{3g+2n-3}$ power behavior of the amplitude $F^{(n,g)}$ to the 
conformal leading  $X^{g+n-1}$ power behavior of Table \ref{TabX}. 
This is easily checked  for the various $SU(2)$ Argyres-Douglas 
points. The simplest one appears for $N_f=1$ at $m_1=\frac{3}{4} \Lambda$ and $u=\frac{3}{4}\Lambda^2$. 
We can see from (\ref{NF=1:11},\ref{NF=1:02})  
that the vanishing indeed occurs. Here, we follow the conventions of~\cite{HK2}, where more about the local expansion of $F^{(0,g)}$ 
around these points can be found. We have performed similar checks for 
$N_f=2$, $m_1=m_2=\pm \Lambda$ and $u=\frac{1}{6}\Lambda^2$ and  $N_f=3$ at $m_1=m_2=m_3=\Lambda$ and $u=\frac{1}{32}\Lambda^2$.  
The vanishing of the  
\begin{equation}  
p_k^{(n,g)}=0 \qquad {\rm for } \ \  g+n-1<k \le 3g+2n-3 
\end{equation}  
in (\ref{generalformfng}) provides an increasing number of  
consistency checks on the prescription for fixing the holomorphic ambiguity presented in section \ref{fixingtheambiguity}.

\subsection{Comparison with Nekrasov's formula at weak coupling} 
We can compare our results against the Nekrasov instanton counting formula (\ref{NK1}) which is valid at weak coupling. The latter is expressed in terms of the flat coordinate $a$ at the weak coupling point. The two relations (\ref{relation1}, \ref{relation2}) allow for the computation of the parameters $q=e^{2\pi i \tau}$ and $a$ in terms of asymptotic power series expansions in the complex modulus $u$ near the infinity and monopole/dyon points, and vice versa.  To perform the comparison, we hence invoke (\ref{relation1}) and (\ref{relation2}) to extract asymptotic series around $a \sim \infty$ from our formulae, while expanding the Nekrasov expressions in accord with (\ref{expansion1}). Upon proper fixing of normalizations, the results match; explicitly, for $N_f=0, 1, 2, 3$, we have checked up to instanton number $5, 5, 3, 3$. Predictions for higher instanton number are easily extracted from our formulae. 
 
Recall that one of the boundary conditions we impose when determining the holomorphic ambiguity is that they be regular at large $u$. The explicit evaluation of the Nekrasov functions at a given $h= g+n$ shows that, in fact, these vanish in this limit, scaling as $a^{2-2h}\sim u^{1-h}$. Assuming that this scaling holds for a given $h$, the holomorphic anomaly equations allow us to conclude that the anholomorphic contributions (i.e. the coefficients of positive powers of $X$) to all $F^{(n,g)}$ at $g+n=h+1$ also obey this scaling, hence implying that the scaling holds separately for the anholomorphic contributions and the holomorphic ambiguity. Finding a physical rationale for the scaling $\sim u^{1-h}$ at large $u$ (rather than extracting it as an experimental observation from Nekrasov's expressions) would hence provide over-determined boundary conditions on the holomorphic ambiguity, and thus a consistency check on the computation.

\section{Seiberg-Witten theory: The conformal cases}  \label{conformal_cases} 
In \cite{SW2}, the deformations of two conformal theories were discussed, $SU(2)$ with 4 massive fundamental flavors, and mass-deformed $N=4$ theory, referred to as $N=2^*$. These theories require slightly different ideas compared to the asymptotically free cases, as due to conformal invariance in the massless limit, they depend on a massless coupling constant $\tau_{uv}$ rather than a dynamically generated scale. The definition of the UV coupling is ambiguous. We will revisit the curves proposed by Seiberg and Witten \cite{SW2} in this section, paying special attention to this ambiguity. What is more, both conformal theories together with their massive deformations are conjectured to exhibit an invariance with regard to an $SL(2,\IZ)$ action on the UV parameters of the theory \cite{SW1,SW2}. The direct integration method in contrast relies on the symmetry of the theory under the monodromy group $\Gamma \subset SL(2,\IZ)$ acting on the complex structure $\tau$ of the curve, which corresponds to the effective coupling of the theory, hence an IR parameter. We will establish the relation between these two group actions in this section.

\subsection{The $SU(2)$ $N=4$ theory and its deformation to $N=2^*$}  
Coupling $N=2$ $SU(2)$ super Yang-Mills theory to a massless hypermultiplet in the adjoint representations yields $N=4$ $SU(2)$ super Yang-Mills theory, which is expected to be  
exactly conformal. To apply the holomorphic anomaly equations to solve the theory in general $\Omega$-backgrounds, we are required to study the mass-deformed case to have recourse to the gap condition to fix the holomorphic ambiguity. The Seiberg-Witten curve for this theory has been identified in~\cite{SW2} as follows,  
\begin{equation}  
 y^2=(x-e_1 u -\frac{1}{4} e_1^2 \, m^2)  (x-e_2  u -\frac{1}{4} e_2^2 \,m^2)(x- 
e_3 u -\frac{1}{4} e_3^2 \,m^2) \,.  
\label{N=4curve1} 
\end{equation} 
The $e_i$ form a modular vector of weight 2 and are sometimes referred to as half-periods in the literature.\footnote{\label{half-periods} The name stems from the fact that they can be expressed in terms of the Weierstrass $\wp$-function of periods $\omega_1$, $\omega_2$ as 
\be 
e_1 = \wp\left(\frac{\omega_1}{2}\right) \,, \quad e_2 = \wp\left(\frac{\omega_2}{2}\right) \,, \quad e_3 = \wp\left(\frac{\omega_1+\omega_2}{2}\right) \,.  
\ee 
They satisfy the relation $e_1 + e_2 + e_3 = 0$, and can be expressed via $\theta$-functions as follows: 
\ban \label{defei} 
e_1-e_2 &=& \left(\frac{\pi}{\omega_1}\right)^2\theta_3^4(0|\tau) \,, \\ 
e_3-e_2 &=& \left(\frac{\pi}{\omega_1}\right)^2\theta_2^4(0|\tau) \,, \nn \\ 
e_1-e_3 &=& \left(\frac{\pi}{\omega_1}\right)^2\theta_4^4(0|\tau) \,.\nn 
\ean 
Note that of the three equations (\ref{defei}), only two are independent due to the relation $\theta_3^4(0|\tau)=\theta_2^4(0|\tau)+\theta_4^4(0|\tau)$. The convention chosen by \cite{SW2} is $\omega_1 = \pi$, hence $\omega_2 = \pi \tau$.}  
Aside from the weight, the $SL(2,\IZ)$ action on the half-periods is via a homomorphism 
\be 
\phi: SL(2,\IZ) \rightarrow S_3   
\ee 
into the permutation group on three objects $\{s,v,c\}$. This map is given by the $\rm{mod}\,\, 2$ reduction of the $SL(2,\IZ)$ matrices; one explicit realization is given by 
\be \label{sl2zintos3} 
\begin{pmatrix} 
0 & 1 \\ 
1 & 0 
\end{pmatrix} 
: v \leftrightarrow s  \,, \quad 
\begin{pmatrix} 
1 & 1 \\ 
1 & 0 
\end{pmatrix} 
: v \leftrightarrow c  \,, \quad 
\begin{pmatrix} 
1 & 0 \\ 
1 & 1 
\end{pmatrix} 
: s \leftrightarrow c  \,, \quad 
\ee 
with the following action on the half-periods, 
\ban 
T_{v \leftrightarrow s}:  && e_1 \leftrightarrow e_2  \,,  \nn\\ 
T_{v \leftrightarrow c}:  && e_2 \leftrightarrow e_3  \,,  \nn\\ 
T_{s \leftrightarrow c}:  && e_1 \leftrightarrow e_3 \,.  \label{triality_on_ei} 
\ean 
In the massless limit, the curve (\ref{N=4curve1}) is parametrized by its complex structure, i.e. the argument $\tau$ of the half-periods coincides with the tau parameter of the curve being parametrized by the equation (\ref{N=4curve1}). This is possible as the complex structure for the massless curve is independent of $u$. To see this, note that a shift of $x$ and $y$ removes the $u$-dependence of the curve all together, or simply note the $u$-independence of the $J$-function for this curve. The $u$-independence of the complex structure was a main justification for identifying the $m=0$ limit of (\ref{N=4curve1}) with the Seiberg-Witten curve of the massless $N=4$ theory, as it implements the constancy of the effective coupling along the RG flow. The deformation away from $m=0$ breaks conformal invariance and yields a curve whose $J$-function does depend on $u$, as well as on the mass parameter $m$ and the argument of the half-periods. If we wish to keep these parameters independent (as we should, since the mass deformation breaks $N=4$ to $N=2$ supersymmetry, enough supersymmetry to forbid a potential for $u$), we can hence no longer identify the argument of the half-periods with the complex structure of the curve, and we shall henceforth denote the latter by $\tau_{uv}$. We therefore have two possible $SL(2,\IZ)$ actions in the theory, one on $\tau_{uv}$ and one on the $\tau$ parameter of the curve. We will refer to the former as $SL(2,\IZ)_{uv}$ in this section. Let us examine the relation between these two actions. 
 
Away from the massless limit, the curve as given in (\ref{N=4curve1}) is no longer invariant under $SL(2,\IZ)_{uv}$ if one assumes that $u$ and $m$ are invariant, as the half-periods, in addition to (\ref{triality_on_ei}), also transform via a weight factor $(c \tau_{uv}+d)^2$. To correct for this, it is often assumed that $u$ also transforms as a weight 2 modular form \cite{SW2,DW,MNW}. We will here make this proposal more concrete. We ask how $\tau$ transforms under $SL(2,\IZ)_{uv}$. As the curve (\ref{N=4curve1}) is invariant under this action (assuming the transformation of $u$), so must its complex structure be (we can verify this explicitly below after calculating the $J$-invariant), hence $\tau$ must remain within its $SL(2,\IZ)$ orbit under an action of $SL(2,\IZ)_{uv}$ on $(\tau_{uv},u)$. Given that in the massless limit, $\tau$ and $\tau_{uv}$ can be taken to coincide, the discreteness of the group action makes it natural to require that $\tau_{uv}$ and $\tau$ in fact transform identically under $SL(2,\IZ)_{uv}$.\footnote{ In fact, \cite{MNW} invoke this property to derive the recursion relation for the prepotential in an expansion in $m^2$ which we juxtaposed with the generalized holomorphic anomaly equations of the conformal theories at $g=0$ in section \ref{conf_limit}.} Assuming this in turn enables us to exhibit a weight 2 expression for $u$ explicitly. To this end, let us compute the $J$-invariant of the Seiberg-Witten curve. It proves possible and computationally highly beneficial to replace the dependence on the five parameters $(u,m,e_1,e_2,e_3)$  
by the dependence on three parameter $(\tilde{u}, \tilde{m}, \q)$, given by 
\begin{equation}  
\tilde{u} = u+ \frac{1}{2} e_2\, m^2\,,\quad\tilde{m}^2 = (e_1-e_2) m^2 \,, \quad \q=\frac{e_3-e_2}{e_1-e_2}  \,. 
\label{newparamdef} 
\end{equation} 
Upon shifting $x$ and rescaling both $x$ and $y$, the curve (\ref{N=4curve1}) can be expressed as 
\begin{equation}  
 y^2=x(x+\tu+\frac{1}{4} \tm^2)(x+\tu \q +\frac{1}{4} \q^2\tm^2) \,. 
\label{N=4curve2} 
\end{equation} 
The $J$-invariant of this curve coincides with that of (\ref{N=4curve1}), and is given by  
\be 
J(\tau)=\frac{4 \left(\tm^4 \left(1-\q^2+\q^4\right)+4 \tm^2 \left(2-\q-\q^2+2 \q^3 \right) \tu+16 \left(1-\q+\q^2\right) \tu^2\right)^3}{27 (-1+\q)^2 \q^2 \left(\tm^2+4 \tu \right)^2 
\left(\tm^4 \q (1+\q)+16 \tu^2+4 \tm^2 (\tu+2 \q \tu)\right)^2}  \,.\label{n4j} 
\ee 
Equating this expression with the universal $J$-invariant (\ref{jexp}) now yields a sixth order equation for $u$. Of the six solutions, the one given by 
\be  \label{u_n4} 
u = \frac{m^2}{4} \frac{\aaa^2 (\BBB-\CCC) + \bbb^2(\CCC-\AAA)+\ccc^2(\AAA-\BBB)}{ \aaa (\BBB-\CCC) + \bbb(\CCC-\AAA)+\ccc(\AAA-\BBB)} 
\ee 
is distinguished, in that it indeed transforms as a weight 2 modular form in $\tau_{uv}$ under a simultaneous $SL(2,\IZ)_{uv}$ transformation of both $\tau$ and $\tau_{uv}$.  
 
Note that the monodromy group $\Gamma$ of the theory is {\it not} the full modular group $SL(2,\IZ)$. $\Gamma$ acts only on $\tau$, not on $\tau_{uv}$, and the explicit expression (\ref{u_n4}) demonstrates that $u$ is not invariant under this action.

\subsubsection{Calculating the amplitudes from the curve} 
\label{N=4amplitudesfromcurve} 
In the massless limit, the $J$-invariant is independent of $u$ (hence of $a$), therefore $\tau$ must be likewise, and we obtain the prepotential by trivially performing two $a$ integrations,  
\begin{equation}  
F^{(0,0)} \sim \int^a da \int^a da \,\tau=\frac{1}{2} a^2 \,\tau \,. 
\label{fclassic} 
\end{equation} 
As $\tau$ here is identified with the UV coupling $\tau_{uv}$, we thus conclude that instanton corrections are absent in this case. 
When we break $N=4$ supersymmetry by giving a mass to the adjoint scalar, the corresponding prepotential does exhibit instanton contributions. To calculate these invoking in particular (\ref{nonlogperiod}), we must take care to correctly identify the Seiberg-Witten differential. The rescaling of $x$ and $y$ that transforms (\ref{N=4curve1}) into (\ref{N=4curve2}) rescales the Seiberg-Witten differential to 
\be 
\frac{d \lambda}{du} = \frac{1}{\sqrt{e_2-e_1}} \frac{\sqrt{2}}{4\pi}\frac{dx}{y}  \,. 
\ee 
Note that the factor $\frac{1}{\sqrt{e_2-e_1}}$ is crucial for maintaining the symmetry (\ref{triality_on_ei}) in the amplitudes $F^{(n,g)}$. 
 
We can compute the instanton corrections to the effective coupling $\tau$ by equating (\ref{jdefu}) and (\ref{relation1}), and invoking (\ref{relation2}) (with the constant $c_1$ fixed by the boundary condition $a \sim \sqrt{2u}$ at $u\rightarrow \infty$) to express $u$ as a function of $a$. The first few terms are reproduced here,  
\be  \label{f00forN=4}  
2 \pi i \tau = \log q  + 2 \log \frac{m^2 + 2 a^2}{2a^2}  + 6 \frac{m^4}{a^4}\, q + \frac{3 m^4 (24 a^4+80 a^2 m^2 +35m^4)}{4 a^8}\, q^2+ O(q^3)  \,. 
\ee 
The prepotential (\ref{fclassic}) is obtained by integrating twice with regard to $a$.  
The instanton corrections agree with the ones computed in~\cite{MNW}. As was pointed out in section  
\ref{conf_limit}, the coefficients of the mass deformation of $F^{(0,0)}(a,m)$ satisfy  
the same holomorphic anomaly equations as the $F^{(n,0)}$ in the conformal limit.  
More precisely, defining 
\begin{equation}  
\label{warner} 
F^{(0,0)}(a,m)=\frac{1}{2}a^2\tau + \frac{m^2}{8 \pi i} \hat f_0  \log\left(\frac{2a}{m}\right)- 
\frac{m^2}{4 \pi i}\sum_{n=1} \frac{\hat f_n(\tau)}{2n} \left(\frac{m}{2a}\right)^{2n}\,,    
\end{equation} 
the $\hat f_n(\tau)$ satisfy (\ref{gen_hol_ano-cf1}) with $c_0=\frac{1}{2}$. The first few read 
\begin{equation} 
\hat f_0=2,\quad \hat f_1=\frac{E_2}{3},\quad \hat f_2=\frac{1}{45}\left(5 E_2^2+E_4\right), \quad \hat f_3= \frac{175 E_2^3+84E_2E_4+11 E_6}{3780}\ .  
\end{equation}

Solving recursively for the amplitudes $F^{(n,g)}$ proceeds as before. Note   
that the discriminant $\Delta$ of (\ref{N=4curve1}) is a perfect square given by  
the denominator of (\ref{n4j}).  The one loop $\beta$-function for the mass  
deformed theory at a singular point comes from one hypermultiplet with mass  
$m\sim (u-u_0)$ running in the loop, where $u_0$ is a simple root of the  
discriminant. This determines the leading behavior $F^{(1,0)}\sim \frac{1}{24} \log (u-u_0)$,  
$F^{(0,1)}\sim -\frac{1}{12} \log (u-u_0)$ and $F^{(n,g)}\sim (u-u_0)^{-2(g+n)+2}$. It follows   
that the formulae for the boundary behavior (\ref{gap}) and in particular (\ref{genus1a},\ref{genus1b})  
apply after defining $\Delta_{there}=\sqrt{\Delta}$. This adjustment is necessary because in spite of the discriminant having a double zero at $u_0$, a single particle is becoming massless here. The boundary behavior then fixes the  
ambiguity completely as discussed above. 
   
To emphasize the symmetry properties of the amplitudes, we will express them in the untilded variables $u, m, e_1,e_2,e_3$, though our computation proceeded with the tilded variables. We can then express the $F^{(n,g)}$ as 
\begin{equation} \label{expansionY} 
F^{(n,g)}=\frac{1}{\tilde{\Delta}^{2(g+n)-2}(u)} \sum_{k=0}^{3g+2n-3} Y^k p^{(n,g)}_k(u) \,, 
\end{equation} 
where  
\be 
Y = (e_2-e_1) X 
\ee 
and $\tilde \Delta$ denotes a reduced discriminant  
\ba 
\tilde \Delta &=&\left(\tm^2+4 \tu\right) \left(\tm^2 \q+4 \tu\right) \left(\tm^2 \q+\tm^2+4 \tu\right) \nn \\ 
&=& (4 u -e_1 m^2) (4u-e_2 m^2) (4u-e_3 m^2) \,. 
\ea 
Both $Y$ and $\tilde \Delta$ are separately invariant under the permutations of the half-periods $e_i$ under the $SL(2,\IZ)_{uv}$ action as described above. $Y$ is hence a modular form of weight 4, $\tilde \Delta$ is modular of weight 6.
 
As a sample, we reproduce our results for $n+g=2$:
\ban 
\nonumber 
p^{(2,0)}_0&=&\frac{37  {E_4}^3 m^{10}-11232  {E_4}^2 m^6 u^2-96  {E_4} \left(7  {E_6} m^8 u+2376 m^2  u^4\right)-4  {E_6} m^4 \left(13  {E_6} m^6+20736 u^3\right)}{116640} \,, \\ \nn 
p^{(2,0)}_1&=&  -\frac{\left( {E_4} m^4-144 u^2\right)^2}{432}  \,, \\ \nonumber  
p^{(1,1)}_0 &=& \frac{m^2 \left(- {E_4}^3 m^8+216  {E_4}^2 m^4 u^2+6  {E_4} \left( {E_6} m^6 u+1728 u^4\right)+ {E_6} m^2 \left( {E_6} m^6+2592 u^3\right)\right)}{2430} \,,\\ \nn 
p^{(1,1)}_1 &=&  \frac{1}{108} \left(5  {E_4}^2 m^8+288  {E_4} m^4 u^2+96  {E_6} m^6 u+20736 u^4\right) \,, \\ \nn 
p^{(1,1)}_2 &=& \frac{1}{2} \left(144 m^2 u^2- {E_4} m^6\right)  \,, \\ \nn 
p^{(0,2)}_0&=&\frac{m^2 \left(4  {E_4}^3 m^8+216  {E_4}^2 m^4 u^2+18  {E_4} \left(7  {E_6} m^6 u-5184 u^4\right)+ {E_6} m^2 \left( {E_6} m^6-14688 u^3\right)\right)}{43740}  \,, \\ \nn 
p^{(0,2)}_1&=&  - \frac{1}{54} \left( {E_4}^2 m^8+144  {E_4} m^4 u^2+24  {E_6} m^6 u\right) \,, \\ \nn 
p^{(0,2)}_2&=&\frac{3}{2} m^2 \left( {E_4} m^4-144 u^2\right)  \,, \\   
p^{(0,2)}_3 &=&  -45 m^4  \,. \label{n+g=2} 
\ean 
Note that the argument of the Eisenstein series $E_4$ and $E_6$ occurring in the polynomials $p_i^{(n,g)}$ is the UV parameter $\tau_{uv}$, whereas the argument of the Eisenstein series and half-periods hidden in the variable $Y$ is the tau parameter $\tau$ of the curve. One easily verifies that the weight factors under $SL(2,\IZ)_{uv}$ transformations cancel in (\ref{expansionY}) for each $k$.

\subsubsection{The massless limit of the $N=2^*$ theory} 
\label{N=4massless} 
 
The massless limit is very interesting as the theory becomes conformal and the qualitative  
dependence of the amplitudes on the anholomorphic generator changes. 
 
We can determine the $F^{(n,g)}$ for $n+g=1$ from the massless limit of (\ref{genus1a}) and (\ref{genus1b}), 
\begin{equation}  \label{massless0110} 
F^{(1,0)}(a)=\frac{1}{4} \log(a)+\mbox{$a$-independent terms,}\qquad F^{(0,1)}(a)=0 +\mbox{$a$-independent terms.}
\end{equation} 
The latter equation expresses the fact that the gravitational anomaly cancels  
for the $N=4$ spectrum. 
 
As we have emphasized throughout, the scaling of the highest power of $E_2 \sim X$ in $F^{(n,g)}$ with $n+g$ is modified in conformal theories, from $F^{(n,g)}\sim E_2^{3g+2 n-3}$ to 
$F^{(n,g)}\sim E_2^{g+n-1}$, see Table \ref{TabX}. Explicitly in this example, this can be seen from the  
$m\rightarrow 0$ limit of (\ref{n+g=2}). We immediately obtain the $n+g=2$ results from  
(\ref{massless0110}) and (\ref{deffng},{\ref{gen_hol_ano-cf2}), 
\begin{equation} 
 F^{(2,0)}=\frac{E_2}{3 \cdot 2^6 a^2},\qquad  F^{(1,1)}=-\frac{E_2}{3 \cdot 2^4 a^2},\qquad F^{(0,2)}=0,  
\end{equation} 
since there can be no holomorphic ambiguity, due to the absence of  
holomorphic weight 2 modular forms. For $n+g>2$, the ambiguity must be   
fixed by the gap condition for the mass-deformed case. We obtain up to $g+n=4$ 
\ban 
F^{(3,0)}&=&-\frac{1}{2^{9} 3^2 5 a^4}\left(5E_2^2+13E_4\right),\ \   
F^{(2,1)}=\frac{1}{2^{8} 3^2 5 a^4}\left(25 E_2^2+29E_4\right)\,, \nonumber \\
F^{(1,2)}&=&-\frac{1}{2^{6} 3 \cdot 5 a^4}\left(5 E_2^2+E_4\right),\ \  F^{(0,3)}=0\, , 
\ean 
as well as  
\ban 
F^{(4,0)}&=& \frac{1}{2^{12} 3^4 5\cdot  7  a^6}\left(175 E_2^3+1092 E_2 E_4+ 3323 E_6\right),\nonumber \\  
F^{(3,1)}&=&-\frac{1}{2^{9} 3^4 a^6}\left(11 E_2^3+\frac{2\cdot3\cdot 47}{5} E_2 E_4+ \frac{2231}{5\cdot 7}E_6\right), \nonumber \\  
F^{(2,2)}&=& \frac{1}{2^{8} 3^4 a^6}\left(2^5 E_2^3+\frac{3 \cdot151}{5} E_2 E_4+ \frac{1199}{5\cdot 7}E_6\right), \nonumber \\  
F^{(1,3)}&=&  -\frac{1}{2^{6} 3^3 a^6}\left(5 E_2^3+3 E_2 E_4+ \frac{2^2}{7}E_6\right), \ \  F^{(0,4)}=0 \ . 
\ean 
The argument of the Eisenstein series here is the $\tau$ parameter of the Seiberg-Witten curve, which in the $N=4$ theory coincides with the UV coupling of the theory.
 
The vanishing of $F^{(0,g)}=0$ for all $g$ follows from the holomorphic anomaly equations  
in conjunction with the fact that the one-loop amplitude $F^{(0,1)}$ has no $a$-dependence: $F^{(0,2)}=0$, as modularity rules out a holomorphic ambiguity of weight 2. More generally, assume that $F^{(0,g')}=0$ holds for $g'<g$. It then follows easily from the holomorphic anomaly equations that $F^{(0,g)}$ vanishes as well, up to possibly the holomorphic ambiguity. To fix the ambiguity, we must mass deform the theory. Upon mass deformation, we have $p_{k}^{0,g}\sim m^l$ with $l>0$ for $k \le 1$. One can argue that the gap condition then implies $p_{0}^{0,g}\sim m^p$ with $p>0$. Therefore, the ambiguity vanishes in the massless limit as well and we can conclude  
inductively that $F^{(0,g)}=0$ in the massless limit. 
 
The amplitudes match Nekrasov's expressions  (\ref{NK1adjoint}) in the massive case after  
the shift $m \rightarrow m+(\epsilon_1+\epsilon_2)/2$, followed by setting $m=0$.  
The massless limit from the point of view of (\ref{NK1adjoint}) is  
hence $m=(\epsilon_1+\epsilon_2)/2$.

\subsubsection{Comparison with Nekrasov's formula at weak coupling} 
Matching our results at weak coupling with Nekrasov's requires rescaling various parameters. Specifically, 
\be  \label{higherng} 
F^{(n,g)}(s,g_s^2,a,m) = F^{(n,g)}_{Nekrasov}\!\left(2s,2 g_s^2,\frac{a}{\sqrt{2}}, m \right) 
\ee 
for $n+g>1$. In addition to these rescalings, the cases $n+g \le 1$ require special attention as it is the derivatives of these amplitudes with regard to $a$ that is physical. The $a$ independent terms that are required to obtain a matching are the following at $n+g=0$, 
 
\be
F^{(0,0)}_{Nekrasov}(a,m) =  F^{(0,0)} (\sqrt{2} a, m)  + \pi i \tau_{uv} a^2 -m^2  \log \left(\frac{\eta(\tau_{uv})}{q_{uv}^{\frac{1}{24}}} \right) + const. \,, 
\ee 
where $const.$ indicates an $m$ dependent constant. 
 
At $n+g=1$, we must add simple $q$ dependent terms to our general formulae (\ref{genus1a}) and (\ref{genus1b}) to obtain a match with Nekrasov's results as follows, 
\begin{equation} \label{01nek} 
F^{(1,0)}_{Nekrasov}(a,m)=F^{(1,0)}( \sqrt{2} a, m) +\log\left(\frac{\theta_3(\tau_{uv})}{q_{uv}^\frac{1}{48}}\right)  + const.
\end{equation} 
and 
\begin{equation} \label{10nek} 
F^{(0,1)}_{Nekrasov}(a,m)=F^{(0,1)}(  \sqrt{2} a, m)     -\log\left(\frac{\eta(\tau_{uv})}{q_{uv}^\frac{1}{12}}\right)  + const. \,. 
\end{equation} 
 
We have performed the check of (\ref{higherng}) up to $g+n=4$ and to instanton number 3, and found agreement. Note that the recursion leading to these expressions takes $F^{(1,0)}$, $F^{(0,1)}$ as obtained from (\ref{genus1a}) and (\ref{genus1b}) (with $\Delta_{there} = \sqrt{\Delta}$, as explained above) as its starting point, not (\ref{01nek}) and (\ref{10nek}).

\subsection{The $SU(2)$ $N_f=4$ theory} 
Motivated by the vanishing of the beta function, Seiberg and Witten argued that the UV coupling of the massless $SU(2)$ $N_f=4$ theory receives neither perturbative nor non-perturbative corrections, and should hence be identified with the IR-coupling, as in the $N=4$ case. Indeed, for the massless case, they propose that the two theories share the same Seiberg-Witten curve (the Seiberg-Witten differentials differing slightly in their normalization). Discrepancies between this ansatz and explicit instanton computations were pointed out soon after \cite{SW1,SW2} appeared, e.g. in \cite{dorey_nf4}. These were traced to a freedom in defining the UV coupling and modulus $u$ of the theory in \cite{dorey_res,AP}.  
 
In \cite{GKMW}, by matching calculations in the field theory limit of the type IIA string compactification on the Enriques Calabi-Yau to the results of Nekrasov for amplitudes for massless $N_f=4$ gauge theory, an exact functional relation $q(\tau)$ between the UV coupling $q$ which serves as the instanton expansion parameter and the effective coupling $\tau$ of the theory was conjectured,
\be \label{def_qtau}
q(\tau) = \frac{e_3-e_2}{e_1-e_2}(\tau) = \frac{\theta_2^4(\tau)}{\theta_3^4(\tau)}  \,.
\ee
As realized in \cite{gaiotto}, this choice of UV coupling fits into a larger framework. For a large class of quiver conformal gauge theories based on the $N_f=4$ theory as a building block, \cite{gaiotto} identified parameters on the moduli space of marginal couplings with coordinates on Teichm\"uller spaces of punctured Riemann surfaces. For the case of $SU(2)$ $N_f=4$ theory, the relevant surface is a 4-punctured Riemann sphere, and the natural coordinate is a cross-ratio $q$ of the location of the punctures. Under conformal transformations, these punctures are permuted, resulting in the following transformations of $q$, 
\be \label{s3action} 
q\,, \frac{1}{q}\,, \frac{1}{1- q}\,, 1-q\,, \frac{q}{q-1}\,, \frac{q-1}{q} \,. 
\ee 
$q(\tau)$ as defined in (\ref{def_qtau}) indeed transforms as (\ref{s3action}) upon an $SL(2,\IZ)$ action on $\tau$. Upon turning on masses, the argument $\tau$ in (\ref{def_qtau}) is no longer the effective IR coupling of the theory. When required for clarity, we therefore refer to it as $\tau_{aux}$ in the following. In the next subsection, we review the curve proposed by Seiberg and Witten for $SU(2)$ SYM with $N_f=4$ massive flavors, and discuss its symmetry properties. This curve depends on $\tau_{aux}$. Then, in subsection \ref{deriving_nf4}, we offer a rederivation of this curve as a function of $q$, taking the transformation properties of $q$ as a defining property of the UV coupling. The curve we thus arrive at proves to be equivalent to the curve proposed by Seiberg and Witten.

\subsubsection{The curve of Seiberg and Witten} 
Seiberg and Witten propose the following curve to describe the massless $N_f=4$ theory, 
\be \label{swnf4massless} 
y^2 = (x - u e_1) (x-u e_2) (x-u e_3) \,. 
\ee 
As noted above, this coincides with the massless limit of the $N=2^*$ curve given in (\ref{N=4curve1}). The argument of the half-periods coincides with the complex structure $\tau$ of the curve, and the $SL(2,\IZ)$ action on this parameter, upon invoking the modularity properties of the half-periods reviewed above, clearly leaves the curve invariant.  
 
To obtain the massive theory, Seiberg and Witten describe how to deform the massless case such that the deformation parameters $m_i$, the masses of the fundamental matter, appear as the residues of the Seiberg-Witten differential,  
\be \label{reslambda1} 
{\rm Res}\,\lambda=\sum_{i=1}^4\frac{n_i m_i}{2 \pi i \sqrt{2}}   \,, \quad n_i\in \mathbb{Z}  \,. 
\ee 
They thus obtain the curve 
\be \label{swnf4massive} 
y^2 = W_1 W_2 W_3 + A \left(W_1 T_1 (e_2-e_3) + W_2 T_2 (e_3-e_1) + W_3 T_3 (e_1 - e_2) \right) - A^2 N   \,, 
\ee 
with  
\be 
W_i = x - e_i u - e_i^2 R \,, \quad A=(e_1-e_2)(e_2-e_3)(e_3-e_1) \,,  
\ee 
and 
\ban 
R&=& \frac{1}{2} \sum_i m_i^2 \,,  \nn \\ 
T_{1,3}&=& \pm \frac{1}{2} \prod_i m_i - \frac{1}{24} \sum_{i>j} m_i^2 m_j^2 + \frac{1}{48} \sum_i m_i^4\,, \nn \\ 
T_2&=& \frac{1}{12} \sum_{i>j} m_i^2 m_j^2 - \frac{1}{24} \sum_i m_i^4\,,  \nn \\ 
N&=&  \frac{3}{16} \sum_{i>j>k} m_i^2 m_j^2 m_k^2 - \frac{1}{96} \sum_{i \neq j} m_i^2 m_j^4 + \frac{1}{96}\sum_i m_i^6  \,. \label{mass_terms} 
\ean 
The argument of the half-periods $e_i$ is $\tau_{aux}$. Upon considering $m_i \neq 0$, it can clearly no longer be identified with the complex structure of the curve, as this depends in addition on the value of the masses $m_i$, as well as on $u$. The fact that the argument of the half-periods is equal to the effective coupling of the theory only for vanishing masses is analogous to the $N=4$ case studied above. Unlike that case, the argument of the half-periods in the massive theory is not identified with $\tau_{uv}$. Rather, $q= \exp 2 \pi i \tau_{uv}$ in (\ref{def_qtau}), with the argument $\tau$ in that equation given by $\tau=\tau_{aux}$.

Note that upon setting $m_1\!=\!m_2\!=\!\frac{1}{2} m$, $m_3\!=\!m_4\!=\!0$, the $N_f=4$ curve (\ref{swnf4massive}) reduces to the $N=4$ curve (\ref{N=4curve1}), hence both theories have the same discriminant with double zeros. In the $N_f=4$ case, two particles are becoming massless at these singularities (which can be separated by breaking the degeneracy in the bare masses). In the $N=4$ theory, as pointed out above, a single particle is becoming massless here. Therefore, the starting points of the direct integration of the holomorphic anomaly equations, $F^{(1,0)}$ and $F^{(0,1)}$, differ, explaining the difference of even the massless limit of the two theories.
 
$SL(2,\IZ)_{uv}$ (following the same nomenclature as in the $N=2^*$ case) acting on $\tau_{aux}$ is clearly no longer a symmetry of the curve. As Seiberg and Witten point out, this is in accord with the physics of the $N_f=4$ theory, whose spectrum is not quite $SL(2,\IZ)$ symmetric. The quarks $(n_m,n_e)=(0,1)$, monopoles $(n_m,n_e)=(1,0)$, and dyons $(n_e,n_m)=(1,1)$, which lie in the same $SL(2,\IZ)$ orbit if one takes the tuple $(n_m,n_e)^T$ to transform in the fundamental, lie in different representations of the flavor group $\Spin(8)$; they transform in the $v$, $s$, and $c$ representation respectively (the spinor representations arise by quantizing fermion zero modes of the quark fields in the monopole background).\footnote{The attentive reader will have noted that the $SL(2,\IZ)$ action in question here is the IR $SL(2,\IZ)$ that acts on the complex structure $\tau$ of the curve, rather than on the UV parameter $\tau_{aux}$; the justification will be as in the $N=2^*$ theory, as we discuss below.} The representations of other stable dyons with coprime $(n_m,n_e)$ are determined by the reduction of the charges modulo 2. At best, one can hence hope that $SL(2,\IZ)_{uv}$ combined with a permutation of the representations of $\Spin(8)$, $SL(2,\IZ) \ltimes \Spin(8)$, be a symmetry of the theory (we drop the subscript in the semi-direct product, as it is only the UV $SL(2,\IZ)$ that occurs thus). Permutations of the three fundamental representations $v$, $s$, $c$, of $\Spin(8)$ induce all outer automorphisms of the group, i.e. $\Aut(\Spin(8)) \cong S_3$. The homomorphism $SL(2,\IZ) \rightarrow S_3$ required to define the semi-direct product $SL(2,\IZ) \ltimes \Spin(8)$ is the one given in (\ref{sl2zintos3}), based on reduction $\rm{mod}\,\,2$. This action on automorphisms of $\Spin(8)$ in fact induces a transformation of the masses $m_i$ under $S_3$. To see this, consider the central charge formula 
\be 
Z= n_e a + n_m a_D + \sum_i S_i \frac{m_i}{\sqrt{2}} \,. 
\ee 
The masses generically break the flavor symmetry down to $U(1)^4$, with the $S_i$ the corresponding $U(1)$ flavor charges. Identifying these with the Cartan generators $H_i$ of the full flavor group $\Spin(8)$, we see that the $m_i$ must transform inversely to $H_i$. The Cartan generators in turn transform inversely to the dual basis $e_i(H_j) = \delta_{ij}$, $i,j=1, \ldots, 4$ in which the weights of $\Spin(8)$ can be expanded. The $m_i$ hence transform as the basis vectors $e_i$. In terms of this basis, the weights for the vector representation $v$ are  
$\{\pm e_i\}$, while for the spinors representations, they are $\{\frac{1}{2}\sum \eta_i e_i\}$ with $\eta_i = \pm 1$ and 
$\prod_{i=1}^4\eta_i=1$ for the $s$ and  $\prod_{i=1}^4 \eta_i=-1$  
for the $c$ representation. The transformations of order 2, two of which suffice to generate $S_3$, act as follows on the masses \cite{SW2}, 
\ban 
T_{v \leftrightarrow s}:&& m'_1=\frac{1}{2}(m_1+m_2+m_3+m_4) \,,  \nn\\ 
&&m'_2=\frac{1}{2}(m_1+m_2-m_3-m_4) \,, \nn\\ 
&&m'_3=\frac{1}{2}(m_1-m_2+m_3-m_4) \,, \nn \\ 
&&m'_4=\frac{1}{2}(m_1-m_2-m_3+m_4) \,,  \nn \\ 
T_{v \leftrightarrow c}:&& m'_1=\frac{1}{2}(m_1-m_2-m_3-m_4) \,,  \nn\\ 
&&m'_2=\frac{1}{2}(-m_1+m_2-m_3-m_4) \,, \nn\\ 
&&m'_3=\frac{1}{2}(-m_1-m_2+m_3-m_4) \,, \nn \\ 
&&m'_4=\frac{1}{2}(-m_1-m_2-m_3+m_4) \,,  \nn \\  
T_{s \leftrightarrow c}:  && m_4'=-m_4 \mbox{ and } m'_j=m_j \mbox{ for } j=1,\dots 3 \,. \nn \\ \label{triality_on_masses} 
\ean 
Indeed, the linear combinations of masses $R, T_i, N$ were chosen by Seiberg and Witten such that $R$ and $N$ are invariant under the action of the triality group on the masses (\ref{triality_on_masses}), while the $T_i$ are permuted in the same fashion as the half-periods. The invariance of (\ref{swnf4massive}) under $SL(2,\IZ) \ltimes \Spin(8)$ is thus manifest, up to one concern. The half-periods are not merely permuted under the action of $SL(2,\IZ)_{uv}$, but also transform modularly with weight 2. To accommodate this behavior, we must hence demand that $u$ transform as a weight 2 modular object as well, just as in the $N=4$ case. With somewhat more work, an analogous expression to $(\ref{u_n4})$ should be derivable which transforms as a weight two modular form upon $SL(2,\IZ) \ltimes \Spin(8)$ action on the UV parameters combined with the induced action on the tau parameter of the curve. Following the same logic as in the $N=2^*$ case, we can argue that $\tau$ should transform in the same manner as $\tau_{aux}$ under an action of $SL(2,\IZ) \ltimes \Spin(8)$. This was indeed required by the consistency of the above argument leading to the identification of the $S_3$ factor in the symmetry group, as it is the $SL(2,\IZ)$ transformation of $\tau$ which acts directly on the spectrum by acting on the periods of $\lambda$. Motivated by the 
AGT conjecture, \cite{Tai:2010im} presents an analysis of the action of the triality group on the solutions of the Picard-Fuchs differential equations.     
 
\subsubsection{The curve in terms of the UV coupling} \label{deriving_nf4} 
{\it The massless curve} 
 
We will parametrize our curve in terms of the UV coupling\footnote{Contrary to previous sections, where we distinguished between UV and IR coupling by introducing the notation $q_{uv}$ and $q$ respectively, we will speak of $q$ and $q_{ir}$ in this section, in the interest of economizing subscripts.} 
\be \label{quv} 
q= \exp[ 2\pi i \tau_{uv}] = \exp[2 \pi i \left( \frac{\theta}{2\pi } + i \frac{4\pi}{g^2} \right)_{UV} ]  \,, 
\ee 
The normalization of $\tau_{uv}$ is chosen to yield the canonical form of the SYM Lagrangian with $\theta$ multiplying the integer-valued instanton contribution $\frac{1}{32 \pi^2} \int F \wedge *F$. Following \cite{SW2}, we choose the normalization of the IR coupling such that the induced electric charge of a magnetic monopole of charge $n_m$ is $\frac{n_m \theta_{ir}}{\pi}$. This normalization convention introduces a relative factor of two between $\tau_{uv}$ and $\tau_{ir}$ at weak coupling, i.e. for $q\rightarrow 0$, which is the point in moduli space where the two can be compared directly, 
\be  \label{tauIR} 
\tau_{ir} = 2 \frac{1}{2 \pi i} \log q + \ldots \,. 
\ee 
This implies a monodromy by $T^2$ acting on the periods of the holomorphic 1-form at weak coupling. 
 
A curve with monodromy $T^n$ around $t=0$ can be brought into the form $(x-1) (x^2 - t^n)$ \cite{SW2}, hence $n$ can be read off from the order of vanishing of the discriminant $\Delta = (1- t^{n/2})(1+t^{n/2}) 4 t^n$ at $t=0$. A $T^2$ monodromy at $q=0$ hence implies $\Delta \sim q^2$. Imposing weak-strong duality of $q$ as defined in (\ref{quv}) while maintaining the interpretation of $q$ as a cross-ratio \cite{gaiotto}, we are led to require the symmetry $q \rightarrow 1-q$, yielding $\Delta \sim (1-q)^2 q^2$.\footnote{Note that of the 6 forms (\ref{s3action}) the cross-ratio takes under permutation of three elements, given $q$ as in (\ref{quv}), only $1-q$ continues to correspond to real gauge coupling $g$, which is why we only demand symmetry of the discriminant under the action $q \rightarrow 1-q$. We will rediscover the full action of the permutation group $S_3$ as a symmetry of the full theory.} The simplest polynomial with this discriminant is $F(x) = x (x-1)(x-q)$. We can now fix the $u$ dependence by requiring the relation $ a \sim \sqrt{u}$ at weak coupling, thus arriving at 
\be \label{nf4massless} 
y^2 = x(x-u)(x- u q)  \,. 
\ee 
The $g_2$ and $g_3$ functions for this curve are 
\ban \label{g2_g3_nf4} 
g_2 &=&  \frac{1}{12} \left(q^2-q+1\right) u^2 \,, \\ 
g_3 &=& \frac{1}{432} \left(2 q^3-3 q^2-3 q+2\right) u^3 \,, \nn 
\ean 
yielding the $J$-invariant  
\be   \label{jinv} 
J= \frac{4(q^2-q+1)^3}{27 (q-1)^2 q^2}  \,. 
\ee 
In contrast to the curve (\ref{nf4massless}), its $J$-invariant is invariant under the full action of $S_3$ on $q$.  
 
Note that $J$ is independent of $u$. Hence, $q$ is a fixed function of the infrared coupling, the tau parameter of the curve. To obtain this function, we first invert (\ref{relation1}) around weak coupling,  
\be 
2 \pi i\tau = -\log J - 3\log 12 + \frac{31}{72J} +\ldots \,. 
\ee 
Inserting the $J$-invariant (\ref{jinv}) yields 
\be  \label{qirquv} 
q_{ir} = \frac{q^2}{256} + \frac{q^3}{256} + \frac{29 q^4}{8192} + \ldots  \,, 
\ee 
where we have defined $q_{ir} = \exp(2 \pi i \tau)$. As was first observed in \cite{GKMW}, the inverse function of this series is given by a quotient of Jacobi theta functions,  
\begin{eqnarray}  \label{qtheta} 
q=\frac{\theta_2(q_{ir})^4}{\theta_3(q_{ir})^4}  \,. 
 \end{eqnarray} 
From (\ref{qirquv}), we see that to obtain (\ref{tauIR}) without a constant term, we should rescale $q \rightarrow 16 q$. We opt for retaining the normalization of $q$ which transforms simply under $S_3$. 
 
Invoking the relations (\ref{defei}) between $\theta$-functions and the half-periods, we can rewrite (\ref{qtheta}) as  
\be  \label{q_of_e} 
q(\tau) = \frac{e_3 - e_2}{e_1 - e_2}(\tau)  \,. 
\ee 
It is then not hard to see that our curve (\ref{nf4massless}) is intimately related to (\ref{swnf4massless}), via a shift of $x$ together with a rescaling of both $x$ and $y$. 
 
From (\ref{q_of_e}), we easily conclude 
\ban  \label{trans_q} 
T_{v \leftrightarrow s}:  &&q \mapsto 1-q \,, \nn \\ 
T_{v \leftrightarrow c}:  &&q \mapsto  \frac{q}{q-1}\,,\nn \\ 
T_{s \leftrightarrow c}:  &&q \mapsto \frac{1}{q} \,.  
\ean 
Note that unlike the behavior of the $e_i$, the weight factors upon an $SL(2,\IZ)$ transformation cancel in the definition of $q$. The transformations (\ref{trans_q}) are evidently not a symmetry of the curve (\ref{nf4massless}). But as shifts and rescalings of $x$ and $y$, relating (\ref{swnf4massless}) to our curve, do not alter the $J$-invariant of a curve, the $J$-invariant of (\ref{nf4massless}) must be invariant under (\ref{trans_q}), and one quickly checks that this is the case. 
 
Before concluding that $SL(2,\IZ)$ is a symmetry of the theory, we must specify the Seiberg-Witten differential $\lambda$ and study its modular transformations. A natural guess for its defining equation would be 
\be  \label{trial_lambda} 
\omega_0=\frac{d \lambda_0}{du} = \frac{\sqrt{2}}{8 \pi} \frac{dx}{y} \,. 
\ee 
To determine the periods of $\lambda_0$, it is easiest to use the relation between our curve and (\ref{swnf4massless}). This yields 
\be 
\int_{\Sigma_A} \omega_0 = \frac{\sqrt{2/u}}{4} \sqrt{e_1-e_2} \,,\quad \int_{\Sigma_B} \omega_0 = \frac{\sqrt{2/u}}{4} \sqrt{e_1-e_2} \,\tau \,, 
\ee 
and hence 
\be 
a_0=\int_{\Sigma_A} \lambda_0 = \frac{\sqrt{2u(e_1-e_2)}}{2} \,,\quad a_{D,0}=\int_{\Sigma_B}\lambda_0 = \frac{\sqrt{2u(e_1-e_2)}}{2} \,\tau \,. 
\ee 
Note that the transformation of the vector $(a_{D,0},a_0)^T$ is complicated due to the factor of $\sqrt{e_1-e_2}$. Canceling this factor, and taking into account that $u$ is to transform with weight 2, yields a vector under $SL(2,\IZ)$, as expected. We hence modify our proposal of the Seiberg-Witten differential as follows,  
\be \label{nf4lambda} 
\omega=\frac{d \lambda}{du} = \frac{\sqrt{2}}{8 \pi} \frac{dx}{y}  \frac{1}{\sqrt{e_1 - e_2}} \,. 
\ee

Unlike the case of the curve given by Seiberg and Witten, (\ref{swnf4massless}), this is the first instance in the present formulation in which the half-periods $e_i$ enter explicitly, i.e. not in the form of a cross-ratio identified with $q$. 
 
The choice (\ref{nf4lambda}) defining $\lambda$ is in fact the one that follows from rewriting the proposal of \cite{SW2} in the variable $q$. Thus, the only difference between the discussion in \cite{SW2} and the current one turns out to be the identification of the UV parameter. 
 
\hspace{1cm} 
 
{\it The massive curve} 
 
We next wish to determine the deformation of the curve (\ref{nf4massless}) in the presence of non-vanishing masses. One can easily follow the procedure outlined in the last section of \cite{SW2} based on identifying the masses with the residues of the Seiberg-Witten differential.

This procedure yields 
\begin{equation} 
\begin{array}{rl}   
y^2= &\displaystyle{x(x-u)(x-q u)- x^2 (1-q)^2 \sum_{i=1}^4 \tilde{m}_i^2}\\ [3 mm] 
     &\displaystyle{-4x(1-q)q \left(2 (1+q) \prod_{i=1}^4 \tilde{m}_i + (1-q) \sum_{i< j} \tilde{m}_i^2 \tilde{m}_j^2\right)}\\ [3 mm] 
&\displaystyle{+ 16 (1-q) q^2 \left( u\prod_{i=1}^4 \tilde{m}_i- (1-q) \sum_{i<j<k} \tilde{m}_i^2 \tilde{m}_j^2 \tilde{m}_k^2\right) \,,}    
\end{array}  
\label{massive} 
\end{equation}  
with $\tilde{m}_i = \frac{1}{2} m_i \sqrt{e_1-e_2}$. The argument of the half-periods $e_i$ is obtained by inverting (\ref{q_of_e}) for a fixed parameter $q$. In contradistinction to the $N=2^*$ case, it is this parameter rather than $q$ that will play merely an auxiliary role in the following. 
 
As to be expected following our discussion above, this curve coincides with that given by Seiberg and Witten, upon a redefinition of the $u$-parameter,\footnote{\label{frame_for_masses}In fact, the choice of frame for the masses, different frames related by the transformations (\ref{triality_on_masses}) (while keeping all other parameters fixed), must also be adjusted between the two curves. The definition of the $T_i$ given in (\ref{mass_terms}) is permuted with regard to the definition in \cite{SW2} to this end.}  
\be \label{id_parameters} 
u_{SW}= u_{us} - \frac{1}{2} e_2 R  \,. 
\ee 
We should hence expect to find the same $SL(2,\IZ) \ltimes \Spin(8)$ symmetry underlying the curve (\ref{massive}). As an explicit check, taking into account the transformation of $u$ that follows from (\ref{id_parameters}), 
\ban 
T_{v \leftrightarrow s}: \quad  u &\rightarrow& u + \frac{1}{2} (e_1-e_2) R  \,, \\ 
T_{v \leftrightarrow c}: \quad  u &\rightarrow& u+ \frac{1}{2} \,q R  \,, \\ 
T_{s \leftrightarrow c}: \quad   u &\rightarrow& u  \,,   \label{triality_on_u}  
\ean 
one can easily verify the invariance of the $J$-invariant of (\ref{massive}) under the $SL(2,\IZ) \ltimes \Spin(8)$ action given by (\ref{trans_q}), (\ref{triality_on_u}), and (\ref{triality_on_masses}). 
 
Note that not only the curve, but also the Seiberg-Witten differential $\lambda$ transform under $SL(2,\IZ) \ltimes \Spin(8)$.

\subsubsection{Calculating the amplitudes from the curve} 
 
As the massless curves for the $N=4$ and the $N_f=4$ theories are identical, we again obtain 
\begin{equation}  
F^{(0,0)} \sim \frac{1}{2} a^2 \,\tau  
\end{equation} 
for the massless theory, with $\tau$ the low energy effective coupling and $a$ independent of $\tau$. Here, however, we do not conclude that instanton corrections are absent, as unlike the $N=4$ case, we have identified the UV coupling for the $N_f=4$ theory not with $\tau$, but with $\log q$, with $q$ given in (\ref{q_of_e}). Inserting the expansion (\ref{qirquv}) yields   
\be 
F^{(0,0)} \sim \frac{1}{2} a^2 \left(  \log \frac{q^2}{256} + q + \frac{13 q^2}{32} + \cO(q^3) \right) \,. 
\ee 
Deforming the theory by considering $m_i \neq 0$ breaks conformal invariance and introduces $u$ dependence into the $J$-invariant and thus into the effective coupling. The computation of the prepotential then proceeds as usual via integration of the effective coupling obtained by equating equations (\ref{relation1}) and (\ref{jdefu}) and substituting $u(a)$ as obtained from (\ref{relation2}).\footnote{In the context of D-brane instanton calculations as well as the AGT conjecture, the expansion of the prepotential in masses is given to rather high order in \cite{billo2}.}
 
The computation of the partition functions $F^{(n,g)}$ via the holomorphic anomaly equations proceeds exactly as in the asymptotically free cases. The only difficulty that arises is the excessive computation time due to the large number of parameters for the massive $N_f=4$ case. One can cut down on computation time and solve the general problem by invoking the permutation symmetry in the four mass parameters, as we describe in the next subsection. To make the direct calculation feasible, one can consider special cases, such as setting all masses equal, or some to zero. These cases are in fact non-generic, in that the discriminant of the curve acquires multiple roots, requiring the modifications in determining the holomorphic ambiguity described at the end of section \ref{completeness}. As a proof of principle, we here reproduce the partition function $F^{(1,1)}$ for the case $m_1 \neq 0$, $m_2=m_3=m_4=0 $: 
\ba 
F^{(1,1)} = \frac{e_1-e_2}{\tilde{\Delta}^2} \Big(-6 \hspace{-0.3cm} &X^2&\hspace{-0.3cm}  \left[\tilde{m}_1^6 \left(q^2+5 q-4\right)+2 \tilde{m}_1^4 (5 q+2) 
   u+24 \tilde{m}_1^2 u^2\right] \\ 
+\hspace{-0.4cm} &2X&\hspace{-0.4cm} \big[4 \tilde{m}_1^4 \left(13 q^2+20 
   q-8\right) u^2+4 \tilde{m}_1^6 \left(2 q^3+9 q^2-10 q+4\right) 
   u \\ 
&&\hspace{-0.4cm} +\frac{1}{2} \tilde{m}_1^8 \left(q^4+8 q^3+10 q^2-40 
   q+24\right)+32 \tilde{m}_1^2 (5 q+2) u^3+192 
   u^4\big] \\ 
&&\hspace{-1cm} -\frac{1}{270} \tilde{m}_1^2 \Big(32 \tilde{m}_1^2 
   \left(197 q^3+15 q^2-411 q+410\right) u^3 \\ 
&&\hspace{-1cm} +24 \tilde{m}_1^4 
   \left(95 q^4+158 q^3-142 q^2-414 q+412\right) u^2 \\ 
&&\hspace{-1cm} +6 
   \tilde{m}_1^6 \left(61 q^5+258 q^4-156 q^3-84 q^2-556 
   q+552\right) u\\ 
&&\hspace{-1cm} +\tilde{m}_1^8 \left(22 q^6+183 q^5+204 q^4-628 
   q^3+252 q^2-420 q+416\right)+6528 \left(q^2-q+1\right) 
   u^4\Big) \Big)   \,, 
\ea 
with $\tilde{\Delta}$ the reduced determinant 
\be 
\tilde{\Delta}= (2u + \tilde{m}_1^2) (16 u^2 + 8 q \tilde{m}_1^2 u + (q^2 + 4q-4) \tilde{m}_1^4) \,. 
\ee 
\subsubsection{The case of 4 generic mass parameters} 
For the case of $N_f=4$ with generic mass parameters, it turns out to be computationally intensive  
to fix the holomorphic ambiguity with the gap conditions directly. Instead, the symmetries of the  
theory in the mass parameters can be used to fix the holomorphic ambiguity by computing the anomaly for various  
one parameter mass deformations. More precisely, one must fix the coefficients of the powers  
of the following symmetric polynomials in the masses, 
\begin{eqnarray} \label{0823massp} 
G_2=\sum_{i=1}^4 \tilde m_i^2, ~~~ G_4=\prod_{i=1}^4 \tilde m_i, ~~~F_4=\sum_{i<j} \tilde m_i^2\tilde m_j^2, ~~~ 
G_6=\sum_{i<j<k} \tilde m_i^2\tilde m_j^2\tilde m_k^2 \,. 
\end{eqnarray} 
These are the polynomials that appear in the Seiberg-Witten curve (\ref{massive}) parametrized in terms of the UV parameter $q$. 
Now consider the general form of a higher genus amplitude as given in (\ref{generalformfng}). Singularities of the theory on the $u$-plane occur at zeros of the discriminant. The numerator $p^{(n,g)}_k(u,\tilde m,q)$ must hence be a polynomial in $u$. We shall make the ansatz that it is also a polynomial in the symmetric polynomials (\ref{0823massp})~\footnote{We should mention that it is not obvious  
from the gap conditions why the $p^{(n,g)}_k(u,\tilde m,q)$ cannot be rational functions of the mass polynomials. Our polynomial ansatz however leads to consistent expressions that coincide with Nekrasov's results at weak coupling.} and a rational function in $q$. Keeping in mind that $F^{(n,g)}$ has mass dimension $2-2n-2g$ and $u$ has   
mass dimension $2$, dimensional analysis leads to the following ansatz   
\begin{eqnarray} 
p^{(n,g)}_0=\sum_{2n+2i_1+4(i_2+i_3)+6i_4=22(n+g-1)} u^n G_2^{i_1} G_4^{i_2} F_4^{i_3} F_6^{i_4}   \cdot d_{n,i_1,i_2,i_3,i_4}(q) \,,  
\end{eqnarray} 
where the $d_{n,i_1,i_2,i_3,i_4}(q)$ need to be determined by the gap condition.   
 
For example, for $(n=0,g=2)$, there are 45 unknown coefficients  $d_{0,i_1,i_2,i_3,i_4}$,  
so we need to compute the holomorphic ambiguity for at least 45 independent one mass parameter deformations of the conformal theory  
to fix them. We can choose $(\tilde m_1,\tilde m_2,\tilde m_3,\tilde m_4)= (0,0,0,m), (0,0,m, n_1 m), (0,m,n_1 m, n_2 m), (m, n_1m, n_2m, n_3m)$,  
where $n_1, n_2, n_3$ are positive integers such that $1\leq n_1\leq n_2\leq n_3 \leq 5$.  
 
Using this method, we are able to fix the genus two amplitude for 4 generic mass parameters,  
and successfully compare with Nekrasov's results at low instanton number.   
Let us reproduce here e.g. the coefficient of $X^3$ in $F^{(0,2)}$, whose vanishing is a necessary condition on the conformal points in parameter space. To emphasize the symmetry properties of this expression, we rewrite it in terms of the Seiberg-Witten parameters (\ref{mass_terms}), and introduce the variable $Y = (e_1-e_2) X$, as in the $N=2^*$ theory above. $Y$ transforms with weight 4 under the action of $SL(2,\IZ) \ltimes \Spin(8)$. We then obtain in the notation of (\ref{expansionY}) 
\be 
p_3^{(0,2)} = 45 q^4 (1-q)^4 p^2  \,,   \label{decomp} 
\ee 
with 
\ba 
p&=& 9\left[\left(e_1^2+e_2^2+e_3^2\right) \left(e_1 
   T_1^2+e_2 T_2^2+e_3 T_3^2\right)-3 e_1 e_2 e_3 \left(T_1^2+T_2^2+T_3^2\right) \right]\\ 
&&+  9\left(e_1^2+e_2^2+e_3^2\right) N u -6 u^2 \left(e_1 T_1+e_2 T_2+e_3 T_3\right) \\ 
&& +R \left(27 e_1 e_2 e_3 N-3 u \left(e_1^2 T_1+e_2^2 T_2+e_3^2 T_3\right)+u^3\right) \\ 
&&+\frac{1}{2}R^2 \left(e_1^2+e_2^2+e_3^2\right)  \left(e_1 T_1+e_2 T_2+e_3 T_3\right) -\frac{1}{2} R^3\left(e_1^2+e_2^2+e_3^2\right)u -R^4 e_1 e_2 e_3  \,. 
\ea 
$p$ is clearly invariant under simultaneous permutation of the $e_i$ and $T_i$, and thus transforms with weight 6 under $SL(2,\IZ) \ltimes \Spin(8)$. The discriminant takes the form 
\be 
\Delta = q^2 (1-q)^2 \tilde{\Delta}  \,, 
\ee 
with $\tilde{\Delta}$ of weight 12 under $SL(2,\IZ) \ltimes \Spin(8)$. As required by our analysis above, the total weight thus vanishes. The expression for the discriminant as well as for the other $p_k^{(n,g)}$ are too lengthy to reproduce here, but can be supplied upon request.

\subsubsection{The massless limit of the $N_f=4$ theory } 
\label{Nf=4massless} 
As in the $N=4$ case, the massless limit of the $N_f=4$ case is conformal,  
and again, we can write down modular expressions for the partition functions  
solely in terms of the IR parameters. As the massless curves for the $N=4$  
and $N_f=4$ theories coincide, one might have expected their partition functions,  
expressed in IR parameters, to coincide as well. Interestingly, this  
turns out to not be the case as far as the $F^{(n,g\neq 0)}$ sector is concerned. As pointed out above, fixing the holomorphic ambiguity requires mass deforming the two theories in different ways. The massless limits of the mass deformed amplitudes do not coincide in general. 
However, in the Nekrasov-Shatashvili limit, the amplitudes are related by a rescaling,  
\begin{equation}  \label{NSconformal} 
F^{(n,0)}_{N=4}(a,m=0) =\frac{1}{2} F^{(n,0)}_{N_f=4}(2 a,m=0)\ .  
\end{equation}  
 
The $a$-independence of the coupling $\tau$ in the conformal limit leads,  
just as in the $N=4$ case studied above, to a weaker anholomorphicity  
of the amplitudes than in non-conformal theories, given by $F^{(n,g)} \sim E_2^{n+g-1}$.  
These conformal amplitudes obey the recursion relations (\ref{gen_hol_ano-cf2}), using (\ref{deffng}) with  
$f{(1,0)}=-\frac{1}{4}$ and $f{(1,0)}=\frac{1}{4}$. They are exhibited below up  
to $g+n=4$: 
\begin{eqnarray}  
F^{(2,0)} = \frac{E_2}{2^5 3a^2}\,, ~~~ F^{(1,1)} = -\frac{E_2}{2^3 3 a^2}\,, ~~~F^{(0,2)} = \frac{E_2}{2^5 a^2} \,, 
\end{eqnarray}  
\begin{eqnarray} 
F^{(3,0)} & =& -\frac{1}{2^83^25a^4}(5E_2^2+13E_4) \,, ~~~ F^{(2,1)} = \frac{1}{2^63^25a^4}(10E_2^2+17E_4) \,,  \nonumber \\ 
F^{(1,2)} &=& -\frac{1}{2^83^25a^4} (95E_2^2+94E_4) \,, ~~~ F^{(0,3)} = \frac{1}{2^73 a^4} (2E_2^2+E_4) \,, 
\end{eqnarray} 
\begin{eqnarray} 
F^{(4,0)} &=& \frac{1}{2^{11} 3^4 5\cdot 7 a^6}(175E_2^3 + 1092 E_2E_4+ 3323 E_6)\,, \nonumber \\ F^{(3,1)}&=& - \frac{1}{2^83^45\cdot 7 a^6} (280E_2^3 + 1533 E_2E_4+ 2777 E_6)\,, \nonumber \\  
F^{(2,2)} &=&  \frac{1}{2^{10}3^45\cdot 7 a^6} (5075E_2^3 + 21084 E_2E_4+ 22431 E_6)\,,  \nonumber \\  
\quad F^{(1,3)}\! &=&\!\!\! - \frac{11(70E_2^3 + 189 E_2E_4+ 131 E_6)}{2^83^3 5\cdot 7 a^6}\,, \   F^{(0,4)} \!=\!  \frac{11E_2^3 + 16 E_2E_4+ 7 E_6}{2^{11} 3 a^6}  \,. 
\end{eqnarray} 
The $F^{(0,g)}$ expansions in $E_2,E_4,E_6$ coincide up to a rescaling of $a=\frac{\mu}{\sqrt{2}}$  
with the ones displayed in~\cite{GKMW}. There, an expansion of $F^{(0,g)}$ in terms of modular  
forms was suggested by the gauge theory limit of type II string theory 
on the Enriques Calabi-Yau space. In ~\cite{GKMW}, the coefficients in front  
of the modular forms, however, were fixed in general by comparing with  
Nekrasov's result. Here, we have calculated $F^{(n,g)}$ independently in the  
B-model, which to all orders in $n+g$ yields a definite expansion in  
terms of modular forms. It remains to prove to all orders in  
$q$ that Nekrasov's sum over partitions can be rewritten in a modular way.

\subsubsection{Comparison with Nekrasov's formula at weak coupling} 
To compare our results with the weak coupling results of Nekrasov, we must express them in the same basis for the masses, related to each other via the transformations (\ref{triality_on_masses}), in which Nekrasov's partition functions are given. It turns out that the two frames are related by the action of $T_{v \leftrightarrow s}$ on the masses. At $g+n \neq 1$, we reproduce Nekrasov's results via 
\be 
F^{(n,g)}(s,g_s^2,a,{\bf m})= F^{(n,g)}_{Nekrasov}\!\left(2s, 2g_s^2, \sqrt{2}a ,T_{v \leftrightarrow s}{\bf m} \right) \,.   
\ee 
 
Again, the amplitudes at $n+g \le 1$ match up to $a$ independent terms. Specifically, at $n+g=0$, 
\ba 
2 F^{(0,0)}_{Nekrasov} (a, {\bf m}) &=& F^{(0,0)} \!\left(\frac{a}{\sqrt{2}}, T_{v \leftrightarrow s}{\bf m} \right) + \pi i \tau_{uv} a^2 \nn\\ 
&& + \frac{1}{2} \sum_i m_i^2 \log (\theta_3 \theta_4) + \frac{1}{4} \sum_{i<j} m_i m_j \log(1-q_{uv}) + const.  \,, 
\ea 
where $const.$ indicates an $m$ dependent constant. The argument of the $\theta$-functions is such that the relation (\ref{qtheta}) holds. 
 
At $g+n=1$, we obtain 
\begin{equation} 
F^{(1,0)}_{Nekrasov} (a, {\bf m})    = F^{(1,0)}\!\left(\frac{a}{\sqrt{2}}, T_{v \leftrightarrow s}{\bf m} \right)     +\frac{1}{24} \log \frac{(1-q)^4}{q^2} 
\end{equation}  
and   
\begin{equation} 
F^{(0,1)}_{Nekrasov}  (a, {\bf m})   = F^{(0,1)}\!\left(\frac{a}{\sqrt{2}}, T_{v \leftrightarrow s}{\bf m} \right)   +\frac{1}{6} \log (1-q) q \,, 
\end{equation}  
where $F^{(1,0)}$ and $F^{(0,1)}$ on the right hand side follow from (\ref{genus1b}) and (\ref{genus1a_hol}) with $\Delta$ the discriminant of the curve (\ref{massive}).  
 
We have performed the check of (\ref{higherng}) up to $g+n=2$ and to instanton number 3, and found agreement.

\subsection{The $SU(N)$ $N_f=2N$ curves} 
In section \ref{deriving_nf4}, we derived the Seiberg-Witten curve for the $N_f=4$ $SU(2)$ theory by imposing strong-weak duality for the UV coupling. Our result proved to coincide with that of Seiberg and Witten \cite{SW2}, requiring solely a different identification of the UV coupling. In \cite{Argyres:1995wt}, a curve is presented for $N_f=2N$ $SU(N)$ theories for general $N$, derived using factorization arguments reducing the problem to the $SU(2)$ curve of Seiberg and Witten. By performing the same re-identification of the UV coupling, we hence expect this curve to allow for the generalization of our results to higher $N$, and in particular to reproduce, in the weak coupling limit, the corresponding results of Nekrasov. In this section, we present the curve of \cite{Argyres:1995wt} in terms of the UV parameter $q$  (see also \cite{keller}) and identify the parameter redefinitions required in the $N=2$ case to relate to our results above. 
 
The curve of \cite{Argyres:1995wt} in a convenient normalization, and with the dependence on the UV coupling $q$ made explicit, is 
\begin{equation} \label{APS} 
y^2=(\PN(x,u))^2 - \frac{4 q}{(1+q)^2} \prod_{i=1}^{N_f}(x+  \tm_i-\frac{2q}{1+q}  \mu) \,, 
\end{equation} 
with  
\begin{equation}  
 \PN=\det(x-\langle \Phi \rangle)=x^N-\sum_{i=0}^{N-2} u_{N-i} x^i \,,\quad  
\mu=\frac{1}{N_f}\sum_{i=1}^{N_{f}} \tm_i \,. 
\end{equation} 
$\Phi$ is the scalar field in the $N=2$ vector multiplet, such that the $u_i$ are symmetric polynomials in the diagonal elements of the VEV of this adjoint valued field (recall that the potential of the $N=2$ theory implies that this field can be diagonalized by gauge transformations in a vacuum). The first term in this curve is based on the pure $SU(N)$ generalization~\cite{Klemm:1994qs}\cite{Argyres:1994xh}  
of~\cite{SW1}, while the mass dependent terms are fixed by requiring the correct residua of the Seiberg-Witten differential. This is identified as 
\begin{equation} 
\lambda =\frac{x-\frac{2q}{1+q}\mu }{2 \pi i}d\left[\log\left(\frac{\PN+y}{\PN-y}\right)\right]  \,,   \label{lambda_aps} 
\end{equation} 
such that  
\be 
\frac{d \lambda}{d u_i}=\frac{1}{2\pi i} \frac{dP_N}{du_i} \frac{dx}{y} = \frac{1}{2\pi i}\frac{x^{N-i}dx}{y} \,,  
\ee 
and the residua of $\lambda$ at its poles $x_i$ are 
\be 
\Res \lambda |_{x_i} = \pm \frac{\tm_i }{2\pi i} \,. 
\ee 
For the $SU(2)$ case, a redefinition of masses via triality, given explicitly by $T_{v\leftrightarrow s} \circ T_{ s \leftrightarrow c}$, together with a redefinition of the $u$ parameter given by 
\be 
u_{our\,\, curve} = - 4(q+1) u_{here} + (1+q) \sum_{i=1}^4 \tm_i^2 - \frac{q}{1+q} \sum_{i,j=1}^4 \tm_i \tm_j 
\ee 
equates the $J$ function of the curve (\ref{APS}) to that of (\ref{massive}). Notice that the masses $\tm_i$ are those defined below (\ref{massive}), hence $T_{v\leftrightarrow s}$ acts in addition to (\ref{triality_on_masses}) by multiplication by a factor $i$ due to (\ref{triality_on_ei}). As before, to ensure triality invariance of the theory, we should modify the definition (\ref{lambda_aps}) of $\lambda$ by a factor of $\frac{1}{\sqrt{e_1-e_2}}$.

\subsection{Superconformal field theories with $E_n$ global symmetry} 
\label{E_n} 
At a singular point of the moduli space where the discriminant of the Seiberg-Witten curve vanishes, a charged particle becomes massless. We may tune the mass parameters in Seiberg-Witten theory to special values such that some roots of the discriminant collide. If the massless particles at the colliding roots are mutually local, then the gap condition at the degenerate root is still valid and our method for solving the higher genus amplitudes applies. However, if the massless particles at the colliding roots are mutually non-local, the gap condition fails. To solve the model, we must then first deform by mass parameters to split the degenerate root, then recover the original theory by considering the limit of degenerate root of the mass deformed higher genus amplitudes.  
 
It is expected that some non-trivial superconformal theories appear at singular points with mutually non-local massless particles. These theories do not have a local Lagrangian description in terms of proper physical degrees of freedom. Some of these superconformal field theories were found in~\cite{APSW}. They exhibit $U(1)$ gauge symmetry and global $A_0, A_1, A_2, D_4$ symmetries, and can be obtained by taking special values of mass parameters in $SU(2)$ Seiberg-Witten theory with fundamental matter. Since we can solve the higher genus amplitudes for $SU(2)$ Seiberg-Witten theory with generic mass parameters, we can obtain the higher genus amplitudes for these superconformal theories as well by taking the appropriate limits.  
 
It turns out that there exist other $U(1)$ superconformal theories that have global $E_6, E_7, E_8$ symmetries, and that cannot be obtained from conventional Seiberg-Witten theories. The elliptic curves describing these theories without mass deformations are \cite{MN1, MN2} 
\begin{eqnarray} \label{masslessENcurve} 
E_6: && ~~ y^2=x^3-\rho^4 \nonumber \,,\\ 
E_7: && ~~ y^2=x^3-2\rho^3x \nonumber \,, \\ 
E_8: && ~~ y^2=x^3-2\rho^5 \,,  
\end{eqnarray} 
where $\rho$ is the expectation value of a scalar field that lives on a complex plane identified with the moduli space of the theory. The Seiberg-Witten 1-form $\lambda_{SW}$ satisfies, as usual,  
\begin{eqnarray} 
\frac{ d\lambda_{SW}}{d\rho} \sim \frac{dx}{y}  \,. 
\end{eqnarray} 
The higher genus amplitudes of the theory can be computed as for the conventional gauge theories above by integrating the holomorphic anomaly equations and imposing the gap condition, provided that there are no common roots between $g_2(\rho)$, $g_3(\rho)$ and the discriminant $\Delta(\rho)$. This should coincide with the condition that the massless particles appearing at any root of the discriminant are mutually local. It is easy to see that the discriminants of the above massless curves (\ref{masslessENcurve}) have only one degenerate root at $\rho=0$, which is also the root of $g_2(\rho)$ or $g_3(\rho)$. To solve the theory, we should hence turn on mass deformations that split this degenerate root.  
 
The relevant mass deformations of the curves are described in~\cite{MN1, MN2}. It is quite complicated to solve the model with all possible generic  
mass deformations. We shall here content ourselves with studying some simple deformations sufficient for ensuring that $\Delta(\rho)$, $g_2(\rho)$, and $g_3(\rho)$  
not have common roots. Let us illustrate the idea for the $E_6$ curve. In this case, the parameter $\rho$ has mass dimension 3. We consider the subgroup $U(1)\times SO(10) \subset E_6$ and deform by two mass parameters $T_2$ and $T_4$ which are the degree $2$ and $4$ symmetric polynomials of the masses $m_i$, $i=1,\ldots,5$,  
in the Cartan algebra of $SO(10)$ as well as the scale $\lambda$ of the $U(1)$~\cite{MN1}.  
The deformed curve is  
\begin{eqnarray} \label{massE6curve} 
y^2&=& x^3 - (\rho ^2 \left(12 \lambda ^2+ T_2\right)+\frac{T_4^2}{3}+8 \lambda  \rho  T_4)x- 
(\rho ^3 \left(4 \lambda  T_2-16 \lambda ^3+\rho \right) \nonumber \\  
&& +\frac{1}{3} \rho ^2 T_4 \left(60 \lambda^2+T_2\right)+\frac{2 T_4^3}{27}+\frac{8}{3} \lambda  \rho  T_4^2) \,. 
\end{eqnarray} 
When either $\lambda=0$ or $T_2=0$, $g_2(\rho)$ and $g_3(\rho)$ no longer have common roots, and our methods apply. For example, for $\lambda=0$ we can use the reduced discriminant 
\begin{equation}  
\Delta= \rho  \left(27 \rho^4-4 \rho^2 T_2^3-T_2^2 T_4^2+18 \rho^2 T_2 T_4+4 T_4^3\right) 
\end{equation} 
in (\ref{genus1a}) to set up the direct integration procedure. Setting also $T_4=0$ for notational simplicity, we obtain for $n+g=2$\footnote{The $T_4$ dependence on this slice as well as   
the results on the other subslice $T_2=0$ have been checked for consistency up to $n+g=5$, and are  
available on request.}       
\ban 
\nonumber 
p^{(2,0)}_0&=& -\frac{1}{20} \rho ^6 T_2^2 \left(63 \rho^2+17 T_2^3\right) \,, \quad p^{(2,0)}_1= -\frac{3}{4} \rho ^4 \left(T_2^3-9 \rho ^2\right)^2\,, \\ \nonumber  
p^{(1,1)}_0 &=&\frac{1}{20} \rho ^6 T_2^2 \left(207 \rho ^2+38 T_2^3\right) \,, \quad p^{(1,1)}_1 = \frac{3}{4} \rho ^4 \left(243 \rho^4+4 T_2^6-45 \rho ^2 T_2^3\right)\,, \\ \nn  
p^{(1,1)}_2 &=&\frac{27}{8} \rho ^4 T_2 \left(9 \rho ^2-T_2^3\right)\,, \quad  
p^{(0,2)}_0=    -\frac{1}{20} \rho ^6 T_2^2 \left(96 \rho ^2+19 T_2^3\right)  \,, \\ \nn   
p^{(0,2)}_1&=& -\frac{9}{8} \rho ^4 \left(108 \rho ^4+2 T_2^6-17 \rho ^2 T_2^3\right) \,, \quad p^{(0,2)}_2=  -\frac{27}{32} \rho ^4 T_2 \left(63 \rho ^2-2 T_2^3\right)   \,, \\   
p^{(0,2)}_3 &=&  -\frac{405}{64} \rho ^4 T_2^2 \,. \label{e6:n+g=2} 
\ean 
These expressions obviously satisfy the consistency conditions for the conformal limit $T_2=0$. It would be interesting to compare them against the $W_3$-conformal  
field theory that arises in the AGT conjecture for $SU(3)$ quivers.     
 
Let us discuss the expected behaviour of the theory at  $\rho\sim \infty$ further.   
As in the Seiberg-Witten case, the elliptic parameter $\tau$  of the curve can be obtained from the $J$-function of the curve. However, unlike the  
Seiberg-Witten theories studied in previous sections, the $\rho\sim \infty$ point does  
not correspond to weak coupling $\tau\sim i\infty$ where $J(\tau)\sim \infty$.  
Instead, at $\rho\sim \infty$  we find for the $E_6$ curve (\ref{massE6curve}) at $\lambda=0$ 
\begin{eqnarray} 
J(\tau) \sim \rho^{-2} \sim 0 \,. 
\end{eqnarray} 
The coupling of the curve $\tau$ hence lies at a zero of $J(\tau)$, given by $\tau_0=e^{\frac{\pi i}{3}}$ and $\tau_0=e^{\frac{2\pi i}{3}}$. $\tau_0$ is a simple root of $E_4$, not a root or pole of $E_2$ and $E_6$, and a triple root of $J(\tau)$. The scaling behaviors near $\rho\sim \infty$ are  
\begin{eqnarray} 
&& J(\tau) \sim (\tau-\tau_0)^3 \sim \rho^{-2} \sim 0, \nonumber \\ 
&& E_4(\tau)\sim (\tau-\tau_0)\sim  \rho^{-\frac{2}{3}} \,. 
\end{eqnarray} 
The two periods of the curve $\Omega_0$ and $\Omega_1$ can be found by solving a Picard-Fuchs differential equation for the differential $\frac{dx}{y}$, and as in the Seiberg-Witten case, we find two linearly independent solutions    
 \begin{eqnarray} 
 \frac{d\Omega_0}{d\rho} \sim \sqrt{\frac{g_2}{g_3}\frac{E_6}{E_4}},~~~~ \frac{d\Omega_1}{d\rho} \sim \tau \sqrt{\frac{g_2}{g_3}\frac{E_6}{E_4}}  \,. 
 \end{eqnarray} 
 Both $\Omega_0$ and $\Omega_1$ have the scaling behavior $\Omega_{0,1} \sim  \rho^{\frac{1}{3}}$. We can take a linear combination of two solutions with smaller scaling exponent of $\rho$ at $\rho\sim \infty$ as  
 \begin{eqnarray} 
 \frac{d t}{d\rho} \sim (\tau-\tau_0) \sqrt{\frac{g_2}{g_3}\frac{E_6}{E_4}},  
 \end{eqnarray} 
 such that the coordinate $t\sim \rho ^{-\frac{1}{3}}$. From previous experience with the expansion of topological string amplitudes around orbifold points in Calabi-Yau moduli space, we expect that this parameter $t$ should be the flat coordinate around $\rho\sim \infty$, hence the expansion of the higher genus amplitudes of the elliptic curves in terms of $t$ should give rise to interesting topological invariants. 
 
Since no effective weak coupling local Lagrangian description at the $\rho\sim \infty$ point for elliptic curves with $E_n$ global symmetry is known, we cannot test our results by comparing their expansion around $\rho\sim \infty$  with known perturbative and instanton calculations. On the other hand, this challenge is also an opportunity, as our method can provide previously unknown results regarding these superconformal theories at $\rho\sim \infty$. We will pursue this study in a future publication~\cite{workinprogress4}.

\subsection{The ambiguity of the UV coupling in the light of geometric engineering}  
 
While the $N=4$ and the $N_f=4$ conformal theory have the same Seiberg-Witten curve, we have identified two different bare coupling parameters for these theories above,\footnote{In conformal theories, the nomenclature {\it UV couplings} is ill-advised. One can speak of bare couplings, or of coordinates on the space of marginal deformations of theory. Both nomenclatures emphasize the ambiguity in the choice of these parameters.}  the tau parameter or effective coupling of the theory in the $N=4$ case, and the logarithm of the modular function of the tau parameter given by $q$ in (\ref{qtheta}) for the $N_f=4$ theory. The choice of the bare parameters remains ambiguous \cite{AS, gaiotto}. In the case of $N=4$ supersymmetry, our identification can be justified by the non-renormalization of the coupling. For the $N_f=4$ theory, the space of marginal deformations more generally of superconformal $SU(2)$ quiver theories can be naturally identified with the Teichm\"uller space of an $n$-punctured sphere \cite{gaiotto}, on which a natural choice of coordinates is given by cross-ratios of the puncture coordinates. In the case of $N_f=4$, $n=4$ and the transformation of the cross-ratio under $SL(2,\IZ)$ is that of $q$ in (\ref{qtheta}). A final justification for these choices is the comparison to Nekrasov's results, and this indeed is how the expression for $q$ was originally divined in \cite{GKMW}. 
 
In this subsection, we wish to outline another perspective on resolving this ambiguity. The $N=4$ and the $N_f=4$ theory can be obtained as different limits of a string theory compactification on the Enriques Calabi-Yau. We will argue that the A-model flat coordinate (corresponding to $\tau$) is the appropriate choice in the geometry yielding the $N=4$ theory, while the B-model algebraic coordinate (corresponding to $q$) is the appropriate choice in the $N_f=4$ limit. 
 
A slight modification of the algebraic coordinate, which can also be motivated from the geometric engineering approach and was presented in \cite{keller}, allows us to rewrite our results derived above for the $N_f=4$ theory to obtain the corresponding amplitudes for the $N=2$ $N_f=4$ gauge theory with gauge group $Sp(1)$. 
 
\subsubsection{The Enriques Calabi-Yau}   
 
The Enriques Calabi-Yau manifold $ECY$ is a $K3$ fibration over $\mathbb{P}^1$, obtained as a $\IZ_2$ quotient of $K3 \times T^2$, with the $\IZ_2$ involution acting by inversion on the coordinates of $T^2$ and via the Enriques involution on the $K3$. Heterotic/type IIA duality, which identifies the heterotic string compactified on $T^4 \times T^2$ with the type IIA string compactified on $K3 \times T^2$, survives the quotienting~\cite{Ferrara:1995yx}. The $\IZ_2$ involution acts on the five summands of the Narain lattice of the heterotic string on $T^4 \times T^2$, 
\begin{equation}  
\Gamma^{6,22}= 
[\Gamma^{1,1}\oplus E_8(-1)]\oplus [\Gamma^{1,1}\oplus E_8(-1)]\oplus \Gamma_g^{1,1} \oplus \Gamma_s^{1,1} \oplus \Gamma^{2,2}   \,, 
\end{equation}   
which is identified with the cohomology ring of $K3 \times T^2$, via  
\begin{equation}  
|p_1,p_2,p_3,p_4,p_5\rangle \mapsto e^{\pi i \delta\cdot p_4}  
|p_2,p_1,-p_3,p_4,-p_5\rangle \ . 
\label{z2} 
\end{equation}  
In the IIA picture, $\Gamma_s^{1,1}=H_0(K3,\mathbb{Z})\oplus H_4(K3,\mathbb{Z})$ and the $\Gamma^{2,2}$ lattices encodes the cohomology of $T^2$ (see e.g.~\cite{GKMW} for more detailed explanations). 
 
As usual in heterotic/Type IIA duality, the complexified K\"ahler parameter 
of the base $\IP^1$, $t_{\mathbb{P}^1}$, is identified with the heterotic dilaton $S$. In the gauge theory limit, it becomes the complexified infrared gauge coupling  
\begin{equation} 
t_{\mathbb{P}^1}=\tau_{ir}=\frac{\theta}{\pi} +\frac{8 \pi i}{g^2}\ . 
\label{tauir} 
\end{equation}  
 
The $ECY$ has Euler number zero and is self-mirror.  
Since there are no worldsheet instanton contributions at genus zero, as proven  
in~\cite{MP2006}, its K\"ahler moduli space ${\cal M}_{ECY}$ is a symmetric space. It factors into ${\cal M}_{T^2}\times {\cal M}_E$, where the $T^2$ factor is  
${\cal M}_{T^2}={\rm SL}(2,\mathbb{Z})\backslash {\rm SL}(2,\mathbb{R})/SO(2)$ 
and the Enriques factor ${\cal M}_E={\rm Gr}(\Pi^{(2)},\Gamma_E^{2,10})/O(\Gamma_E)$ is  
related to its total cohomology lattice  
\begin{equation}  
\Gamma^{2,10}_E=\Gamma_s^{1,1}\oplus \Gamma_g^{1,1}(2)\oplus E_8(-2)\ , 
\end{equation} 
the invariant part of the $K3$ lattice under (\ref{z2}). Another  
consequence of the absence of instantons is that the gauge field theory  
limits are conformal\footnote{The gravitational  
$\beta$ functions for the coupling between the self-dual curvature and  
the graviphoton do not vanish, but seem even in generic directions to break  
the conformal symmetry less than normal, reflected in the slower than generic growth of   
the power of the anholomorophic generator in the fiber direction as reproduced in Table \ref{TabX}. 
In the $N=4$ (gauge) theories, these couplings vanish.}, which is reflected in the growth of the power of  
the anholomorophic generator in the base, cf. Table ~\ref{TabX}. 
 
Gauge group enhancement occurs when representatives in homology classes  
defined by the lattice vectors $e$ with $e^2=-2$ shrink to zero size. In terms of the moduli  
described by the symmetric space ${\rm Gr}(\Pi^{(2)},\Gamma_E^{2,10})$, this  
means that the two plane $\Pi^{(2)}$ is rotated to be orthogonal to the $e$'s.  
Two principal situations occur: 
 
If the $e$'s are chosen in the lattice of $2$-cycles $\Gamma_g^{1,1}(2)\oplus E_8(-2)$, an  
$N=4$ theory with maximal gauge group $SU(2)\times E_8$ can be  
engineered~\cite{Ferrara:1995yx}. Shrinking 2-cycles does not invalidate the geometric description of the compactification manifold. One can hence remain in the type IIA picture and describe the physics in terms of the flat coordinate (\ref{tauir}).~\footnote{In the calculation of~\cite{GKMW}, the $F^{(g)}$  
come from {\sl geometric reduction} in which the reduction vector  
is embedded in $\Gamma_s^{1,1}$ and becomes zero in the field theory limit.}  
 
The second situation is that $e\in \Gamma_s^{1,1}$.  This choice can  
only lead to $SU(2)$ enhancement, with conformal  
matter spectrum $N_f=4$, as argued in~\cite{Ferrara:1995yx}. Since the volume of the whole $K3$  
fibre collapses, it is now appropriate to use the mirror type IIB picture and describe the physics in terms of the algebraic mirror coordinates. In particular, the base $\IP^1$ of the ECY can be described as the $\IZ^2$ quotient under $y \rightarrow -y$ of a hyperelliptic $T^2$, 
 \begin{equation}  
y^2=\prod_{i=1}^4(x-f_i)  \,. 
\end{equation} 
Non-redundant complex structure variables yielding the algebraic coordinate for the base are given by cross-ratios  
$\frac{(f_i-f_j)(f_k-f_l)}{(f_i-f_l)(f_k-f_j)}$. A choice of cross-ratio $q=\frac{\theta_2^4(\tau_{ir} )  }{\theta_3^4(\tau_{ir})}$  was identified by \cite{GKMW} as the UV coupling with the correct weak coupling behavior $\tau_{ir}=\frac{2}{2 \pi i} \log(q/2^4)+{\cal O}(q)$.\footnote{The calculation of the $F^{(g)}(\tau_{ir})$ can be performed   
in the field theory limit of {\sl Borcherds reduction}, in which the reduction vector  
is embedded into $\Gamma_g^{1,1}$, up to low $g$. After using the inverse   
mirror map in $F^{(g)}(\tau_{ir})$, one reproduces Nekrasov's results.} This is the relation that entered into the AGT  
correspondence~\cite{AGT}.

\subsubsection{The $Sp(1)$ $N_f=4$ instanton sum}  
Above we referred to ``$Su(2)$ instantons'' as the result that 
is obtained by decoupling of the $U(1)$ part inside $U(2)$, 
which is straightforward in the formalism of~\cite{Nekrasov},
see e.g.~\cite{GKMW}.

However instantons of the conformal $Su(2)=Sp(1)$ theory were also 
calculated directly by methods similar to~\cite{Nekrasov} in~\cite{keller}.  
This calculation is much more involved and could be done only for low   
instanton numbers, but it was concluded that relative to 
the ``$Su(2)$ instantons'' the difference is due to a simple change 
in the identification of the UV coupling. Relative to the $Su(2)$ case 
the relation between IR- and the UV couplings relevant for the $Sp(1)$ 
instantons is obtained by doubling both of them, i.e.  
$q_{Sp(1)}^2=16 \frac{\theta_2^4(2\tau_{ir})}{\theta_3^4(2\tau_{ir})}$. 
Using the doubling identities $\theta^2_{2}(2\tau)=\frac{1}{2}(\theta_3^2-\theta_4^2)$,  
$\theta^2_{3}(2\tau)=\frac{1}{2}(\theta_3^2+\theta_4^2)$  
and the Jacobi identity, this can be written as   
\begin{equation}  
q_{SU(2)}=\frac{q_{Sp(1)}}{(1+\frac{q_{Sp(1)}}{4} )^2 }\ , 
\label{su2sp1rel} 
\end{equation} 
with $q_{SU{2}}$ given by (\ref{qtheta}). \cite{keller} also argue for 
this relation by demonstrating that the $SU(2)$ Seiberg-Witten curve 
that arises naturally from a brane construction of the theory is the 
double cover of the corresponding $Sp(1)$ curve, with this 
identification of the parameters. Such relations between SW-curves 
is also natural from the geometric engineering point of view, where $SO$  
and $Sp$ gauge groups are engineered by $\mathbb{Z}_2$ monodromy 
actions on the homology of ALE fibres (non-compact limits of K3) 
when moving along closed curves in the base.

The result of \cite{keller} in particular implies that our expressions in section \ref{Nf=4massless} 
reproduces the  $Sp(1)$  instatons sums in the conformal limit upon substituting (\ref{su2sp1rel}). 
Hence, the amplitudes of the $Sp(1)$ theory are determined by the 
B-model approach, i.e.  (\ref{hae},\ref{f1anomaly}) and the 
boundary  conditions described in chapter~\ref{completeness}.

\section{The Nekrasov-Shatashvili limit} 
\label{NSLimit}  
In \cite{NS}, Nekrasov and Shatashvili discuss the limit $\epsilon_1=0$, $\epsilon_2 \ll 1$ of the partition function (\ref{expansion1}), 
and they conjecture that the corresponding free energy in the small $\epsilon_2$ expansion is a quantization  
of the Seiberg-Witten prepotential (in a sense we make precise below), with $\epsilon_2$ playing the role of Planck constant $\hbar$. 
This so-called Nekrasov-Shatashvili limit has subsequently been studied in a series of papers~\cite{ACDKV, MT, MM}. Independently, it was already considered previously in the mathematics literature~\cite{MR2214246}\footnote{This paper establishes 
the equivalence of the Nekrasov partition function in the NS limit with the spectrum of an associated 
Schroedinger operator with Toda potential and generalizes the calculation from the partition function 
to correlators of surface operators.}. \cite{MM} implement the conjecture in the case of pure $SU(2)$ gauge theory to obtain differential 
equations for the amplitudes $F^{(n,0)}$ in terms of $u$-derivative of the Seiberg-Witten periods. With the 
methods described in this paper, we are able to check these equations and thus the conjecture 
of~\cite{NS} exactly to a given order in $n$. We have performed this check up to $n=2$, as we present in this section. 
 
The starting point of these considerations, as explained in more detail in \cite{MM}, is relating pure $SU(2)$ Seiberg-Witten theory to the one-dimensional sine-Gordon model 
\begin{eqnarray} 
S=\int (\frac{1}{2}\dot{\phi}^2-\gamma \cos\phi ) ~dt \,. 
\end{eqnarray} 
The Schr\"odinger equation of this theory is the following, 
\be \label{schrod} 
\left( - \frac{\hbar^2}{2} \frac{\partial^2}{\partial \phi^2} + \gamma \cos \phi \right) \Psi(\phi) = E \,\Psi(\phi)  \,. 
\ee 
The connection to Seiberg-Witten theory arises by relating the eigenvalues $E$ to the periods of the Seiberg-Witten curve. Specifically, writing the eigenvector to eigenvalue $E$ as 
\begin{eqnarray}  \label{wkb} 
\Psi(E,\phi) = \exp \left(\frac{i}{\hbar} \int^\phi P(E,\phi) d\phi \right) \,, 
\end{eqnarray} 
the function $P(E,\phi)$ can be calculated in a WKB series expansion in small $\hbar$. One now introduces the quantized periods  
\begin{eqnarray} \label{0518period} 
\tilde{a}=\oint_A P(E,\phi) d\phi, ~~~~\tilde{a}_D=\oint_B P(E,\phi) d\phi \,, 
\end{eqnarray} 
for appropriately chosen contours $A$ and $B$. The exact eigenvalues of (\ref{schrod}) are obtained by solving the equation 
\be 
\tilde{a} = 2 \pi \hbar \left( n + \frac{1}{2} \right)  
\ee 
for $E$. To leading order in the WKB expansion,  
\begin{eqnarray} \label{0822-7.5} 
P(E,\phi) = \sqrt{2(E-\gamma \cos\phi)} +\mathcal{O}(\hbar) \,, 
\end{eqnarray} 
and one can show that the periods $a$ and $a_D$ that follow from (\ref{0518period}) to this order coincide with the periods of pure $SU(2)$ Seiberg-Witten gauge theory, upon identifying the energy eigenvalue $E$ with the modular parameter $u$ of the Seiberg-Witten theory, and the parameter $\gamma$ with the energy scale $\Lambda^2$, $E=u$ and $\gamma=\Lambda^2$.\footnote{One noteworthy aspect of this formal identification is that the energy spectrum of the sine-Gordon equation is discrete, while the modular parameter $u$ in Seiberg-Witten theory is continuous.} The quantized prepotential is introduced in analogy to Seiberg-Witten theory as  
\be  \label{def_quant_f} 
\tilde{a}_D= \frac{\partial F(\tilde{a}|\hbar)}{\partial \tilde{a}}  \,. 
\ee 
The conjecture of Nekrasov and Shatashvili is that in the WKB expansion, the quantized prepotential coincides with the $\epsilon_1=0$ limit of (\ref{expansion1}), with $\epsilon_2$ identified with $\hbar$,\footnote{In fact, we identify $\epsilon_2 = \frac{\hbar}{2}$ to match our conventions.} 
\begin{eqnarray}  \label{ns_conj} 
F(\tilde{a}|\hbar) = \sum_{n=0}^\infty F^{(n,0)}(\tilde{a})\left(\frac{\hbar}{2}\right)^{2n}   \,. 
\end{eqnarray} 
 
Denoting the periods (\ref{0518period}) collectively as $\Pi$, and to leading order as $\Pi^{(0)}$, one can derive a Picard-Fuchs equation for $\Pi^{(0)}$ by inspection of (\ref{0822-7.5}) \cite{MM} 
\begin{eqnarray} \label{pfns} 
[\gamma (\partial_E^2 + \partial_\gamma^2) + 2E\partial_E \partial_\gamma] \Pi^{(0)} =0  \,. 
\end{eqnarray} 
In the conventions of \cite{MM}, the period $\frac{\Pi^{(0)}}{\sqrt{\gamma}}$ is a function of $\frac{E}{\gamma}$, implying the relation $\partial_\gamma (\frac{\Pi^{(0)}}{\gamma}) =- \frac{E}{\gamma} \partial_E (\frac{\Pi^{(0)}}{\sqrt{\gamma}})$, and thus 
\begin{eqnarray} 
\partial_\gamma \Pi^{(0)} &=& \frac{\Pi^{(0)}}{2\gamma} -\frac{E}{\gamma}\partial_E \Pi^{(0)}  \nonumber \,, \\ 
\partial_E\partial_\gamma \Pi^{(0)} &=& -\frac{\partial_E\Pi^{(0)}}{2\gamma} -\frac{E}{\gamma}\partial^2_E \Pi^{(0)}  \nonumber \,,\\ 
\partial^2_\gamma \Pi^{(0)} &=& -\frac{\Pi^{(0)}}{4\gamma^2} +\frac{E}{\gamma^2}\partial_E \Pi^{(0)}   
+\frac{E^2}{\gamma^2}\partial^2_E \Pi^{(0)}  \,. 
\end{eqnarray} 
Substituting these relations into (\ref{pfns}) and expressing the result in terms of Seiberg-Witten variables at $\Lambda^2=1$, the Picard-Fuchs equation becomes  
\begin{eqnarray} 
4(1-u^2)\partial^2_u\Pi^{(0)}= \Pi^{(0)} \,. 
\end{eqnarray} 
This coincides with the Picard-Fuchs equation used in \cite{HK1}.  
 
Plugging the ansatz (\ref{wkb}) into the Schr\"odinger equation, one obtains an iterative equation for computing the higher order terms of $P(E,\phi)$ in (\ref {0822-7.5}). It turns out that odd power terms in $\hbar$ do not have a square root cut, and hence yield vanishing contour integrals. As shown in \cite{MM}, the non-vanishing even order sub-leading terms of the periods (\ref{0518period}) in the WKB expansion can be computed by acting with certain differential operators on the leading order period. One finds   
\begin{eqnarray} 
&& \Pi^{(2)} = -\frac{\hbar^2 \gamma}{24} \partial^2_{E\gamma} \Pi^{(0)} =\frac{\hbar^2}{24}(\frac{1}{2}\partial_u\Pi^{(0)}+u\partial_u^2\Pi^{(0)})   \,,\\ 
&& \Pi^{(4)} = \frac{\hbar^4 \gamma}{1152} (-\frac{2}{5}E\partial_E +\gamma\partial_\gamma) \partial^2_E\partial_\gamma\Pi^{(0)}  \nonumber \\ 
&& ~~~~~~ = \frac{\hbar^4 }{23040} (75\partial^2_u \Pi^{(0)} +120u \partial^3_u \Pi^{(0)} +28u^2 \partial^4_u \Pi^{(0)})  \,. 
\end{eqnarray} 
The exact periods to sub-leading orders are therefore 
\begin{eqnarray} 
\tilde{a} &=& a+ \frac{\hbar^2}{24}(\frac{1}{2}\partial_u a+u\partial_u^2 a)+ 
\frac{\hbar^4 }{23040} (75\partial^2_u a +120u \partial^3_u a +28u^2 \partial^4_u a) +\mathcal{O}(\hbar^6)   \,, \nonumber \\ 
\tilde{a}_D &=& a_D+ \frac{\hbar^2}{24}(\frac{1}{2}\partial_u a_D+u\partial_u^2 a_D)  \nonumber \\ 
&&  + 
\frac{\hbar^4 }{23040} (75\partial^2_u a_D +120u \partial^3_u a_D +28u^2 \partial^4_u a_D) +\mathcal{O}(\hbar^6)  \,. 
\end{eqnarray} 
These relations can now be used to eliminate the quantized periods in terms of the classical periods and their derivatives. The definition of the quantized prepotential (\ref{def_quant_f}), assuming the Nekrasov-Shatashvili conjecture (\ref{ns_conj}), simply reproduces at leading order the special geometry relation $\partial_aF^{(0,0)}=a_D$. At order $\hbar^2$, one obtains 
\begin{eqnarray} \label{0518F1eq2} 
\frac{\partial F^{(1,0)}(a)}{\partial a} = \frac{2\pi i \tau} {3} (\frac{1}{2}\frac{\partial a}{\partial u} + u \frac{\partial ^2 a}{\partial u^2} )+\frac{1}{6}  (\frac{1}{2}\frac{\partial a_D}{\partial u} + u \frac{\partial ^2 a_D}{\partial u^2} )  \,, 
\end{eqnarray}  
where we have used the formula $\frac{\partial^2 F^{(0,0)}(a)}{\partial a^2} = -4\pi i \tau$. Similarly, the order $\hbar^4$ equation is  
\begin{eqnarray} \label{0822F2eq} 
\frac{\partial F^{(2,0)}(a)}{\partial a} &=& -\frac{1}{6}(\frac{1}{2}\partial_u a+u\partial_u^2 a) \partial^2_a F^{(1,0)} 
+\frac{\pi i}{18}\frac{\partial \tau}{\partial a} (\frac{1}{2}\partial_u a+u\partial_u^2 a)^2  
 \nonumber \\ 
&&  +\frac{\pi i \tau }{360} (75\partial^2_u a +120u \partial^3_u a +28u^2 \partial^4_u a)  \nonumber \\ 
&& 
+\frac{1}{1440} (75\partial^2_u a_D +120u \partial^3_u a_D +28u^2 \partial^4_u a_D)  \,. 
\end{eqnarray} 
We can check the veracity of these relations explicitly. The periods and the $u$ parameter of pure $SU(2)$ gauge theory can be expressed in terms of the IR gauge coupling $\tau$, see e.g. \cite{HK1},  
\begin{eqnarray} 
a &=& \frac{E_2(\tau) +\theta_3^4(\tau)+\theta_4^4(\tau) }{3\theta_2^2(\tau)} \,,  
\nonumber \\ 
u &=& \frac{\theta_3^4(\tau)+ \theta_4^4(\tau)}{\theta_2^4(\tau)} \,, \nonumber \\ 
a_D &=& -4\pi i \tau a -2\partial_a u \,. 
\end{eqnarray} 
The formula for $a_D$ is derived by taking the derivative with respect to $a$ of the Matone relation $F^{(0,0)}-\frac{1}{2}a \frac{\partial F^{(0,0)}}{\partial a}+u=0$. The formulae for $F^{(1,0)}$ and  $F^{(2,0)}$, determined by the methods described in this paper in \cite{HK3}, are   
\begin{eqnarray} 
F^{(1,0)} &=& \frac{1}{24} \log(u^2-1) \,, \nonumber \\ 
F^{(2,0)} &=& \frac{1}{4320 (u^2-1)^2} (10u^2\frac{E_2(\tau)}{\theta_2^4(\tau)} +u^3-75 u) \,. 
\end{eqnarray} 
Using these formulae together with derivative identities for the Jacobi theta functions and Eisenstein series, we can easily check equations (\ref{0518F1eq2}) and (\ref{0822F2eq}) by invoking theta function identities. For example, we can express both sides of the order 2 relation (\ref{0518F1eq2}) as follows, 
\begin{eqnarray}  
 \frac{\partial F^{(1,0)}(a)}{\partial a} &=& \frac{2\pi i \tau} {3} (\frac{1}{2}\frac{\partial a}{\partial u} + u \frac{\partial ^2 a}{\partial u^2} )+\frac{1}{6}  (\frac{1}{2}\frac{\partial a_D}{\partial u} + u \frac{\partial ^2 a_D}{\partial u^2} ) 
\nonumber  \\ &=& \theta_2^2(\tau) \frac{\theta_3^4(\tau)+\theta_4^4(\tau)}{12\theta_3^4(\tau)\theta_4^4(\tau)}  \,. 
\end{eqnarray}

\section{General $\Omega$-background for the ${\cal O}(-K)\rightarrow \mathbb{P}^2$ geometry }   
\label{DT} 
  
The formalism developed in section \ref{directintegration} can readily be applied to non-compact Calabi-Yau  
manifolds. Our aim is to calculate refined BPS invariants, mathematically known as motivic  
Donaldson Thomas invariants, and further to present refined orbifold invariants. We consider the simple yet non-trivial  
A-model geometry ${\cal O}(-3)\rightarrow \mathbb{P}^2$. The B-model geometry of the mirror curve has been studied e.g. in~\cite{Haghighat:2008gw} in the context of direct integration of the usual 
topological string. Some consistency checks for the refined model have been performed in~\cite{HK3} by comparing with the results of~\cite{IKV} 
at large radius.  In~\cite{ABK,Haghighat:2008gw,Alim:2008kp}, one can find a  more complete discussion of the properties of the periods at various 
points. The mirror curve ${\cal C}_0$ can be determined by standard methods to be  
\be  
H(x,y;z)= y^2 + x y + y+z x^3=0 \ , 
\label{mirrorp2} 
\ee 
with $z$ the modulus of the geometry. The meromorphic differential can be written  
as $\lambda = \log(y) \frac{d x}{x}$. We notice that the parameters in  
this context are such that $z \frac{d}{d z} \lambda= \omega+exact$.   
From (\ref{mirrorp2}), it is straightforward to calculate\footnote{We have chosen $g_2,g_3$  
so that (\ref{nonlogperiod}) gives the correct behavior at infinity, as determined via the Picard-Fuchs equations, with $c_1=1$.} $g_2=3^3(1+24z)$ and $g_3=3^3(1+36 z+216z^2)$  
and the $J$-invariant as  
\be  
J=\frac{1}{q}+744+196884 q+21493760 q^2+O\left(q^3\right)=-\frac{(1+24 z)^3}{1728 z^3 (1+27z) }  \,. 
\ee  
The parametrization is related to the more symmetric cubic $\sum_{i=1}^3 x_i^3- 3 \psi \prod_{i=1}^{3} x_i=0$  
in $\mathbb{P}^2$ by $z=\frac{1-\psi^3}{27 \psi^3}$. The Picard-Fuchs operator for the periods of $\lambda$ is   
\begin{equation}  
{\cal D}=\theta^3+3 z \left(3\theta-1\right)\left(3\theta-2\right)\theta={\cal L} \theta \ , 
\label{pfp^2} 
\end{equation} 
where $\theta=z \frac{d}{dz}$ and  ${\cal L}$ is the Picard-Fuchs operator annihilating the  
holomorphic periods  $\int_{\Sigma_i} \omega$. The singular points of the theory are at $z=0$, corresponding to  
the large radius limit of the $A$-model, $z=-\frac{1}{27}$ the conifold  point, and $1/z=0$ the  
orbifold point. To obtain the prepotential $F^{(0,0)}$ in the conventional A-model normalization, we must choose the constant relating it to the tau parameter of the mirror curve as $F^{(0,0)}=-\frac{1}{9} \int d t \int d t \, \tau$. Near $z=0$, we can express the result of this integration in terms of the mirror coordinate $Q=e^t$ with $z(Q)=Q+6 Q^2+9 Q^3+56 Q^4+O\left(Q^5\right)$,  
yielding the genus 0 instanton expansion 
\begin{equation} 
F^{(0,0)}=-\frac{1}{18} \log ^3(Q)+3 Q-\frac{45 Q^2}{8}+\frac{244 Q^3}{9}-\frac{12333 Q^4}{64} + {\cal O}(Q^5)\ . 
\end{equation} 
Rigid special geometry implies $\partial_t^3  F^{(0,0)}=C_{zzz}(\partial_t z)^3$ with  $C_{zzz}=-\frac{1}{3}\frac{1}{z^3 (1+27 z)}$.  
The amplitudes for $n+g=1$ are  
\begin{equation} 
F^{(0,1)}=-\frac{1}{2}\log\left(G_{z\bar z}|z^7\Delta|^\frac{1}{3}\right),\qquad  F^{(1,0)}=\frac{1}{24} \log( z^{-1} \Delta)\ ,     
\end{equation} 
in terms of $\Delta=(1+27 z)$. The holomorphic limit  of $F^{(0,1)}$ is  
$F^{(0,1)}_{hol} = -\frac{1}{2}\log\left(\frac{d t}{d z}\right) -\frac{1}{12} \log( z^7 \Delta)$. 
 
The recursion proceeds as described in section~\ref{directintegration}. The  
anholomorphic object defined in~\cite{Haghighat:2008gw,HK3}, the propagator $S:=\frac{2}{C_{zzz}}\frac{\partial F^{(0,1)}(z)}{\partial z}$, is related to the one we use in this paper, $X=\frac{g_3 \hat E_2 E_4}{g_2 E_6}$,  
by  
\begin{equation}  
\label{propagatorp2} 
S=\frac{z^2}{4}(9X-1)\ .  
\end{equation} 
For reference, we note that in the real polarization,  
i.e. in the holomorphic limit near $z=0$, $\lim_{{\rm Im}{(\tau)}\rightarrow \infty}X  
=\frac{1}{3}+4 Q+84 Q^3+O\left(Q^4\right)$.  In terms of $X$, we obtain for $g+n=2$ 
\ban 
F^{(2,0)}&=&\frac{7 \chi(M) }{138240}+ \frac{15 X-7776 z^2+288 z-5}{7680 \Delta^2} \,,\nonumber \\ 
F^{(1,1)}&=&-\frac{7 \chi(M) }{34560}+ \frac{135 X^2+60 X (54 z-1)-3888 z^2-936 z+5}{3840 \Delta^2} \,, \nonumber\\ 
F^{(0,2)}&=&\frac{\chi(M) }{5760}+\frac{2025 X^3-1485 X^2+375 X+ 2592 z^2+144 z+35}{7680 \Delta ^2}  \,. 
\label{p2n+g=2}  
\ean

\subsection{Motivic Donaldson-Thomas invariants} 
We solved the recursion described in section~\ref{directintegration} up to $g+n=84$ in order to calculate the BPS numbers  
up to $d=9$, with $d$ the multiplicity of the hyperplane divisor $H$ in $\IP^2$, $\beta= d H$. 
The BPS degeneracies $N^\beta_{j_-,j_+}=:N^d_{j_l,j_r}$ are extracted using (\ref{schwingerloope1e2})\footnote{The $[j_l,j_r]$ notation  
is used here to compare with~\cite{KKV2}}. Apart from their  
integrality and positivity, we note that the highest spin representation occurs at  
\be 
2 j^{max}_l(d)=g^{max}(d)=\frac{1}{2}(d-1)(d-2)\ ,\qquad 2 j^{max}_r(d)=n^{max}(d)=\frac{1}{2}d (3 +d), 
\ee  
with multiplicity one, $N^d_{j^{max}_l(d),j^{max}_r(d)}=1$. This is in perfect accord with the analysis of~\cite{KKV2} and  
a highly non-trivial test of the integrality structure encoded in (\ref{schwingerloope1e2}) and our method of  
fixing the holomorphic ambiguity. Further remarkable properties, e.g. 
\be 
N^d_{j_l,j_r} = 0 \qquad  {\rm if}\ \ 2(j_l+j_r)+d \ {\rm mod}\, 2=0\  , 
\ee 
which implies that the cohomology of the moduli space of the $D2/D0$ is even on  
the local $\mathbb{P}^2$, or the fact that 
\be 
N^d_{j^{max}_l(d)-\frac{i}{2},j^{max}_r(d)-\frac{j}{2}} 
\ee 
is symmetric in $i,j$ and independent of $d$ for $j+i<2 d-4$, will  
be discussed in~\cite{workinprogress}. The refined vertex cannot calculate the refined amplitudes 
for the case at hand directly. It can however calculate the invariants of the Hirzebruch surface ${\cal O}(-K)\rightarrow \mathbb{F}_1$, which 
as a blow up of ${\cal O}(-K)\rightarrow \mathbb{P}^2$ contains the results for the latter 
geometry in the class $H+F$. Our results agree with those obtained in this manner in~\cite{IKV} up to degree 5, 
aside from the multiplicity $N^5_{\frac{1}{2},\frac{9}{2}}=2$, which seems to be missing in~\cite{IKV}.
  
\begin{table}[h!] 
\centering{{ 
\begin{tabular}[h]{|c|c|cccccccccccccccccccccccccccccccccccc|} 
\hline  
d\!&\!\!$j_l\backslash j_r$\!&\!\!0\!&\!\!\!$\frac{1}{2}$\!\!\!&\!\!\!1\!\!\!&\!\!\!$\frac{3}{2}$\!\!\!&\!\!\!2\!\!\!&\!\!\!$\frac{5}{2}$\!\!\!&\!\!\!3\!\!\!&\!\!\!$\frac{7}{2}$\!\!\!&\!\!\!4\!\!\!& 
\!\!\!$\frac{9}{2}$\!\!\!&\!\!\!5\!\!\!&\!\!\!$\frac{11}{2}$\!\!\!&\!\!\!6\!\!\!&\!\!\!$\frac{13}{2}$\!\!\!&\!\!\!7\!\!\!&\!\!\!$\frac{15}{2}$\!\!\!&\!\!\!8\!\!\!&\!\!\!$\frac{17}{2}$\!\!\! 
&\!\!\!9\!\!\!&\!\!\!$\frac{19}{2}$\!\!\!&\!\!\!10\!\!\!&\!\!\!$\frac{21}{2}$\!\!\!&\!\!\!11\!\!\!&\!\!\!$\frac{23}{2}$\!\!\!&\!\!\!12\!\!\!&\!\!\!$\frac{25}{2}$\!\!\!&\!\!\!13\!\!\! 
&\!\!\!$\frac{27}{2}$\!\!\!&\!\!\!14\!\!\!&\!\!\!$\frac{29}{2}$\!\!\!&\!\!\!15\!\!\!&\!\!\!$\frac{31}{2}$\!\!\!&\!\!\!16\!\!\!&\!\!\!$\frac{33}{2}$\!\!\!&\!\!\!17\!\!\!&\!\!\!$\frac{35}{2}$\!\!\!\\ 
\hline 
1\!\!\!&\!\!\!0\!\!\!&\!\!\!\!\!\!&\!\!\!\!\!\!&\!\!\!1\!\!\!&\!\!\!\!\!\!&\!\!\!\!\!\!&\!\!\!\!\!\!&\!\!\!\!\!\!&\!\!\!\!\!\!&\!\!\!\!\!\!&\!\!\!\!\!\!&\!\!\!\!\!\!&\!\!\!\!\!\!&\!\!\!\!\!\!&\!\!\!\!\!\!&\!\!\!\!\!\!&\!\!\!\!\!\!&\!\!\!\!\!\!&\!\!\!\!\!\!&\!\!\!\!\!\!&\!\!\!\!\!\!&\!\!\!\!\!\!&\!\!\!\!\!\!&\!\!\!\!\!\!&\!\!\!\!\!\!&\!\!\!\!\!\!&\!\!\!\!\!\!&\!\!\!\!\!\!&\!\!\!\!\!\!&\!\!\!\!\!\!&\!\!\!\!\!\!&\!\!\!\!\!\!&\!\!\!\!\!\!&\!\!\!\!\!\!&\!\!\!\!\!\!&\!\!\!\!\!\!&\!\!\!\\ 
\hline 
2\!\!\!&\!\!\!0\!\!\!&\!\!\!\!\!\!&\!\!\!\!\!\!&\!\!\!\!\!\!&\!\!\!\!\!\!&\!\!\!\!\!\!&\!\!\!1\!\!\!&\!\!\!\!\!\!&\!\!\!\!\!\!&\!\!\!\!\!\!&\!\!\!\!\!\!&\!\!\!\!\!\!&\!\!\!\!\!\!&\!\!\!\!\!\!&\!\!\!\!\!\!&\!\!\!\!\!\!&\!\!\!\!\!\!&\!\!\!\!\!\!&\!\!\!\!\!\!&\!\!\!\!\!\!&\!\!\!\!\!\!&\!\!\!\!\!\!&\!\!\!\!\!\!&\!\!\!\!\!\!&\!\!\!\!\!\!&\!\!\!\!\!\!&\!\!\!\!\!\!&\!\!\!\!\!\!&\!\!\!\!\!\!&\!\!\!\!\!\!&\!\!\!\!\!\!&\!\!\!\!\!\!&\!\!\!\!\!\!&\!\!\!\!\!\!&\!\!\!\!\!\!&\!\!\!\!\!\!&\!\!\!\\ 
\hline 
3\!\!\!&\!\!\!0\!\!\!&\!\!\!\!\!\!&\!\!\!\!\!\!&\!\!\!\!\!\!&\!\!\!\!\!\!&\!\!\!\!\!\!&\!\!\!\!\!\!&\!\!\!1\!\!\!&\!\!\!\!\!\!&\!\!\!\!\!\!&\!\!\!\!\!\!&\!\!\!\!\!\!&\!\!\!\!\!\!&\!\!\!\!\!\!&\!\!\!\!\!\!&\!\!\!\!\!\!&\!\!\!\!\!\!&\!\!\!\!\!\!&\!\!\!\!\!\!&\!\!\!\!\!\!&\!\!\!\!\!\!&\!\!\!\!\!\!&\!\!\!\!\!\!&\!\!\!\!\!\!&\!\!\!\!\!\!&\!\!\!\!\!\!&\!\!\!\!\!\!&\!\!\!\!\!\!&\!\!\!\!\!\!&\!\!\!\!\!\!&\!\!\!\!\!\!&\!\!\!\!\!\!&\!\!\!\!\!\!&\!\!\!\!\!\!&\!\!\!\!\!\!&\!\!\!\!\!\!&\!\!\!\\ 
\!\!\!&\!\!\!$\frac{1}{2}$\!\!\!&\!\!\!\!\!\!&\!\!\!\!\!\!&\!\!\!\!\!\!&\!\!\!\!\!\!&\!\!\!\!\!\!&\!\!\!\!\!\!&\!\!\!\!\!\!&\!\!\!\!\!\!&\!\!\!\!\!\!&\!\!\!1\!\!\!&\!\!\!\!\!\!&\!\!\!\!\!\!&\!\!\!\!\!\!&\!\!\!\!\!\!&\!\!\!\!\!\!&\!\!\!\!\!\!&\!\!\!\!\!\!&\!\!\!\!\!\!&\!\!\!\!\!\!&\!\!\!\!\!\!&\!\!\!\!\!\!&\!\!\!\!\!\!&\!\!\!\!\!\!&\!\!\!\!\!\!&\!\!\!\!\!\!&\!\!\!\!\!\!&\!\!\!\!\!\!&\!\!\!\!\!\!&\!\!\!\!\!\!&\!\!\!\!\!\!&\!\!\!\!\!\!&\!\!\!\!\!\!&\!\!\!\!\!\!&\!\!\!\!\!\!&\!\!\!\!\!\!&\!\!\!\\ 
\hline 
4\!\!\!&\!\!\!0\!\!\!&\!\!\!\!\!\!&\!\!\!\!\!\!&\!\!\!\!\!\!&\!\!\!\!\!\!&\!\!\!\!\!\!&\!\!\!1\!\!\!&\!\!\!\!\!\!&\!\!\!\!\!\!&\!\!\!\!\!\!&\!\!\!1\!\!\!&\!\!\!\!\!\!&\!\!\!\!\!\!&\!\!\!\!\!\!&\!\!\!1\!\!\!&\!\!\!\!\!\!&\!\!\!\!\!\!&\!\!\!\!\!\!&\!\!\!\!\!\!&\!\!\!\!\!\!&\!\!\!\!\!\!&\!\!\!\!\!\!&\!\!\!\!\!\!&\!\!\!\!\!\!&\!\!\!\!\!\!&\!\!\!\!\!\!&\!\!\!\!\!\!&\!\!\!\!\!\!&\!\!\!\!\!\!&\!\!\!\!\!\!&\!\!\!\!\!\!&\!\!\!\!\!\!&\!\!\!\!\!\!&\!\!\!\!\!\!&\!\!\!\!\!\!&\!\!\!\!\!\!&\!\!\!\\ 
\!\!\!&\!\!\!$\frac{1}{2}$\!\!\!&\!\!\!\!\!\!&\!\!\!\!\!\!&\!\!\!\!\!\!&\!\!\!\!\!\!&\!\!\!\!\!\!&\!\!\!\!\!\!&\!\!\!\!\!\!&\!\!\!\!\!\!&\!\!\!1\!\!\!&\!\!\!\!\!\!&\!\!\!1\!\!\!&\!\!\!\!\!\!&\!\!\!1\!\!\!&\!\!\!\!\!\!&\!\!\!\!\!\!&\!\!\!\!\!\!&\!\!\!\!\!\!&\!\!\!\!\!\!&\!\!\!\!\!\!&\!\!\!\!\!\!&\!\!\!\!\!\!&\!\!\!\!\!\!&\!\!\!\!\!\!&\!\!\!\!\!\!&\!\!\!\!\!\!&\!\!\!\!\!\!&\!\!\!\!\!\!&\!\!\!\!\!\!&\!\!\!\!\!\!&\!\!\!\!\!\!&\!\!\!\!\!\!&\!\!\!\!\!\!&\!\!\!\!\!\!&\!\!\!\!\!\!&\!\!\!\!\!\!&\!\!\!\\ 
\!\!\!&\!\!\!1\!\!\!&\!\!\!\!\!\!&\!\!\!\!\!\!&\!\!\!\!\!\!&\!\!\!\!\!\!&\!\!\!\!\!\!&\!\!\!\!\!\!&\!\!\!\!\!\!&\!\!\!\!\!\!&\!\!\!\!\!\!&\!\!\!\!\!\!&\!\!\!\!\!\!&\!\!\!1\!\!\!&\!\!\!\!\!\!&\!\!\!\!\!\!&\!\!\!\!\!\!&\!\!\!\!\!\!&\!\!\!\!\!\!&\!\!\!\!\!\!&\!\!\!\!\!\!&\!\!\!\!\!\!&\!\!\!\!\!\!&\!\!\!\!\!\!&\!\!\!\!\!\!&\!\!\!\!\!\!&\!\!\!\!\!\!&\!\!\!\!\!\!&\!\!\!\!\!\!&\!\!\!\!\!\!&\!\!\!\!\!\!&\!\!\!\!\!\!&\!\!\!\!\!\!&\!\!\!\!\!\!&\!\!\!\!\!\!&\!\!\!\!\!\!&\!\!\!\!\!\!&\!\!\!\\ 
\!\!\!&\!\!\!$\frac{3}{2}$\!\!\!&\!\!\!\!\!\!&\!\!\!\!\!\!&\!\!\!\!\!\!&\!\!\!\!\!\!&\!\!\!\!\!\!&\!\!\!\!\!\!&\!\!\!\!\!\!&\!\!\!\!\!\!&\!\!\!\!\!\!&\!\!\!\!\!\!&\!\!\!\!\!\!&\!\!\!\!\!\!&\!\!\!\!\!\!&\!\!\!\!\!\!&\!\!\!1\!\!\!&\!\!\!\!\!\!&\!\!\!\!\!\!&\!\!\!\!\!\!&\!\!\!\!\!\!&\!\!\!\!\!\!&\!\!\!\!\!\!&\!\!\!\!\!\!&\!\!\!\!\!\!&\!\!\!\!\!\!&\!\!\!\!\!\!&\!\!\!\!\!\!&\!\!\!\!\!\!&\!\!\!\!\!\!&\!\!\!\!\!\!&\!\!\!\!\!\!&\!\!\!\!\!\!&\!\!\!\!\!\!&\!\!\!\!\!\!&\!\!\!\!\!\!&\!\!\!\!\!\!&\!\!\!\\ 
\hline 
5\!\!\!&\!\!\!0\!\!\!&\!\!\!\!\!\!&\!\!\!\!\!\!&\!\!\!1\!\!\!&\!\!\!\!\!\!&\!\!\!\!\!\!&\!\!\!\!\!\!&\!\!\!1\!\!\!&\!\!\!\!\!\!&\!\!\!1\!\!\!&\!\!\!\!\!\!&\!\!\!2\!\!\!&\!\!\!\!\!\!&\!\!\!2\!\!\!&\!\!\!\!\!\!&\!\!\!2\!\!\!&\!\!\!\!\!\!&\!\!\!1\!\!\!&\!\!\!\!\!\!&\!\!\!\!\!\!&\!\!\!\!\!\!&\!\!\!\!\!\!&\!\!\!\!\!\!&\!\!\!\!\!\!&\!\!\!\!\!\!&\!\!\!\!\!\!&\!\!\!\!\!\!&\!\!\!\!\!\!&\!\!\!\!\!\!&\!\!\!\!\!\!&\!\!\!\!\!\!&\!\!\!\!\!\!&\!\!\!\!\!\!&\!\!\!\!\!\!&\!\!\!\!\!\!&\!\!\!\!\!\!&\!\!\!\\ 
\!\!\!&\!\!\!$\frac{1}{2}$\!\!\!&\!\!\!\!\!\!&\!\!\!\!\!\!&\!\!\!\!\!\!&\!\!\!\!\!\!&\!\!\!\!\!\!&\!\!\!1\!\!\!&\!\!\!\!\!\!&\!\!\!1\!\!\!&\!\!\!\!\!\!&\!\!\!2\!\!\!&\!\!\!\!\!\!&\!\!\!2\!\!\!&\!\!\!\!\!\!&\!\!\!3\!\!\!&\!\!\!\!\!\!&\!\!\!2\!\!\!&\!\!\!\!\!\!&\!\!\!1\!\!\!&\!\!\!\!\!\!&\!\!\!\!\!\!&\!\!\!\!\!\!&\!\!\!\!\!\!&\!\!\!\!\!\!&\!\!\!\!\!\!&\!\!\!\!\!\!&\!\!\!\!\!\!&\!\!\!\!\!\!&\!\!\!\!\!\!&\!\!\!\!\!\!&\!\!\!\!\!\!&\!\!\!\!\!\!&\!\!\!\!\!\!&\!\!\!\!\!\!&\!\!\!\!\!\!&\!\!\!\!\!\!&\!\!\!\\ 
\!\!\!&\!\!\!1\!\!\!&\!\!\!\!\!\!&\!\!\!\!\!\!&\!\!\!\!\!\!&\!\!\!\!\!\!&\!\!\!\!\!\!&\!\!\!\!\!\!&\!\!\!\!\!\!&\!\!\!\!\!\!&\!\!\!1\!\!\!&\!\!\!\!\!\!&\!\!\!1\!\!\!&\!\!\!\!\!\!&\!\!\!2\!\!\!&\!\!\!\!\!\!&\!\!\!2\!\!\!&\!\!\!\!\!\!&\!\!\!2\!\!\!&\!\!\!\!\!\!&\!\!\!1\!\!\!&\!\!\!\!\!\!&\!\!\!\!\!\!&\!\!\!\!\!\!&\!\!\!\!\!\!&\!\!\!\!\!\!&\!\!\!\!\!\!&\!\!\!\!\!\!&\!\!\!\!\!\!&\!\!\!\!\!\!&\!\!\!\!\!\!&\!\!\!\!\!\!&\!\!\!\!\!\!&\!\!\!\!\!\!&\!\!\!\!\!\!&\!\!\!\!\!\!&\!\!\!\!\!\!&\!\!\!\\ 
\!\!\!&\!\!\!$\frac{3}{2}$\!\!\!&\!\!\!\!\!\!&\!\!\!\!\!\!&\!\!\!\!\!\!&\!\!\!\!\!\!&\!\!\!\!\!\!&\!\!\!\!\!\!&\!\!\!\!\!\!&\!\!\!\!\!\!&\!\!\!\!\!\!&\!\!\!\!\!\!&\!\!\!\!\!\!&\!\!\!1\!\!\!&\!\!\!\!\!\!&\!\!\!1\!\!\!&\!\!\!\!\!\!&\!\!\!2\!\!\!&\!\!\!\!\!\!&\!\!\!1\!\!\!&\!\!\!\!\!\!&\!\!\!1\!\!\!&\!\!\!\!\!\!&\!\!\!\!\!\!&\!\!\!\!\!\!&\!\!\!\!\!\!&\!\!\!\!\!\!&\!\!\!\!\!\!&\!\!\!\!\!\!&\!\!\!\!\!\!&\!\!\!\!\!\!&\!\!\!\!\!\!&\!\!\!\!\!\!&\!\!\!\!\!\!&\!\!\!\!\!\!&\!\!\!\!\!\!&\!\!\!\!\!\!&\!\!\!\\ 
\!\!\!&\!\!\!2\!\!\!&\!\!\!\!\!\!&\!\!\!\!\!\!&\!\!\!\!\!\!&\!\!\!\!\!\!&\!\!\!\!\!\!&\!\!\!\!\!\!&\!\!\!\!\!\!&\!\!\!\!\!\!&\!\!\!\!\!\!&\!\!\!\!\!\!&\!\!\!\!\!\!&\!\!\!\!\!\!&\!\!\!\!\!\!&\!\!\!\!\!\!&\!\!\!1\!\!\!&\!\!\!\!\!\!&\!\!\!1\!\!\!&\!\!\!\!\!\!&\!\!\!1\!\!\!&\!\!\!\!\!\!&\!\!\!\!\!\!&\!\!\!\!\!\!&\!\!\!\!\!\!&\!\!\!\!\!\!&\!\!\!\!\!\!&\!\!\!\!\!\!&\!\!\!\!\!\!&\!\!\!\!\!\!&\!\!\!\!\!\!&\!\!\!\!\!\!&\!\!\!\!\!\!&\!\!\!\!\!\!&\!\!\!\!\!\!&\!\!\!\!\!\!&\!\!\!\!\!\!&\!\!\!\\ 
\!\!\!&\!\!\!$\frac{5}{2}$\!\!\!&\!\!\!\!\!\!&\!\!\!\!\!\!&\!\!\!\!\!\!&\!\!\!\!\!\!&\!\!\!\!\!\!&\!\!\!\!\!\!&\!\!\!\!\!\!&\!\!\!\!\!\!&\!\!\!\!\!\!&\!\!\!\!\!\!&\!\!\!\!\!\!&\!\!\!\!\!\!&\!\!\!\!\!\!&\!\!\!\!\!\!&\!\!\!\!\!\!&\!\!\!\!\!\!&\!\!\!\!\!\!&\!\!\!1\!\!\!&\!\!\!\!\!\!&\!\!\!\!\!\!&\!\!\!\!\!\!&\!\!\!\!\!\!&\!\!\!\!\!\!&\!\!\!\!\!\!&\!\!\!\!\!\!&\!\!\!\!\!\!&\!\!\!\!\!\!&\!\!\!\!\!\!&\!\!\!\!\!\!&\!\!\!\!\!\!&\!\!\!\!\!\!&\!\!\!\!\!\!&\!\!\!\!\!\!&\!\!\!\!\!\!&\!\!\!\!\!\!&\!\!\!\\ 
\!\!\!&\!\!\!3\!\!\!&\!\!\!\!\!\!&\!\!\!\!\!\!&\!\!\!\!\!\!&\!\!\!\!\!\!&\!\!\!\!\!\!&\!\!\!\!\!\!&\!\!\!\!\!\!&\!\!\!\!\!\!&\!\!\!\!\!\!&\!\!\!\!\!\!&\!\!\!\!\!\!&\!\!\!\!\!\!&\!\!\!\!\!\!&\!\!\!\!\!\!&\!\!\!\!\!\!&\!\!\!\!\!\!&\!\!\!\!\!\!&\!\!\!\!\!\!&\!\!\!\!\!\!&\!\!\!\!\!\!&\!\!\!1\!\!\!&\!\!\!\!\!\!&\!\!\!\!\!\!&\!\!\!\!\!\!&\!\!\!\!\!\!&\!\!\!\!\!\!&\!\!\!\!\!\!&\!\!\!\!\!\!&\!\!\!\!\!\!&\!\!\!\!\!\!&\!\!\!\!\!\!&\!\!\!\!\!\!&\!\!\!\!\!\!&\!\!\!\!\!\!&\!\!\!\!\!\!&\!\!\!\\ 
\hline 
6\!\!\!&\!\!\!0\!\!\!&\!\!\!\!\!\!&\!\!\!1\!\!\!&\!\!\!\!\!\!&\!\!\!1\!\!\!&\!\!\!\!\!\!&\!\!\!3\!\!\!&\!\!\!\!\!\!&\!\!\!2\!\!\!&\!\!\!\!\!\!&\!\!\!6\!\!\!&\!\!\!\!\!\!&\!\!\!4\!\!\!&\!\!\!\!\!\!&\!\!\!8\!\!\!&\!\!\!\!\!\!&\!\!\!5\!\!\!&\!\!\!\!\!\!&\!\!\!7\!\!\!&\!\!\!\!\!\!&\!\!\!2\!\!\!&\!\!\!\!\!\!&\!\!\!2\!\!\!&\!\!\!\!\!\!&\!\!\!\!\!\!&\!\!\!\!\!\!&\!\!\!\!\!\!&\!\!\!\!\!\!&\!\!\!\!\!\!&\!\!\!\!\!\!&\!\!\!\!\!\!&\!\!\!\!\!\!&\!\!\!\!\!\!&\!\!\!\!\!\!&\!\!\!\!\!\!&\!\!\!\!\!\!&\!\!\!\\ 
\!\!\!&\!\!\!$\frac{1}{2}$\!\!\!&\!\!\!\!\!\!&\!\!\!\!\!\!&\!\!\!1\!\!\!&\!\!\!\!\!\!&\!\!\!2\!\!\!&\!\!\!\!\!\!&\!\!\!3\!\!\!&\!\!\!\!\!\!&\!\!\!5\!\!\!&\!\!\!\!\!\!&\!\!\!6\!\!\!&\!\!\!\!\!\!&\!\!\!9\!\!\!&\!\!\!\!\!\!&\!\!\!9\!\!\!&\!\!\!\!\!\!&\!\!\!10\!\!\!&\!\!\!\!\!\!&\!\!\!7\!\!\!&\!\!\!\!\!\!&\!\!\!5\!\!\!&\!\!\!\!\!\!&\!\!\!1\!\!\!&\!\!\!\!\!\!&\!\!\!1\!\!\!&\!\!\!\!\!\!&\!\!\!\!\!\!&\!\!\!\!\!\!&\!\!\!\!\!\!&\!\!\!\!\!\!&\!\!\!\!\!\!&\!\!\!\!\!\!&\!\!\!\!\!\!&\!\!\!\!\!\!&\!\!\!\!\!\!&\!\!\!\\ 
\!\!\!&\!\!\!1\!\!\!&\!\!\!\!\!\!&\!\!\!\!\!\!&\!\!\!\!\!\!&\!\!\!1\!\!\!&\!\!\!\!\!\!&\!\!\!1\!\!\!&\!\!\!\!\!\!&\!\!\!3\!\!\!&\!\!\!\!\!\!&\!\!\!3\!\!\!&\!\!\!\!\!\!&\!\!\!7\!\!\!&\!\!\!\!\!\!&\!\!\!7\!\!\!&\!\!\!\!\!\!&\!\!\!11\!\!\!&\!\!\!\!\!\!&\!\!\!9\!\!\!&\!\!\!\!\!\!&\!\!\!9\!\!\!&\!\!\!\!\!\!&\!\!\!4\!\!\!&\!\!\!\!\!\!&\!\!\!2\!\!\!&\!\!\!\!\!\!&\!\!\!\!\!\!&\!\!\!\!\!\!&\!\!\!\!\!\!&\!\!\!\!\!\!&\!\!\!\!\!\!&\!\!\!\!\!\!&\!\!\!\!\!\!&\!\!\!\!\!\!&\!\!\!\!\!\!&\!\!\!\!\!\!&\!\!\!\\ 
\!\!\!&\!\!\!$\frac{3}{2}$\!\!\!&\!\!\!\!\!\!&\!\!\!\!\!\!&\!\!\!\!\!\!&\!\!\!\!\!\!&\!\!\!\!\!\!&\!\!\!\!\!\!&\!\!\!1\!\!\!&\!\!\!\!\!\!&\!\!\!1\!\!\!&\!\!\!\!\!\!&\!\!\!3\!\!\!&\!\!\!\!\!\!&\!\!\!4\!\!\!&\!\!\!\!\!\!&\!\!\!7\!\!\!&\!\!\!\!\!\!&\!\!\!7\!\!\!&\!\!\!\!\!\!&\!\!\!10\!\!\!&\!\!\!\!\!\!&\!\!\!6\!\!\!&\!\!\!\!\!\!&\!\!\!4\!\!\!&\!\!\!\!\!\!&\!\!\!\!\!\!&\!\!\!\!\!\!&\!\!\!\!\!\!&\!\!\!\!\!\!&\!\!\!\!\!\!&\!\!\!\!\!\!&\!\!\!\!\!\!&\!\!\!\!\!\!&\!\!\!\!\!\!&\!\!\!\!\!\!&\!\!\!\!\!\!&\!\!\!\\ 
\!\!\!&\!\!\!2\!\!\!&\!\!\!\!\!\!&\!\!\!\!\!\!&\!\!\!\!\!\!&\!\!\!\!\!\!&\!\!\!\!\!\!&\!\!\!\!\!\!&\!\!\!\!\!\!&\!\!\!\!\!\!&\!\!\!\!\!\!&\!\!\!1\!\!\!&\!\!\!\!\!\!&\!\!\!1\!\!\!&\!\!\!\!\!\!&\!\!\!3\!\!\!&\!\!\!\!\!\!&\!\!\!4\!\!\!&\!\!\!\!\!\!&\!\!\!7\!\!\!&\!\!\!\!\!\!&\!\!\!6\!\!\!&\!\!\!\!\!\!&\!\!\!6\!\!\!&\!\!\!\!\!\!&\!\!\!2\!\!\!&\!\!\!\!\!\!&\!\!\!1\!\!\!&\!\!\!\!\!\!&\!\!\!\!\!\!&\!\!\!\!\!\!&\!\!\!\!\!\!&\!\!\!\!\!\!&\!\!\!\!\!\!&\!\!\!\!\!\!&\!\!\!\!\!\!&\!\!\!\!\!\!&\!\!\!\\ 
\!\!\!&\!\!\!$\frac{5}{2}$\!\!\!&\!\!\!\!\!\!&\!\!\!\!\!\!&\!\!\!\!\!\!&\!\!\!\!\!\!&\!\!\!\!\!\!&\!\!\!\!\!\!&\!\!\!\!\!\!&\!\!\!\!\!\!&\!\!\!\!\!\!&\!\!\!\!\!\!&\!\!\!\!\!\!&\!\!\!\!\!\!&\!\!\!1\!\!\!&\!\!\!\!\!\!&\!\!\!1\!\!\!&\!\!\!\!\!\!&\!\!\!3\!\!\!&\!\!\!\!\!\!&\!\!\!3\!\!\!&\!\!\!\!\!\!&\!\!\!5\!\!\!&\!\!\!\!\!\!&\!\!\!3\!\!\!&\!\!\!\!\!\!&\!\!\!2\!\!\!&\!\!\!\!\!\!&\!\!\!\!\!\!&\!\!\!\!\!\!&\!\!\!\!\!\!&\!\!\!\!\!\!&\!\!\!\!\!\!&\!\!\!\!\!\!&\!\!\!\!\!\!&\!\!\!\!\!\!&\!\!\!\!\!\!&\!\!\!\\ 
\!\!\!&\!\!\!3\!\!\!&\!\!\!\!\!\!&\!\!\!\!\!\!&\!\!\!\!\!\!&\!\!\!\!\!\!&\!\!\!\!\!\!&\!\!\!\!\!\!&\!\!\!\!\!\!&\!\!\!\!\!\!&\!\!\!\!\!\!&\!\!\!\!\!\!&\!\!\!\!\!\!&\!\!\!\!\!\!&\!\!\!\!\!\!&\!\!\!\!\!\!&\!\!\!\!\!\!&\!\!\!1\!\!\!&\!\!\!\!\!\!&\!\!\!1\!\!\!&\!\!\!\!\!\!&\!\!\!3\!\!\!&\!\!\!\!\!\!&\!\!\!3\!\!\!&\!\!\!\!\!\!&\!\!\!3\!\!\!&\!\!\!\!\!\!&\!\!\!1\!\!\!&\!\!\!\!\!\!&\!\!\!\!\!\!&\!\!\!\!\!\!&\!\!\!\!\!\!&\!\!\!\!\!\!&\!\!\!\!\!\!&\!\!\!\!\!\!&\!\!\!\!\!\!&\!\!\!\!\!\!&\!\!\!\\ 
\!\!\!&\!\!\!$\frac{7}{2}$\!\!\!&\!\!\!\!\!\!&\!\!\!\!\!\!&\!\!\!\!\!\!&\!\!\!\!\!\!&\!\!\!\!\!\!&\!\!\!\!\!\!&\!\!\!\!\!\!&\!\!\!\!\!\!&\!\!\!\!\!\!&\!\!\!\!\!\!&\!\!\!\!\!\!&\!\!\!\!\!\!&\!\!\!\!\!\!&\!\!\!\!\!\!&\!\!\!\!\!\!&\!\!\!\!\!\!&\!\!\!\!\!\!&\!\!\!\!\!\!&\!\!\!1\!\!\!&\!\!\!\!\!\!&\!\!\!1\!\!\!&\!\!\!\!\!\!&\!\!\!2\!\!\!&\!\!\!\!\!\!&\!\!\!1\!\!\!&\!\!\!\!\!\!&\!\!\!1\!\!\!&\!\!\!\!\!\!&\!\!\!\!\!\!&\!\!\!\!\!\!&\!\!\!\!\!\!&\!\!\!\!\!\!&\!\!\!\!\!\!&\!\!\!\!\!\!&\!\!\!\!\!\!&\!\!\!\\ 
\!\!\!&\!\!\!\!4\!\!&\!\!\!\!\!\!&\!\!\!\!\!\!&\!\!\!\!\!\!&\!\!\!\!\!\!&\!\!\!\!\!\!&\!\!\!\!\!\!&\!\!\!\!\!\!&\!\!\!\!\!\!&\!\!\!\!\!\!&\!\!\!\!\!\!&\!\!\!\!\!\!&\!\!\!\!\!\!&\!\!\!\!\!\!&\!\!\!\!\!\!&\!\!\!\!\!\!&\!\!\!\!\!\!&\!\!\!\!\!\!&\!\!\!\!\!\!&\!\!\!\!\!\!&\!\!\!\!\!\!&\!\!\!\!\!\!&\!\!\!1\!\!\!&\!\!\!\!\!\!&\!\!\!1\!\!\!&\!\!\!\!\!\!&\!\!\!1\!\!\!&\!\!\!\!\!\!&\!\!\!\!\!\!&\!\!\!\!\!\!&\!\!\!\!\!\!&\!\!\!\!\!\!&\!\!\!\!\!\!&\!\!\!\!\!\!&\!\!\!\!\!\!&\!\!\!\!\!\!&\!\!\!\\ 
\!\!\!&\!\!\!$\frac{9}{2}$\!\!\!&\!\!\!\!\!\!&\!\!\!\!\!\!&\!\!\!\!\!\!&\!\!\!\!\!\!&\!\!\!\!\!\!&\!\!\!\!\!\!&\!\!\!\!\!\!&\!\!\!\!\!\!&\!\!\!\!\!\!&\!\!\!\!\!\!&\!\!\!\!\!\!&\!\!\!\!\!\!&\!\!\!\!\!\!&\!\!\!\!\!\!&\!\!\!\!\!\!&\!\!\!\!\!\!&\!\!\!\!\!\!&\!\!\!\!\!\!&\!\!\!\!\!\!&\!\!\!\!\!\!&\!\!\!\!\!\!&\!\!\!\!\!\!&\!\!\!\!\!\!&\!\!\!\!\!\!&\!\!\!1\!\!\!&\!\!\!\!\!\!&\!\!\!\!\!\!&\!\!\!\!\!\!&\!\!\!\!\!\!&\!\!\!\!\!\!&\!\!\!\!\!\!&\!\!\!\!\!\!&\!\!\!\!\!\!&\!\!\!\!\!\!&\!\!\!\!\!\!&\!\!\!\\ 
\!\!\!&\!\!\!5\!\!\!&\!\!\!\!\!\!&\!\!\!\!\!\!&\!\!\!\!\!\!&\!\!\!\!\!\!&\!\!\!\!\!\!&\!\!\!\!\!\!&\!\!\!\!\!\!&\!\!\!\!\!\!&\!\!\!\!\!\!&\!\!\!\!\!\!&\!\!\!\!\!\!&\!\!\!\!\!\!&\!\!\!\!\!\!&\!\!\!\!\!\!&\!\!\!\!\!\!&\!\!\!\!\!\!&\!\!\!\!\!\!&\!\!\!\!\!\!&\!\!\!\!\!\!&\!\!\!\!\!\!&\!\!\!\!\!\!&\!\!\!\!\!\!&\!\!\!\!\!\!&\!\!\!\!\!\!&\!\!\!\!\!\!&\!\!\!\!\!\!&\!\!\!\!\!\!&\!\!\!1\!\!\!&\!\!\!\!\!\!&\!\!\!\!\!\!&\!\!\!\!\!\!&\!\!\!\!\!\!&\!\!\!\!\!\!&\!\!\!\!\!\!&\!\!\!\!\!\!&\!\!\!\\ 
\hline 
7\!\!\!&\!\!\!0\!\!\!&\!\!\!\!\!\!&\!\!\!\!\!\!&\!\!\!6\!\!\!&\!\!\!\!\!\!&\!\!\!6\!\!\!&\!\!\!\!\!\!&\!\!\!12\!\!\!&\!\!\!\!\!\!&\!\!\!13\!\!\!&\!\!\!\!\!\!&\!\!\!19\!\!\!&\!\!\!\!\!\!&\!\!\!21\!\!\!&\!\!\!\!\!\!&\!\!\!26\!\!\!&\!\!\!\!\!\!&\!\!\!26\!\!\!&\!\!\!\!\!\!&\!\!\!26\!\!\!&\!\!\!\!\!\!&\!\!\!22\!\!\!&\!\!\!\!\!\!&\!\!\!15\!\!\!&\!\!\!\!\!\!&\!\!\!9\!\!\!&\!\!\!\!\!\!&\!\!\!4\!\!\!&\!\!\!\!\!\!&\!\!\!2\!\!\!&\!\!\!\!\!\!&\!\!\!\!\!\!&\!\!\!\!\!\!&\!\!\!\!\!\!&\!\!\!\!\!\!&\!\!\!\!\!\!&\!\!\!\\ 
\!\!\!&\!\!\!$\frac{1}{2}$\!\!\!&\!\!\!\!\!\!&\!\!\!4\!\!\!&\!\!\!\!\!\!&\!\!\!7\!\!\!&\!\!\!\!\!\!&\!\!\!12\!\!\!&\!\!\!\!\!\!&\!\!\!17\!\!\!&\!\!\!\!\!\!&\!\!\!24\!\!\!&\!\!\!\!\!\!&\!\!\!29\!\!\!&\!\!\!\!\!\!&\!\!\!37\!\!\!&\!\!\!\!\!\!&\!\!\!41\!\!\!&\!\!\!\!\!\!&\!\!\!45\!\!\!&\!\!\!\!\!\!&\!\!\!41\!\!\!&\!\!\!\!\!\!&\!\!\!35\!\!\!&\!\!\!\!\!\!&\!\!\!23\!\!\!&\!\!\!\!\!\!&\!\!\!13\!\!\!&\!\!\!\!\!\!&\!\!\!5\!\!\!&\!\!\!\!\!\!&\!\!\!1\!\!\!&\!\!\!\!\!\!&\!\!\!\!\!\!&\!\!\!\!\!\!&\!\!\!\!\!\!&\!\!\!\!\!\!&\!\!\!\\ 
\!\!\!&\!\!\!1\!\!\!&\!\!\!2\!\!\!&\!\!\!\!\!\!&\!\!\!3\!\!\!&\!\!\!\!\!\!&\!\!\!8\!\!\!&\!\!\!\!\!\!&\!\!\!11\!\!\!&\!\!\!\!\!\!&\!\!\!18\!\!\!&\!\!\!\!\!\!&\!\!\!23\!\!\!&\!\!\!\!\!\!&\!\!\!33\!\!\!&\!\!\!\!\!\!&\!\!\!40\!\!\!&\!\!\!\!\!\!&\!\!\!48\!\!\!&\!\!\!\!\!\!&\!\!\!50\!\!\!&\!\!\!\!\!\!&\!\!\!49\!\!\!&\!\!\!\!\!\!&\!\!\!39\!\!\!&\!\!\!\!\!\!&\!\!\!25\!\!\!&\!\!\!\!\!\!&\!\!\!12\!\!\!&\!\!\!\!\!\!&\!\!\!4\!\!\!&\!\!\!\!\!\!&\!\!\!1\!\!\!&\!\!\!\!\!\!&\!\!\!\!\!\!&\!\!\!\!\!\!&\!\!\!\!\!\!&\!\!\!\\ 
\!\!\!&\!\!\!$\frac{3}{2}$\!\!\!&\!\!\!\!\!\!&\!\!\!1\!\!\!&\!\!\!\!\!\!&\!\!\!3\!\!\!&\!\!\!\!\!\!&\!\!\!4\!\!\!&\!\!\!\!\!\!&\!\!\!9\!\!\!&\!\!\!\!\!\!&\!\!\!13\!\!\!&\!\!\!\!\!\!&\!\!\!21\!\!\!&\!\!\!\!\!\!&\!\!\!27\!\!\!&\!\!\!\!\!\!&\!\!\!38\!\!\!&\!\!\!\!\!\!&\!\!\!44\!\!\!&\!\!\!\!\!\!&\!\!\!50\!\!\!&\!\!\!\!\!\!&\!\!\!46\!\!\!&\!\!\!\!\!\!&\!\!\!38\!\!\!&\!\!\!\!\!\!&\!\!\!22\!\!\!&\!\!\!\!\!\!&\!\!\!10\!\!\!&\!\!\!\!\!\!&\!\!\!3\!\!\!&\!\!\!\!\!\!&\!\!\!1\!\!\!&\!\!\!\!\!\!&\!\!\!\!\!\!&\!\!\!\!\!\!&\!\!\!\\ 
\!\!\!&\!\!\!2\!\!\!&\!\!\!\!\!\!&\!\!\!\!\!\!&\!\!\!1\!\!\!&\!\!\!\!\!\!&\!\!\!1\!\!\!&\!\!\!\!\!\!&\!\!\!3\!\!\!&\!\!\!\!\!\!&\!\!\!5\!\!\!&\!\!\!\!\!\!&\!\!\!10\!\!\!&\!\!\!\!\!\!&\!\!\!14\!\!\!&\!\!\!\!\!\!&\!\!\!22\!\!\!&\!\!\!\!\!\!&\!\!\!29\!\!\!&\!\!\!\!\!\!&\!\!\!38\!\!\!&\!\!\!\!\!\!&\!\!\!41\!\!\!&\!\!\!\!\!\!&\!\!\!41\!\!\!&\!\!\!\!\!\!&\!\!\!31\!\!\!&\!\!\!\!\!\!&\!\!\!19\!\!\!&\!\!\!\!\!\!&\!\!\!7\!\!\!&\!\!\!\!\!\!&\!\!\!2\!\!\!&\!\!\!\!\!\!&\!\!\!\!\!\!&\!\!\!\!\!\!&\!\!\!\!\!\!&\!\!\!\\ 
\!\!\!&\!\!\!$\frac{5}{2}$\!\!\!&\!\!\!\!\!\!&\!\!\!\!\!\!&\!\!\!\!\!\!&\!\!\!\!\!\!&\!\!\!\!\!\!&\!\!\!1\!\!\!&\!\!\!\!\!\!&\!\!\!1\!\!\!&\!\!\!\!\!\!&\!\!\!3\!\!\!&\!\!\!\!\!\!&\!\!\!5\!\!\!&\!\!\!\!\!\!&\!\!\!10\!\!\!&\!\!\!\!\!\!&\!\!\!14\!\!\!&\!\!\!\!\!\!&\!\!\!22\!\!\!&\!\!\!\!\!\!&\!\!\!27\!\!\!&\!\!\!\!\!\!&\!\!\!34\!\!\!&\!\!\!\!\!\!&\!\!\!32\!\!\!&\!\!\!\!\!\!&\!\!\!26\!\!\!&\!\!\!\!\!\!&\!\!\!14\!\!\!&\!\!\!\!\!\!&\!\!\!6\!\!\!&\!\!\!\!\!\!&\!\!\!1\!\!\!&\!\!\!\!\!\!&\!\!\!\!\!\!&\!\!\!\!\!\!&\!\!\!\\ 
\!\!\!&\!\!\!3\!\!\!&\!\!\!\!\!\!&\!\!\!\!\!\!&\!\!\!\!\!\!&\!\!\!\!\!\!&\!\!\!\!\!\!&\!\!\!\!\!\!&\!\!\!\!\!\!&\!\!\!\!\!\!&\!\!\!1\!\!\!&\!\!\!\!\!\!&\!\!\!1\!\!\!&\!\!\!\!\!\!&\!\!\!3\!\!\!&\!\!\!\!\!\!&\!\!\!5\!\!\!&\!\!\!\!\!\!&\!\!\!10\!\!\!&\!\!\!\!\!\!&\!\!\!14\!\!\!&\!\!\!\!\!\!&\!\!\!21\!\!\!&\!\!\!\!\!\!&\!\!\!24\!\!\!&\!\!\!\!\!\!&\!\!\!26\!\!\!&\!\!\!\!\!\!&\!\!\!19\!\!\!&\!\!\!\!\!\!&\!\!\!11\!\!\!&\!\!\!\!\!\!&\!\!\!3\!\!\!&\!\!\!\!\!\!&\!\!\!1\!\!\!&\!\!\!\!\!\!&\!\!\!\!\!\!&\!\!\!\\ 
\!\!\!&\!\!\!$\frac{7}{2}$\!\!\!&\!\!\!\!\!\!&\!\!\!\!\!\!&\!\!\!\!\!\!&\!\!\!\!\!\!&\!\!\!\!\!\!&\!\!\!\!\!\!&\!\!\!\!\!\!&\!\!\!\!\!\!&\!\!\!\!\!\!&\!\!\!\!\!\!&\!\!\!\!\!\!&\!\!\!1\!\!\!&\!\!\!\!\!\!&\!\!\!1\!\!\!&\!\!\!\!\!\!&\!\!\!3\!\!\!&\!\!\!\!\!\!&\!\!\!5\!\!\!&\!\!\!\!\!\!&\!\!\!10\!\!\!&\!\!\!\!\!\!&\!\!\!13\!\!\!&\!\!\!\!\!\!&\!\!\!18\!\!\!&\!\!\!\!\!\!&\!\!\!18\!\!\!&\!\!\!\!\!\!&\!\!\!15\!\!\!&\!\!\!\!\!\!&\!\!\!7\!\!\!&\!\!\!\!\!\!&\!\!\!2\!\!\!&\!\!\!\!\!\!&\!\!\!\!\!\!&\!\!\!\!\!\!&\!\!\!\\ 
\!\!\!&\!\!\!4\!\!\!&\!\!\!\!\!\!&\!\!\!\!\!\!&\!\!\!\!\!\!&\!\!\!\!\!\!&\!\!\!\!\!\!&\!\!\!\!\!\!&\!\!\!\!\!\!&\!\!\!\!\!\!&\!\!\!\!\!\!&\!\!\!\!\!\!&\!\!\!\!\!\!&\!\!\!\!\!\!&\!\!\!\!\!\!&\!\!\!\!\!\!&\!\!\!1\!\!\!&\!\!\!\!\!\!&\!\!\!1\!\!\!&\!\!\!\!\!\!&\!\!\!3\!\!\!&\!\!\!\!\!\!&\!\!\!5\!\!\!&\!\!\!\!\!\!&\!\!\!9\!\!\!&\!\!\!\!\!\!&\!\!\!11\!\!\!&\!\!\!\!\!\!&\!\!\!13\!\!\!&\!\!\!\!\!\!&\!\!\!9\!\!\!&\!\!\!\!\!\!&\!\!\!5\!\!\!&\!\!\!\!\!\!&\!\!\!1\!\!\!&\!\!\!\!\!\!&\!\!\!\!\!\!&\!\!\!\\ 
\!\!\!&\!\!\!$\frac{9}{2}$\!\!\!&\!\!\!\!\!\!&\!\!\!\!\!\!&\!\!\!\!\!\!&\!\!\!\!\!\!&\!\!\!\!\!\!&\!\!\!\!\!\!&\!\!\!\!\!\!&\!\!\!\!\!\!&\!\!\!\!\!\!&\!\!\!\!\!\!&\!\!\!\!\!\!&\!\!\!\!\!\!&\!\!\!\!\!\!&\!\!\!\!\!\!&\!\!\!\!\!\!&\!\!\!\!\!\!&\!\!\!\!\!\!&\!\!\!1\!\!\!&\!\!\!\!\!\!&\!\!\!1\!\!\!&\!\!\!\!\!\!&\!\!\!3\!\!\!&\!\!\!\!\!\!&\!\!\!5\!\!\!&\!\!\!\!\!\!&\!\!\!8\!\!\!&\!\!\!\!\!\!&\!\!\!8\!\!\!&\!\!\!\!\!\!&\!\!\!7\!\!\!&\!\!\!\!\!\!&\!\!\!3\!\!\!&\!\!\!\!\!\!&\!\!\!1\!\!\!&\!\!\!\!\!\!&\!\!\!\\ 
\!\!\!&\!\!\!5\!\!\!&\!\!\!\!\!\!&\!\!\!\!\!\!&\!\!\!\!\!\!&\!\!\!\!\!\!&\!\!\!\!\!\!&\!\!\!\!\!\!&\!\!\!\!\!\!&\!\!\!\!\!\!&\!\!\!\!\!\!&\!\!\!\!\!\!&\!\!\!\!\!\!&\!\!\!\!\!\!&\!\!\!\!\!\!&\!\!\!\!\!\!&\!\!\!\!\!\!&\!\!\!\!\!\!&\!\!\!\!\!\!&\!\!\!\!\!\!&\!\!\!\!\!\!&\!\!\!\!\!\!&\!\!\!1\!\!\!&\!\!\!\!\!\!&\!\!\!1\!\!\!&\!\!\!\!\!\!&\!\!\!3\!\!\!&\!\!\!\!\!\!&\!\!\!4\!\!\!&\!\!\!\!\!\!&\!\!\!6\!\!\!&\!\!\!\!\!\!&\!\!\!4\!\!\!&\!\!\!\!\!\!&\!\!\!2\!\!\!&\!\!\!\!\!\!&\!\!\!\!\!\!&\!\!\!\\ 
\!\!\!&\!\!\!$\frac{11}{2}$\!\!\!&\!\!\!\!\!\!&\!\!\!\!\!\!&\!\!\!\!\!\!&\!\!\!\!\!\!&\!\!\!\!\!\!&\!\!\!\!\!\!&\!\!\!\!\!\!&\!\!\!\!\!\!&\!\!\!\!\!\!&\!\!\!\!\!\!&\!\!\!\!\!\!&\!\!\!\!\!\!&\!\!\!\!\!\!&\!\!\!\!\!\!&\!\!\!\!\!\!&\!\!\!\!\!\!&\!\!\!\!\!\!&\!\!\!\!\!\!&\!\!\!\!\!\!&\!\!\!\!\!\!&\!\!\!\!\!\!&\!\!\!\!\!\!&\!\!\!\!\!\!&\!\!\!1\!\!\!&\!\!\!\!\!\!&\!\!\!1\!\!\!&\!\!\!\!\!\!&\!\!\!3\!\!\!&\!\!\!\!\!\!&\!\!\!3\!\!\!&\!\!\!\!\!\!&\!\!\!3\!\!\!&\!\!\!\!\!\!&\!\!\!1\!\!\!&\!\!\!\!\!\!&\!\!\!\\ 
\!\!\!&\!\!\!6\!\!\!&\!\!\!\!\!\!&\!\!\!\!\!\!&\!\!\!\!\!\!&\!\!\!\!\!\!&\!\!\!\!\!\!&\!\!\!\!\!\!&\!\!\!\!\!\!&\!\!\!\!\!\!&\!\!\!\!\!\!&\!\!\!\!\!\!&\!\!\!\!\!\!&\!\!\!\!\!\!&\!\!\!\!\!\!&\!\!\!\!\!\!&\!\!\!\!\!\!&\!\!\!\!\!\!&\!\!\!\!\!\!&\!\!\!\!\!\!&\!\!\!\!\!\!&\!\!\!\!\!\!&\!\!\!\!\!\!&\!\!\!\!\!\!&\!\!\!\!\!\!&\!\!\!\!\!\!&\!\!\!\!\!\!&\!\!\!\!\!\!&\!\!\!1\!\!\!&\!\!\!\!\!\!&\!\!\!1\!\!\!&\!\!\!\!\!\!&\!\!\!2\!\!\!&\!\!\!\!\!\!&\!\!\!1\!\!\!&\!\!\!\!\!\!&\!\!\!1\!\!\!&\!\!\!\\ 
\!\!\!&\!\!\!$\frac{13}{2}$\!\!\!&\!\!\!\!\!\!&\!\!\!\!\!\!&\!\!\!\!\!\!&\!\!\!\!\!\!&\!\!\!\!\!\!&\!\!\!\!\!\!&\!\!\!\!\!\!&\!\!\!\!\!\!&\!\!\!\!\!\!&\!\!\!\!\!\!&\!\!\!\!\!\!&\!\!\!\!\!\!&\!\!\!\!\!\!&\!\!\!\!\!\!&\!\!\!\!\!\!&\!\!\!\!\!\!&\!\!\!\!\!\!&\!\!\!\!\!\!&\!\!\!\!\!\!&\!\!\!\!\!\!&\!\!\!\!\!\!&\!\!\!\!\!\!&\!\!\!\!\!\!&\!\!\!\!\!\!&\!\!\!\!\!\!&\!\!\!\!\!\!&\!\!\!\!\!\!&\!\!\!\!\!\!&\!\!\!\!\!\!&\!\!\!1\!\!\!&\!\!\!\!\!\!&\!\!\!1\!\!\!&\!\!\!\!\!\!&\!\!\!1\!\!\!&\!\!\!\!\!\!&\!\!\!\\ 
\!\!\!&\!\!\!7\!\!\!&\!\!\!\!\!\!&\!\!\!\!\!\!&\!\!\!\!\!\!&\!\!\!\!\!\!&\!\!\!\!\!\!&\!\!\!\!\!\!&\!\!\!\!\!\!&\!\!\!\!\!\!&\!\!\!\!\!\!&\!\!\!\!\!\!&\!\!\!\!\!\!&\!\!\!\!\!\!&\!\!\!\!\!\!&\!\!\!\!\!\!&\!\!\!\!\!\!&\!\!\!\!\!\!&\!\!\!\!\!\!&\!\!\!\!\!\!&\!\!\!\!\!\!&\!\!\!\!\!\!&\!\!\!\!\!\!&\!\!\!\!\!\!&\!\!\!\!\!\!&\!\!\!\!\!\!&\!\!\!\!\!\!&\!\!\!\!\!\!&\!\!\!\!\!\!&\!\!\!\!\!\!&\!\!\!\!\!\!&\!\!\!\!\!\!&\!\!\!\!\!\!&\!\!\!\!\!\!&\!\!\!1\!\!\!&\!\!\!\!\!\!&\!\!\!\!\!\!&\!\!\!\\ 
\!\!\!&\!\!\!$\frac{15}{2}$\!\!\!&\!\!\!\!\!\!&\!\!\!\!\!\!&\!\!\!\!\!\!&\!\!\!\!\!\!&\!\!\!\!\!\!&\!\!\!\!\!\!&\!\!\!\!\!\!&\!\!\!\!\!\!&\!\!\!\!\!\!&\!\!\!\!\!\!&\!\!\!\!\!\!&\!\!\!\!\!\!&\!\!\!\!\!\!&\!\!\!\!\!\!&\!\!\!\!\!\!&\!\!\!\!\!\!&\!\!\!\!\!\!&\!\!\!\!\!\!&\!\!\!\!\!\!&\!\!\!\!\!\!&\!\!\!\!\!\!&\!\!\!\!\!\!&\!\!\!\!\!\!&\!\!\!\!\!\!&\!\!\!\!\!\!&\!\!\!\!\!\!&\!\!\!\!\!\!&\!\!\!\!\!\!&\!\!\!\!\!\!&\!\!\!\!\!\!&\!\!\!\!\!\!&\!\!\!\!\!\!&\!\!\!\!\!\!&\!\!\!\!\!\!&\!\!\!\!\!\!&\!\!\!\!1\\ 
\hline 
d\!&\!\!$j_l\slash j_r$\!&\!\!0\!&\!\!\!$\frac{1}{2}$\!\!\!&\!\!\!1\!\!\!&\!\!\!$\frac{3}{2}$\!\!\!&\!\!\!2\!\!\!&\!\!\!$\frac{5}{2}$\!\!\!&\!\!\!3\!\!\!&\!\!\!$\frac{7}{2}$\!\!\!&\!\!\!4\!\!\!& 
\!\!\!$\frac{9}{2}$\!\!\!&\!\!\!5\!\!\!&\!\!\!$\frac{11}{2}$\!\!\!&\!\!\!6\!\!\!&\!\!\!$\frac{13}{2}$\!\!\!&\!\!\!7\!\!\!&\!\!\!$\frac{15}{2}$\!\!\!&\!\!\!8\!\!\!&\!\!\!$\frac{17}{2}$\!\!\! 
&\!\!\!9\!\!\!&\!\!\!$\frac{19}{2}$\!\!\!&\!\!\!10\!\!\!&\!\!\!$\frac{21}{2}$\!\!\!&\!\!\!11\!\!\!&\!\!\!$\frac{23}{2}$\!\!\!&\!\!\!12\!\!\!&\!\!\!$\frac{25}{2}$\!\!\!&\!\!\!13\!\!\! 
&\!\!\!$\frac{27}{2}$\!\!\!&\!\!\!14\!\!\!&\!\!\!$\frac{29}{2}$\!\!\!&\!\!\!15\!\!\!&\!\!\!$\frac{31}{2}$\!\!\!&\!\!\!16\!\!\!&\!\!\!$\frac{33}{2}$\!\!\!&\!\!\!17\!\!\!&\!\!\!$\frac{35}{2}$\!\!\!\\ 
\hline 
\end{tabular}}} 
\caption{Non vanishing BPS numbers $N^d_{j_l,j_r}$ of local ${\cal O}(-3)\rightarrow \mathbb{P}^2$} 
\end{table}  
 
\subsection{Orbifold and conifold expansions} 
 
As explained in section \ref{wavefunction}, the generalized holomorphic anomaly  
equations (\ref{hae}) can be interpreted as guaranteeing the wave function  
transformation properties of the partition function $Z=e^{F}$. We can therefore  
apply the formalism described in~\cite{ABK} to define the holomorphic 
$A$-model expansion as counting function of possible $A$-model invariants .

\subsubsection{The refined theory near the conifold point} 
\label{conifoldinvariantsp2} 
As mentioned in section \ref{fixingtheambiguity}, $F$ in the strict conifold  
limit is equivalent to a double scaling limit of the  free energy of the   
$c=1$ string compactified on $S^1$, where the flat coordinate  
\begin{equation}  
t_c =\delta +\frac{11 \delta ^2}{18}+\frac{109 \delta ^3}{243}+\frac{9389 \delta ^4}{26244}+O\left(\delta ^5\right)\ , 
\end{equation}  
with $\delta=1 + 27 z$ playing the role of the cosmological constant $\Delta$ and $\epsilon_1=\frac{1}{\beta \mu}$,  
$\epsilon_2=-\frac{1}{2 \beta \mu R}$ in the notation of~\cite{Gross:1990ub}. The topological  
string specialization corresponds to taking the self-dual radius  
$R=\frac{1}{2}$ in units of $\alpha'$. In real polarisation, the holomorphic  
limit of $S_c$ is simpler than that of $X_c$  
and reads    
\begin{equation}  \label{sc} 
S_c= -\frac{1}{2}+\frac{4 t_c}{3}-\frac{103 t_c^2}{54}+\frac{317 t_c^3}{162}+O\left(t_c^4\right)\ . 
\end{equation} 
As (\ref{p2n+g=2}) and (\ref{propagatorp2}) are global relations, we can obtain $X_c$ from the expansion (\ref{sc}) of $S_c$, and substitute this as well as $z(\delta(t_c))$ into the globally defined $F^{n,g}(X,z)$, yielding\footnote{With $a_1=\frac{3 \log(3)+1}{2 \pi i}$, $a_0=-\frac{\pi}{3} -1.678699904 i= 
\frac{1}{i\sqrt{3}\Gamma\left(\frac{1}{3}\right)\Gamma\left(\frac{2}{3}\right)}G^{3\,3}_{2\,2} 
\left({{\frac{1}{3}\ \frac{2}{3} \ 1} \atop {0 \ 0 \ 0}}\biggr| -1\right)$. The constant  
$c_{1,0}= \frac{1}{24} (\pi i  + 3 \log(3))$ and $c_{0,1}$ depends on a further regularization. } 
\ban 
F^{(0,0)}_c&=&c_{0,0} + \frac{a_0}{3} t_c+\left(\frac{a_1}{6}-\frac{1}{12}\right) t_c^2+ t_c^2 \frac{\log (t_c)}{6}-\frac{t_c^3}{324} + \frac{t_c^4}{69984} +O\left(t_c^5\right) \nonumber\\  
F^{(1,0)}_c&=&c_{1,0}+\frac{\log (t_c)}{24}+\frac{7 t_c}{432}+\frac{t_c^2}{46656}-\frac{19 t_c^3}{314928}+\frac{439 t_c^4}{50388480}+O\left(t_c^5\right)\nonumber\\ 
F^{(0,1)}_c&=&c_{0,1}-\frac{\log (t_c)}{12}+\frac{5 t_c}{216}-\frac{t_c^2}{23328}-\frac{5 t_c^3}{157464}+\frac{283 t_c^4}{75582720}+O\left(t_c^5\right)\nonumber\\ 
F^{(2,0)}_c&=&-\frac{7}{1920 t_c^2}+\frac{1906-189 \chi}{3732480}+\frac{1169 t_c}{12597120}-\frac{61303 t_c^2}{3023308800}+\frac{16153 t_c^3}{6122200320}+O\left(t_c^4\right)\nonumber\\ 
F^{(1,1)}_c&=&\frac{7}{480 t_c^2}-\frac{974+189 \chi}{933120}+\frac{631 t_c}{3149280}-\frac{29897 t_c^2}{755827200}+\frac{7247 c^3}{1530550080}+O\left(t_c^4\right)\nonumber\\  
F^{(0,2)}_c&=&-\frac{1}{80 t_c^2}+\frac{9 \chi-26}{51840}+\frac{t_c}{19440}-\frac{3187 t_c^2}{377913600}+\frac{239 t_c^3}{255091680}+O\left(t_c^4\right)\ . 
\ean 
The transformations of $F^{(0,0)}(t_c)$ and $F^{(0,g)}(t_c)$ have been worked out in~\cite{Haghighat:2008gw}.  
 
It remains a challenge to calculate the coefficients of the finite expansion from the correlations  
functions of the $c=1$ string at arbitrary radius.  
 
\subsubsection{The refined theory near the orbifold point} 
\label{orbifoldinvariantsp2} 
Near the orbifold point, the predictions of orbifold Gromov-Witten theory made in~\cite{ABK}  
have subsequently been verified by mathematicians, see \cite{Bouchard:2007nr}  
for a review. One might hope that the refined invariants that can be defined in  
the $B$-model also have an interpretation in terms of the orbifold $A$-model.  
The flat coordinate $t_o$ is given by  
\begin{equation} 
t_o=3w-\frac{w^4}{8}+\frac{4 w^7}{105}-\frac{49 w^{10}}{2700}+\frac{245 w^{13}}{23166}+O\left(w^{14}\right), 
\end{equation} 
where  $w=-\frac{(-1)^\frac{1}{3}}{3 z^\frac{1}{3}}$. Furthermore, the propagator $S_o$ at the orbifold point is simpler than the variable $X$ expanded around this point, and given by 
\begin{equation}  
S_o= \frac{t_o^5}{43740}-\frac{7 t_o^8}{28343520}+\frac{16039 t_o^{11}}{3535570684800}+O\left(t_o^{14}\right)\ .  
\end{equation} 
With these definitions, we extract the orbifold expansions up to $n+g=2$ 
\ban  
F^{(0,0)}_o&=&\frac{t_o^3}{3 \cdot 3!} - \frac{t_o^6}{3^3 6!} + \frac{t_o^9}{3^2 9!} - \frac{1093\,t_o^{12}}{3^6 12!} + \frac{119401\, t_0^{15}}{3^7 15!}+ O\left({t_o}^{18}\right) \nonumber \\ 
F^{(1,0)}_o&=&\frac{\log (3)}{8}+\frac{{t_o}^3}{2^2 3^3 3!}-\frac{5{t_o}^6}{2^2 3^4 6!}+\frac{1319 {t_o}^9}{2^2 3^7 9!}-\frac{114983 {t_o}^{12}}{2^23^8 12!}+O\left({t_o}^{15}\right) \nonumber\\  
F^{(0,1)}_o&=&\frac{{t_o}^6}{3^5 6!}-\frac{14 {t_o}^9}{3^5 9!}+\frac{13007 {t_o}^{12}}{3^8 12!} -\frac{8354164 t_0^{15}}{3^{10}15!}+  O\left({t_o}^{18}\right) \nonumber\\  
F^{(2,0)}_o&=& \frac{7 \chi-192}{138240}+\frac{{t_o}^3}{2^3 3^3 5\cdot 3!}-\frac{79 {t_o}^6}{2^43^6 6!}+\frac{29 {t_o}^9}{2^2 3^25\cdot 9!}-\frac{4656751 {t_o}^{12}}{ 2^4 3^85\cdot 12!}+O\left({t_o}^{15}\right)\nonumber\\  
F^{(1,1)}_o&=&\frac{-7 \chi-48}{34560}-\frac{{t_o}^3}{2^23^45\cdot 3!}+\frac{7 {t_o}^6}{2^2 3^56!}-\frac{8933 {t_o}^9}{2^23^7 5\cdot  9!}+\frac{ 1628851{t_o}^{12}}{ 2^23^8 5\cdot 12!}+O\left({t_o}^{15}\right)\nonumber\\  
F^{(0,2)}_o&=&\frac{3 \chi-8}{17280}+\frac{{t_o}^3}{2^43^{10} 5\cdot 3!}-\frac{13 {t_o}^6}{ 2^4 3^6 6!}+\frac{20693 {t_o}^9}{ 2^4 3^{13}  5\cdot 9!}-\frac{12803923{t_o}^{12}}{ 2^4 3^{14}  5\cdot  12! }+O\left({t_o}^{15}\right) \,. \nn\\ 
\ean 
Note that the $F^{(0,0)}$ coefficients are calculated independently in the A-model in \cite{CIT}. It would be  
interesting to calculate higher genus invariants for the topological string and even more so to understand   
the $\Omega$-deformation at the orbifold in the A-model context or in a brane description.

\section{Conclusion} 
 
In this paper, we gave evidence that the generalized holomorphic anomaly equations (\ref{gen_hol_ano}) hold for  
all 4d rigid $N=2$ theories for which the B-model geometry  
is given by a non-compact Calabi-Yau geometry. The full extent of the latter  
class is not known, but it is possible that they exhaust all rigid $N=2$ theories.  
It does include e.g. all non-compact toric Calabi-Yau manifolds, the 3 dimensional orbifold  
singularities~\cite{MR1169227}, and the canonical affine hypersurface  
singularities classified in~\cite{MR605348}. Many specific examples that have been studied involve additional fibration structures like the singularities  
of elliptic fibrations discussed in F-theory, as well as ADE singularities  
fibered over $\mathbb{P}^1$, familar from the heterotic/type II  
dual~\cite{Klemm:1996bj} construction of gauge theories. Within  
the class of toric singularities, many reproduce $N=2$ gauge theories with  
matter~\cite{KKV} on  
subslices in their moduli space, including in more general cases a  
wide variety of quiver gauge theories~\cite{Katz:1997eq} (arguably all with special unitary gauge groups).       
 
We showed that the boundary conditions given by the gap condition suffice to fix the holomorphic ambiguity  
and therefore determine the refined partition function if deformations exist that split  
all singularities into conifold singularities. Our formalism makes  
strong use of the discrete automorphic groups acting on the  
moduli space of the geometries and organizes the refined  partition  
functions in terms of generators of the latter. In a somewhat similar context, relevant deformations of a conformal Landau-Ginzburg theory  
were recently used to describe generating functions of higher genus  
Gromov-Witten invariants on orbifolds of $\mathbb{P}^1$~\cite{Yongbin}. 
It should be straightforward to calculate the refined invariants  
by (\ref{gen_hol_ano}).  
 
Note that the amplitudes $F^{(n,g)}$ at genus $0$ -- one can think of these as pure $s$-deformations of a classical theory deformed along the $(s,g_s)$ plane -- can be solved independently by  
a specialization of the holomorphic anomaly equations, together with  
the gap condition specialized to the genus 0 sector. This subsector of the theory should have a  
universal meaning in the integrable models related to the  
Nekrasov-Shatashvili limit. If one further restricts to the conformal cases, the pure $s$-deformation satisfies the same holomorphic anomaly equations (\ref{gen_hol_ano-cf1}) as the mass deformations  
of $F^{(0,0)}$. This together with the observation that unlike all 
other directions in deformation space, it by (\ref{NSconformal}) appears to only depend on the monodromy group in the  
conformal limit suggests that this direction might be described  
by an isomonodromic deformation of the geometry. 
 
As a general feature in our study, we saw that the leading power in the anholomorphic generator  
grows more slowly in conformal than in non-conformal theories. In this sense,  
the breaking of holomorphicity is weaker in conformal theories, suggesting a  
relation between its breaking and the breaking of conformal invariance. Note that the anholomorphicity of the amplitudes $F^{(n,g)}$ has not yet found a  
satisfactory explanation in the literature from a purely field theoretic vantage point.  
The 4d/2d correspondence implies that the holomorphic anomaly equations 
(\ref{gen_hol_ano}) govern 2d theories as well, and are perhaps easier to understand  
in this context. In related works~\cite{Jan1,Jan2}, such anholomorphicities  
are seen to arise due to regularization prescriptions of the path integral.  
The softer anholomorphicities we find for conformal theories, which tend to have  
better convergence properties, are consistent with this finding.

\section{Acknowledgements} 
It is a pleasure to thank Gaetan Borot, Sheldon Katz, Christoph Keller, Marcos Mari\~no,  
Yongbin Ruan, Jake Solomon, Yuji Tachikawa, Jan Troost, and Johannes Walcher for discussions.  
Further, we would like to thank Gaetan Borot, Sheldon Katz, Christoph Keller, Rahul  
Pandharipande, and Cumrun Vafa for correspondences and Thomas Wotschke  for reading  
the draft. 

AK would like to thank the IPMU and LPTENS for their great hospitality. AKKP would 
like to thank the University of Bonn for its kind hospitality. MH would like to
thank the HCM Bonn, Zhejiang University, University of Science and Technology 
of China for hospitality where parts of this work was conducted. The work of AK is supported 
by DFG contract KL 2271/1-1. 

\appendix

\section{The instanton partition function} \label{formalismsec}

The Nekrasov partition function is computed by localization of integrals over the moduli space of instantons. It can be written as a sum over 2d Young tableaux.  A 2d Young tableau $Y$ can be represented by a sequence of non-negative non-increasing integers $Y_{,1}\geq Y_{,2} \geq \cdots \geq 0$, with the total number of boxes $|Y|\equiv \sum_{i=0}^{\infty} Y_{,i} $ finite.  Denote
\begin{eqnarray} 
E^{Y_1,Y_2}_{i,j}(a)  \equiv  a+\epsilon_1 (Y_{1,j}^T-i+1) -\epsilon_2 (Y_{2,i}-j)  \,,
\end{eqnarray}
where $Y^T$ is the transpose of the Young tableau, and $\epsilon_1$ and $\epsilon_2$ are the deformation parameters of the $\Omega$-background. For the $SU(2)$ case, we have a single period $a$ which is the flat coordinate in the large modulus limit. The instanton part of the Nekrasov function can be written as sums over the boxes of Young tableaux as follows,
\begin{eqnarray}  \label{NK1}
&& Z_{\textrm{instanton}} (a,\epsilon_1,\epsilon_2)  
= \sum_{Y_1,Y_2}   \lambda^{|Y_1|+|Y_2|}  \\ 
&&\times  \prod_{(i,j)\in Y_1} \frac{\prod_{k=1}^{N_f} (a+\epsilon_1(i-1)+\epsilon_2(j-1)+\hat{m}_k)}{E^{Y_1,Y_1}_{i,j}(0)(\epsilon-E^{Y_1,Y_1}_{i,j}(0))E^{Y_1,Y_2}_{i,j}(2a)(\epsilon-E^{Y_1,Y_2}_{i,j}(2a))}  \nonumber \\ 
&&\times  \prod_{(i,j)\in Y_2} \frac{\prod_{k=1}^{N_f} (-a+\epsilon_1(i-1)+\epsilon_2(j-1)+\hat{m}_k)}{E^{Y_2,Y_2}_{i,j}(0)(\epsilon-E^{Y_2,Y_2}_{i,j}(0))E^{Y_2,Y_1}_{i,j}(-2a)(\epsilon-E^{Y_2,Y_1}_{i,j}(-2a))}  \nonumber    \,,
\end{eqnarray}
where $\epsilon=\epsilon_1+\epsilon_2$, and the $\hat{m}_k$'s are the mass parameters of massive flavors in the fundamental representation, related to the masses $m_k$ that appear in the Seiberg-Witten curve via the shift (\ref{shift}). The $n$-instanton contributions are given by the sum over Young tableaux $Y_1$ and $Y_2$ whose total number of boxes is $n$, $|Y_1|+|Y_2|=n$.  The parameter $\lambda$ keeps track of the instanton number. It serves as the expansion parameter in the computation of the free energy $F=\log(Z)$. In the asymptotically free cases of $N_f=1,2,3$, the parameter $\lambda$  is a dimensionful parameter proportional to a power of the strong coupling scale. For convenience, we can set it to a fixed numerical value; it can be recovered easily by dimensional analysis. For conformal theories, it is dimensionless and is related to the bare gauge coupling constant  $\lambda= q_0 =e^{2\pi i\tau_0}$.   

We can also consider the case where there is one hypermultiplet of mass $m$ in the adjoint representation of the gauge group. In this case, the theory is $\mathcal{N}=4$ supersymmetric if the adjoint matter is massless. The mass term breaks the $\mathcal{N}=4$ supersymmetry to $\mathcal{N}=2$, yielding a theory referred to as $N=2^*$. The Nekrasov partition function in this case is 
\begin{eqnarray}  \label{NK1adjoint}
&& Z_{\textrm{instanton}} (a,\epsilon_1,\epsilon_2)  
= \sum_{Y_1,Y_2}   \lambda^{|Y_1|+|Y_2|}  \\  
&&  \prod_{(i,j)\in Y_1} \frac{ E^{Y_1,Y_1}_{i,j}(-\hat{m})(\epsilon-E^{Y_1,Y_1}_{i,j}(\hat{m}))E^{Y_1,Y_2}_{i,j}(2a-\hat{m})(\epsilon-E^{Y_1,Y_2}_{i,j}(2a+\hat{m}))  }{E^{Y_1,Y_1}_{i,j}(0)(\epsilon-E^{Y_1,Y_1}_{i,j}(0))E^{Y_1,Y_2}_{i,j}(2a)(\epsilon-E^{Y_1,Y_2}_{i,j}(2a))}  \nonumber \\ 
&& \prod_{(i,j)\in Y_2} \frac{E^{Y_2,Y_2}_{i,j}(-\hat{m})(\epsilon-E^{Y_2,Y_2}_{i,j}(\hat{m}))E^{Y_2,Y_1}_{i,j}(-2a-\hat{m})(\epsilon-E^{Y_2,Y_1}_{i,j}(-2a+\hat{m})) }{E^{Y_2,Y_2}_{i,j}(0)(\epsilon-E^{Y_2,Y_2}_{i,j}(0))E^{Y_2,Y_1}_{i,j}(-2a)(\epsilon-E^{Y_2,Y_1}_{i,j}(-2a))}  \nonumber  \,. 
\end{eqnarray} 

The perturbative part of the Nekrasov function $Z_{\textrm{pert}}=e^{F_\textrm{pert}}$ with massive fundamental flavors is computed via
\begin{eqnarray} \label{NK2}
F_\textrm{pert} &=& -\gamma(2a)-\gamma(-2a)+\sum_{i=1}^{N_f} [\gamma(a+\hat{m}_i) + \gamma(-a+\hat{m}_i) ],   \nonumber \\
\gamma(x) &\equiv &   \frac{d}{ds} \frac{1}{\Gamma(s)} \int_0^\infty dt \frac{t^{s-1}e^{-tx}}
{(e^{-\epsilon_1t}-1) (e^{-\epsilon_2t}-1)  } \arrowvert_{s=0} \,,
\end{eqnarray}
while in the case of adjoint matter of mass $m$, the perturbative part is 
\begin{eqnarray}  \label{NK2adjoint}
F_\textrm{pert} &=& -\gamma(2a)-\gamma(-2a)+ \gamma(2a+\hat{m}) + \gamma(-2a+\hat{m}) \,.
\end{eqnarray}

The total Nekrasov partition function includes the perturbative and the instanton parts $Z=Z_{\textrm{pert}}Z_{\textrm{instanton}}$.

\bibliography{hkk}
\bibliographystyle{utcaps}

\end{document}